**Magnetically Guided Sequential Multi-Filler Alignment Enables Anisotropic Heat Pathways in Stretchable Composites for Soft Electronics**

*Altynay Kaidarova, Viktor Naenen, Isidro Florenciano, Ruben Windey, Xiaoyan Qing, Laurens Snels, Juhyung Park, Tom Van der Donck, Jeroen Soete, David Seveno and Francisco Molina-Lopez**

A. Kaidarova, V. Naenen, I. Florenciano, R. Windey, X. Qing, L. Snels, J. Park, T. Van der Donck, J. Soete, D. Seveno, F. Molina-Lopez
Department of Materials Engineering, KU Leuven, Kasteelpark Arenberg 44, 3001, Leuven, Belgium.
E-mail: francisco.molinalopez@kuleuven.be

Funding: A. K. received funding from the European Union's Horizon Europe research and innovation programme under the Marie Skłodowska-Curie grant agreement No 101153244. Views and opinions expressed are however those of the author(s) only and do not necessarily reflect those of the European Union or the European Research Executive Agency. Research was also supported by the European Research Council (ERC) with the H2020-3DALIGN project (grant number: 948922). V.N supported by FWO Research Foundation – Flanders, project number 11K9223N. Neither the European Union nor the granting authority can be held responsible for them. I.F. acknowledges the support of the Internal Funds KU Leuven, C1 project C14/21/078. X.Q. acknowledges the support of the Internal Funds KU Leuven, PDMT2/25/063. L.S acknowledge the FWO Research Foundation–Flanders, project S008023N. J.P received funding from the European Union's Horizon Europe research and innovation programme under the Marie Skłodowska-Curie grant agreement No 101204463.

Keywords: thermal interface materials, magnetically aligned composites, liquid metal, nickel particles, anisotropic thermal conductivity, thermoelectric generators, skin electronics.

Stretchable, skin-conformal electronics demand thermal interface materials that combine high through-plane heat transport with mechanical softness and lateral electrical insulation. In particle-filled composites, these requirements are typically antagonistic. Heat transport is governed not only by the intrinsic thermal conductivity of the fillers but, more critically, by thermal resistance at the interparticle interface. As the interparticle distance is reduced by increasing the particle content, mechanical reinforcement and in-plane electrical conductivity become unavoidable. The existing particle alignment processing strategies result in inefficient uncoupling of multifunctionality. Here, we present a hierarchical two-step process strategy to induce strong conductivity anisotropy in a soft composite. First, an align-before-blend process forms a scaffold of magnetically aligned nickel (Ni) microparticles in air before polymer infiltration. The absence of the viscous elastomer matrix during the particle alignment facilitates handling large amounts of particles, which assemble into highly ordered and densely packed vertical heat-conduction pathways. Second, a liquid metal (LM)/Ni mixture is infiltrated into the aligned scaffold to bridge interparticle gaps, minimizing interparticle thermal resistance. The composite reaches 7.5 W/(m·K) through-plane thermal conductivity with $<10^{-12}$ S/m lateral electrical insulation and ultrasoft mechanical compliance (~160 kPa), enabling coupling layers in thermoelectric generators and finger-mounted heat-spreading pads for wearable technologies.

## 1. Introduction

As electronic systems continue to miniaturize and operate at higher power densities, thermal management has emerged as a key factor in ensuring device reliability, with thermal failures responsible for over half of reported malfunction[1,2]. The increasing complexity of multilayered architectures also leads to localized heat accumulation, necessitating effective thermal pathways to dissipate energy from active regions and maintain stable temperature gradients across device layers [3,4]. Directional heat transport is especially critical in multilayer circuit boards[5,6], light-emitting[7–9] and laser diodes[8,9], thermoelectric generators[10,11] and 5G communication modules[12,13], where vertical thermal conduction is essential for performance and reliability. A distinct set of requirements applies to soft and hybrid electronics, where materials must simultaneously deliver efficient heat dissipation, stretchability and skin compatibility[14–16]. For skin-conformal electronics, recommended matrix moduli are comparable to or below the reported human-skin range (approximately 140–600 kPa), minimizing mechanical mismatch and enabling synchronous deformation with the skin[17].

Thermal interface materials (TIMs) are widely employed to minimize contact resistance and facilitate heat transfer between heat sources and heat spreaders. While conventional TIMs, such as greases[18], pastes [19], gels[20], and adhesive tapes[21–23], perform well in rigid electronic permanent assemblies, where standardized thermal resistance, bond-line thickness, and pressure stability dominate qualification, they are incompatible with soft, reusable, and repeatedly deformable systems[24]. Their susceptibility to mechanical failure under modest strain[21–23], and lack of elasticity and intrinsically isotropic thermal conductivity compromise reliability under repeated deformation and the ability to support directional heat transport [24].

Thermal interface elastic materials (TIEMs) have emerged as promising candidates to address the limitations of conventional TIMs in stretchable electronics[25–27]. TIEMs are composed of cross-linked elastomeric matrices with embedded thermally conductive fillers, enabling composite thermal conductivities ($\kappa_c$) up to an order of magnitude higher than the base polymer[28,29]. While increasing the filler volume fraction ($\phi$) beyond the thermal percolation threshold ($\phi_c$) enhances $\kappa_c$, it also leads to increased stiffness and does not inherently promote anisotropic heat transport[30] if the fillers are simply blended with the matrix[28,29]. In this paper, we will call this conventional approach the blended-only (B) approach. Compared to isotropic blending, inducing filler arrangement into a preferential direction can achieve higher heat transport for similar filler loadings[31,32]. To induce through-plane anisotropy, a variety of processing techniques have been explored, including ice-templating[33,34], electrostatic flocking[35,36], template-assisted 3D network formation[37,38], shear-induced alignment[39,40], direct ink writing[41,42], layer-by-layer lamination[43,44], and filler alignment using electric[45,46] or magnetic fields [47–50].

In composites with magnetically responsive fillers, magnetic field-particle interaction and interparticle dipole–dipole interactions induced by the field drive the self-assembly of particle chains along the field lines, forming continuous networks that promote anisotropic heat transport[51–61]. Compared to other anisotropy-inducing strategies, such as ice-templating and electric-field alignment, magnetic-field-assisted alignment avoids thermal processing, electrical breakdown, and electrode integration while enabling uniform, contactless, and in some cases reversible alignment[62]. However, most implementations adopt the conventional blend–then–align (BA) approach, wherein filler particles are oriented within a viscous polymer precursor[49,50,54–60,62,63]. At high filler loadings, desired to maximize $\kappa_c$, strong interparticle forces coupled with the viscous drag of the matrix hinder free rotation and translation, often resulting in incomplete alignment and disrupted network formation. Furthermore, even when alignment is achieved, the matrix material surrounding the particles makes physical particle-particle contact unlikely, resulting in micro/nanoscale gaps leading to large interfacial thermal resistance, which significantly constrain the through-plane thermal conductivity ($\kappa_{\perp}$).

In this work, we present a two-stage magnetic-field-assisted fabrication strategy that addresses the limitations of conventional alignment methods. Termed ABBA (Align-Before-Blend followed by Blend-Then-Align), the proposed hierarchical strategy decouples magnetic filler alignment from polymer viscous drag and minimizes interfacial thermal resistance by introducing two different fillers: nickel (Ni) and liquid metal (LM). In the innovative align-before-blend (AB) first step, Ni microparticles are magnetically aligned in air before polymer infiltration, enabling the formation of a scaffold composed of highly ordered, vertically continuous, and densely packed pathways, even at high particle loading. To further improve thermal coupling between pre-aligned particles, without compromising alignment fidelity or mechanical compliance, a liquid metal (LM) based particle-bridging strategy is introduced in the next step, during the polymer matrix infiltration. This second step involves the injection of an LM/Ni mixture, blended with elastomer via the conventional BA route, into the pre-aligned AB scaffold. LM offers high thermal conductivity and does not introduce mechanical reinforcement, but it is inherently non-magnetic. Therefore, a small fraction of Ni particles serves as magnetic carriers, guiding the LM droplets toward the scaffold pre-aligned vertical pathways under the applied magnetic field. This guided infiltration allows the morphing LM to selectively localize at Ni interparticle junctions while preserving the particles vertical alignment before the matrix crosslinking, reducing effective interparticle thermal resistance. The ABBA architecture enhances $\kappa_{\perp}$ by nearly an order of magnitude compared to BA or even AB composites, while maintaining in-plane electrical insulation and mechanical softness. The ABBA strategy is not distinguished by magnetic alignment, liquid-metal incorporation, or the use of a soft elastomer matrix individually, because each of these concepts has been reported previously. Instead, its innovation lies in the processing sequence and functional partitioning of Ni and liquid metal, which allows increasing the concentration of Ni filler while maintaining strong vertical alignment and locally bridging interfiller distance. This process maximizes anisotropic transport without introducing mechanical reinforcement. This differentiating combination of high $\kappa_{\perp}$, in-plane insulation, ultrasoft mechanical compliance, and tensile elasticity is key to skin electronics and other stretchable electronic systems. Accordingly, ABBA composites are designed as high-performance TIEMs for low-pressure, stretchable, skin-conformal applications, including thermoelectric modules and skin-mounted heat-spreading layers, rather than as a general replacement for rigid-packaging TIMs.

## 2. Results and Discussion

### 2.1 Fabrication and Structural Characterization

**Figure 1a** schematically outlines the fabrication pathways for the composites produced using the conventional blend-only (B) and blend-then-align (BA) methods, as well as the proposed align-before-blend (AB) new approach. The critical difference among these methods is the sequence of the interactions between the magnetic particles, the polymer precursor, and the externally applied magnetic field, factors that govern the ability of the particles to reorganize into continuous, vertically aligned networks, as shown in the cross-sectional optical images in **Figure 1**.

In the B route, Ni microparticles are directly dispersed into the uncured silicone elastomer Ecoflex matrix, resulting in a random, isotropic microstructure. Without magnetic alignment, the particles remain as isolated inclusions with no preferential arrangement. Since thermal transport relies solely on stochastic particle-to-particle contacts, the B route leads to highly discontinuous and inefficient conductive pathways, as detailed below. The BA route improves upon this by applying a magnetic field to the mixture before curing. However, particle migration occurs within a viscous medium, where high viscous drag and frequent particle–particle collisions, especially at elevated loadings, restrict particle mobility. This results in incomplete and often disrupted vertical networks, as shown in **Figure 1**. In contrast, the AB method first aligns Ni particles in air before polymer infiltration, thereby eliminating viscous drag and allowing for unhindered particle arrangement (**Figure S1**). Time-resolved optical microscopy confirms that the dry aligned scaffold formed swiftly, in a repeatable manner, and remained stable (at least over 24 h) as long as the magnetic field is applied (**Figure S2, Video S1**) and that subsequent Ecoflex precursor infiltration caused only limited (field-constrained) rearrangement without disrupting the columnar architecture (**Figures S3-S6, Videos S2 and S3**). The B method was included in the study as an isotropic reference (**Figure S7**). This drag-free environment promotes the formation of straighter, well-defined vertical columns across all particle sizes tested, 1 μm, 5 μm, and 15 μm in diameter, even at high concentrations.

Finally, **Figure 1b** shows a hierarchical assembly strategy, termed ABBA, that improves AB processing by combining the long-range (i.e., thickness-spanning), vertically aligned architecture of the AB scaffold with a localized, highly conductive eutectic GaIn $(76:24)_{at\%}$ (EGaIn) LM bridging in between pre-aligned Ni particles. In the ABBA process, the pre-aligned Ni scaffold providing the highest $\kappa_{\perp}/\sigma_{||}$ ratio (AB, 10.6 vol%, 5 μm-diameter particles) was infiltrated with a secondary Ni/LM/Ecoflex mixture, prepared via the conventional BA method. Mechanical co-grinding of Ni and LM produced stable Ni–LM agglomerates (**Figures S8–S10;**

**Table S1**). This LM-Ni association is attributed to the dynamic processing conditions during mortar-and-pestle grinding rather than to simple equilibrium wetting between materials, as suggested by contact angle measurements in **Figure S11**. During grinding, the native oxide of LM is repeatedly disrupted, promoting spreading of LM on Ni particles to form agglomerates. A new fresh oxide is likely immediately formed on the LM in the agglomerate, precluding further spread of LM onto any material, unless mechanical force is employed. In the subsequent gentle mixing with Ecoflex and magnetic alignment steps, the oxide-skinned LM remained probably intact, preventing LM from mixing with the polymer matrix. Under the applied magnetic field, the Ni component of the Ni–LM agglomerates served as a magnetic carrier, guiding the associated LM phase toward the pre-aligned Ni columns within the AB scaffold. The LM-Ni agglomerates joined the pre-aligned Ni-rich scaffold regions, including interparticle regions, which the compliant and morphing nature of LM (even with its thin oxide skin) helped connect without disrupting the columnar architecture.

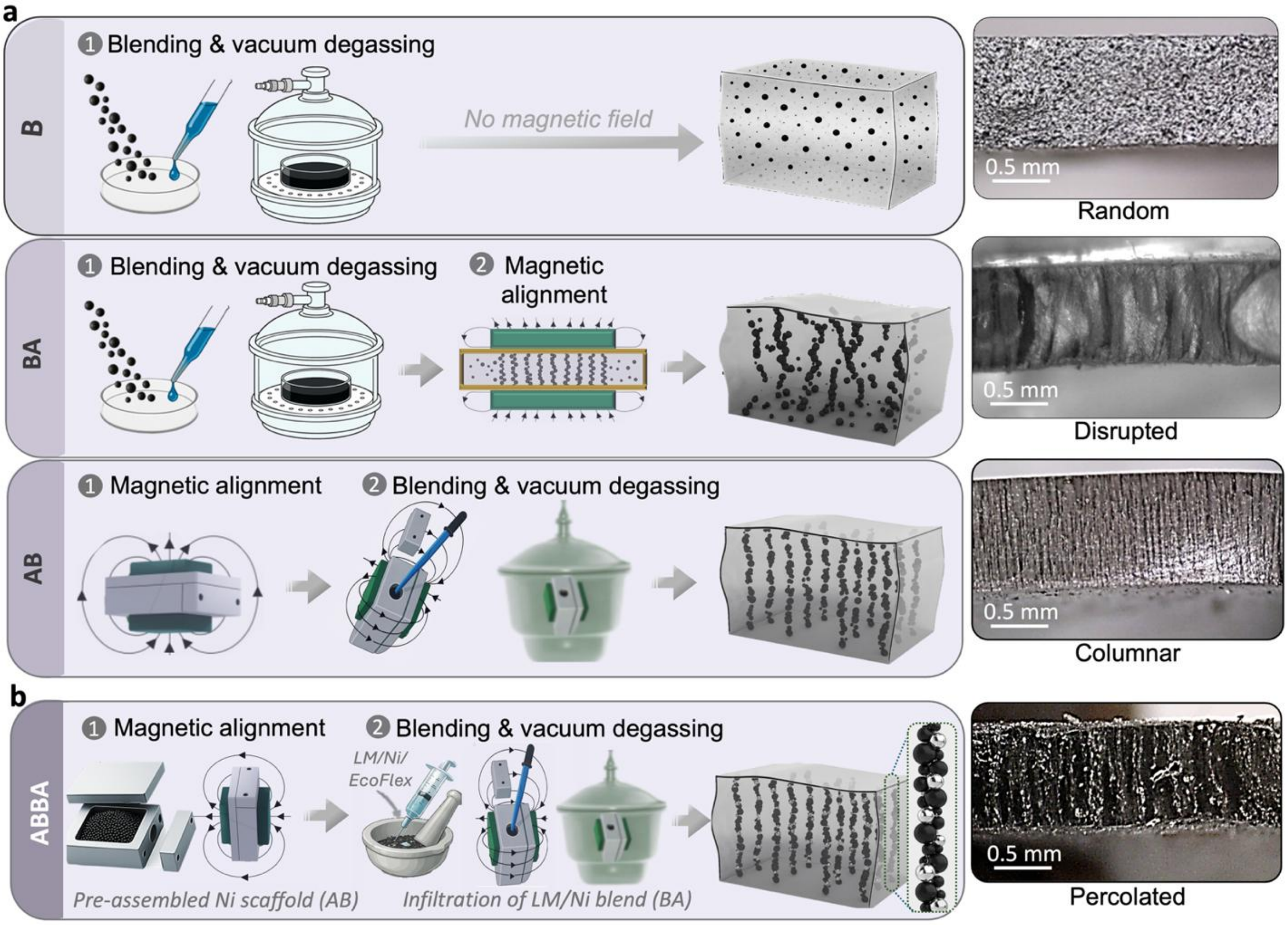

**Figure 1**. (**a**) Schematic illustration of the fabrication routes for blend-only (B), blend-then-align (BA), align-before-blend (AB), and (**b**) ABBA composites, the latter integrating the AB and multi-filler BA strategies. The schematics highlight how processing sequence governs filler organization in thermal interface elastomer materials (TIEMs). Right: cross-sectional optical

micrographs of composites fabricated using 5 μm-diameter Ni particles via the B, BA, AB and ABBA routes.

To quantitatively assess the alignment of AB composites, Fourier transform (FT)–based power spectral density (PSD) analysis (described in Section 4) was applied to optical cross-sections like those in **Figure S7**. The evolution of filler organization with increasing Ni content is summarized in **Figure 2a**, where a large PSD anisotropy ratio corresponds to strong column alignment. The reference B composites consistently exhibited low PSD anisotropy ratios across all loadings, reflecting a randomly connected, isotropic network. In contrast, AB composites display a pronounced increase in anisotropy at low to intermediate Ni concentrations, corresponding to the formation of vertically aligned columns that remain largely disconnected in the in-plane direction. As the Ni content increases further, these columns progressively develop lateral connections (**Figure 2b**), leading to a gradual decrease in PSD anisotropy. This structural transition toward lateral connections coincides with a rise in tensile stiffness (> 0.3MPa, **Figure 2c**) and the onset of in-plane electrical percolation (>$10^{-8}$ S/cm, **Figure 2d**).
The effect of particle size in AB composites was further studied. As expected, larger particles tended to form wider columns, due to their stronger magnetic polarization, which enhances interparticle attraction and fosters clustering (**Figure S7**). This trend was quantitatively supported by chord-length and FT analyses, which confirmed that 5 μm and 15 μm Ni particles exhibited the largest effective column width, inter-space column, and thus pitch *d* at comparable loadings (**Figures S12-S15**). The same analysis reveals that after an initial sharp decrease of *d* at low filler loading, corresponding to the onset of the scaffold formation, *d* decreases gradually with extra loading for all particle sizes. Slow pitch reduction indicates a delayed addition of new columns into the scaffold framework as fillers are added. Having a scaffold with fewer and more spaced columns minimizes the risk of in-plane percolation, which is critical to maintaining anisotropic conduction. This insight, while supporting the potential of magnetic alignment to induce anisotropy, must be nuanced, as not all particle sizes showed the same degree of alignment, and the smaller the particle, the more interfaces it presents for a certain distance *d*, hindering transport.
Among the AB series, the 5 μm-particle system exhibited the highest PSD ratio over the widest filler concentration window. This observation suggests that the scaffold columns remain disconnected even at the high filler loadings (above 10 vol%) necessary to achieve high through-plane conductivity. Such a delayed in-plane connectivity suppressed the undesired abrupt increase in in-plane stiffness (**Figure 2c**) and in-plane electrical conductivity (**Figure S16**) observed at similar loadings in other particle size systems with a lower PSD ratio. Note that the onset of in-plane electrical conduction for the 5 µm-particle system is comparable to that of the 1 µm-particle system despite presenting a few times fewer interfaces between columns.

Consequently, the 5 µm-particle AB system presented the morphology offering the best compromise between high through-plane thermal transport and strong in-plane insulation (**Figure 2e**, **Figure S16**). In particular, for a particle concentration of 10.6 vol%, we observed scaffolds with relatively few but wide (~90 µm, **Figure S14**) columns spaced far apart ($d$ ~ 160 µm, **Figure S14, S15**), and with a strong vertical alignment bolstered by a PSD anisotropy ratio of 22 (**Figure 2a**). As a result, 5 µm-particle composites were selected for the rest of the study. 3D optical reconstructions further revealed uninterrupted vertical architectures in AB composites containing 5 µm Ni particles at 10.6 vol% loading (**Figure 2f**). Consistently, micro-computed tomography (µ-CT) of a 1 mm-thick AB composite (**Figure 2g**) showed a high density of dot-like features on the top view, corresponding to the cross-section of vertically aligned columns. The full 3D reconstruction confirmed that these columns extended continuously through the entire thickness with minimal lateral deviation. The uniformity and structural persistence of these columns demonstrate the efficacy of the AB method in forming

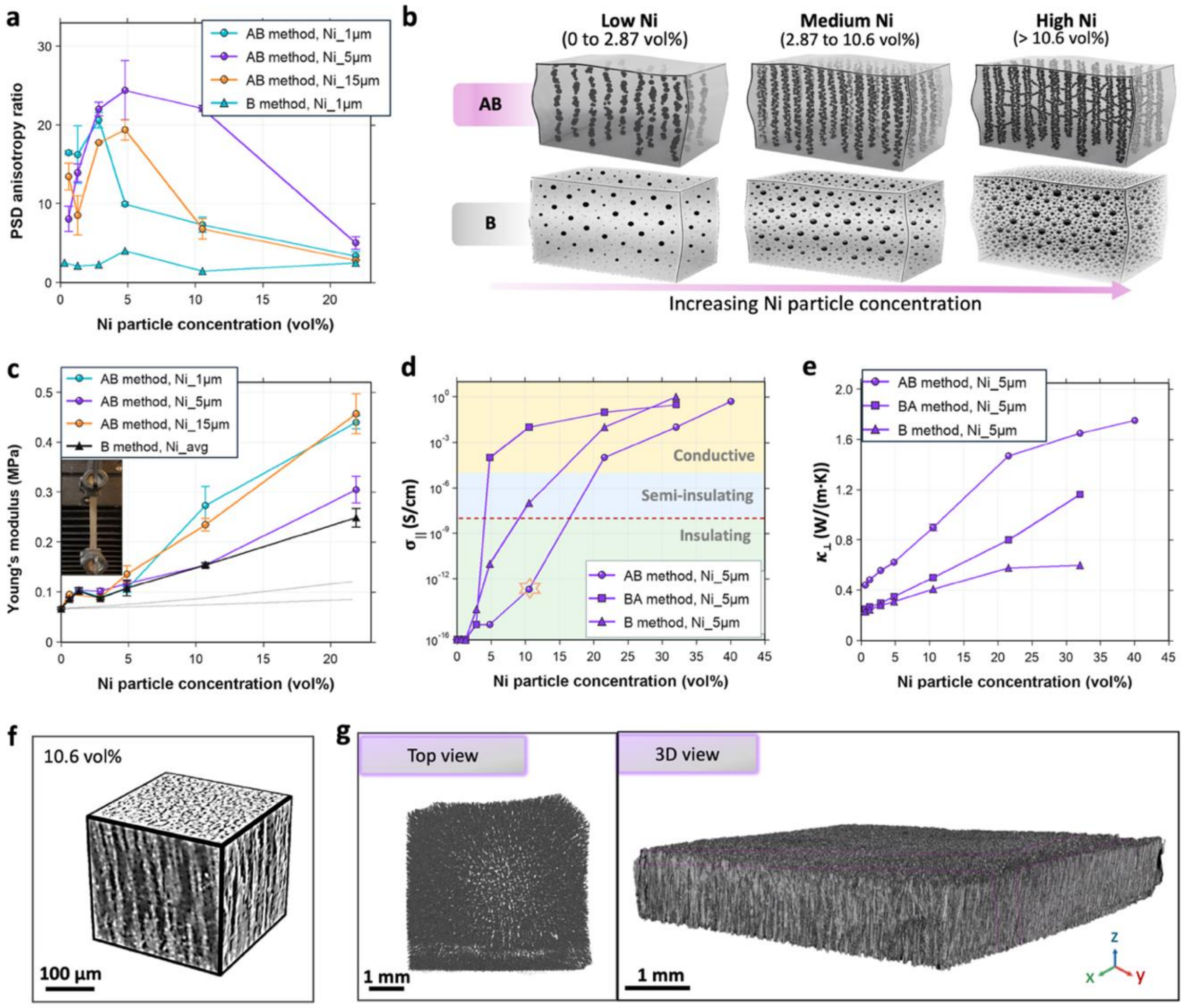

long-range, through-plane thermal conduction networks without in-plane connectivity (**Video S4**).

**Figure 2.** (**a**) PSD anisotropy ratio versus Ni particle concentration for AB and B composites using 1, 5, and 15 μm-diameter Ni particles; (**b**) Schematic illustration of microstructural evolution in AB and B composites with increasing Ni filler concentration. (**c**) In-plane Young's modulus of AB and B composites as a function of Ni particle size and concentration. (**d**) In-plane electrical conductivity and (**e**) through-plane thermal conductivity of composites fabricated by B, BA, and AB routes for 5 μm Ni particles. The star represents the point providing the best trade-off between a high $\kappa_\perp$ and a low $\sigma_{||}$. (**f**) 3D optical reconstruction of an AB composite (5 μm-diameter Ni particles, 10.6 vol%). (**g**) Top view (left) and 3D view (right) of a μ-CT scan of an AB composite (5 μm-diameter Ni particles, 10.6 vol%).

## 2.2 Structure-Property Correlation in Directional Transport

**Figure 2e** presents $\kappa_\perp$ of B, BA, and AB composites containing 5 μm-diameter Ni particles as a function of loading. A consistent hierarchy of AB > BA > B is observed across all loadings, reflecting the progressive enhancement in structural anisotropy and network continuity as the process is optimized. This hierarchy is preserved across all tested particle sizes (1, 5, and 15 μm, **Figure S16**). AB composites exhibit the steepest increase in $\kappa_\perp$ with particle loading. As expected, the AB processing route also supports higher filler loading than B and BA, achieving $\phi_{max}$ values of 24.2 vol% for 1 μm, 40.1 vol% for 5 μm, and 46.4 vol% for 15 μm Ni (**Figure S16**). Composites with such high particle loading were too viscous to be handled using the B and BA processes. We measured the highest $\kappa_\perp = 1.87 \pm 0.03$ W/(m·K) at 46.4 vol% for the 15 μm AB composite, representing a twelvefold enhancement over neat Ecoflex and outperforming the max $\kappa_\perp$ achieved by BA and B counterparts by factors of 1.3 and 2.4, respectively. Nonetheless, no large differences in $\kappa_\perp$ were appreciated for different-sized particles at a fixed loading. This result is, in principle, counterintuitive when considering that the bottleneck for thermal transport should be the interparticle thermal resistance. Accordingly, larger particles should present fewer junctions and a higher $\kappa_\perp$. The observed similar $\kappa_\perp$ for different particle sizes must then be attributed to the better alignment (**Figure 2a**) and/or tighter clustering for smaller particles, which minimizes the interparticle micro/nanoscale gap and compensates for their more numerous junctions. This insight is reflected in the extracted thermal contact-resistivity term, $R_c/D$ (**eq. S21,** see complete model details in **Figure S17-S19**), where $R_c$ represents the effective interparticle thermal contact resistivity considering surface effects and the presence of matrix between adjacent particles, and $D$ is the particle diameter. If $R_c$ were identical for 5 μm and 15 μm particles, the larger 15 μm particles would be expected to show a lower $R_c/D$ contribution because of their larger $D$ and reduced number of junctions.

However, the comparable $R_c/D$ values shown in **Figure S19a** imply that the effective contact resistivity $R_c$ increases for larger particles, likely due to differences in scaffold morphology and local packing quality. According to **Figure 2a**, the 5 µm system exhibited the highest PSD anisotropy ratio over the widest concentration window, meaning that it preserved strong vertical alignment and suppressed lateral connectivity at the loadings required for efficient through-plane heat transport. This alignment can be observed in **Figure S20**. The 15 µm system might have benefited from fewer junctions per through-plane path, but this advantage was partially offset by lower scaffold alignment leading to larger interparticle distances, explaining why both systems show comparable $\kappa_\perp$.

TIM materials should dissipate heat through-plane while preventing an electrical shortcut in-plane. Hence, the in-plane electrical conductivity ($\sigma_{||}$) must be kept as low as possible. As shown in **Figure 2d**, AB composites remain insulating ($\sigma_{||}$ below $10^{-8}$ S/cm) over a broad loading range, up well above 10 vol%. This loading is comparatively higher than for B and BA composites (4.81-10 vol%), indicating effective suppression of lateral conduction, due to strong vertical alignment that precludes lateral filler contacts (**Figure 2a**). B composites show a smooth (close to exponential) increase, which is characteristic of isotropic percolation behavior.

Among all tested AB composites (**Figure S16**), the composite with 5 µm Ni at 10.6 vol% (marked with a star in **Figure 2d,e)** offers the most balanced anisotropic performance, achieving $\kappa_\perp \approx 0.95 \pm 0.03$ W/(m·K) while maintaining excellent electrical insulation underpinned by a $\sigma_{||}$ below $10^{-12}$ S/cm. Given the inherent trade-offs whereby increasing filler content simultaneously enhances both $\kappa_\perp$ and $\sigma_{||}$, further improvement of $\kappa_\perp$, without boosting $\sigma_{||}$, cannot be warranted by adjusting filler content or particle size alone. Instead, it requires precise junction-level engineering to selectively bridge through-plane micro/nanoscale interparticle gaps and reduce through-plane interfacial thermal resistance.

## 2.3 Liquid-Metal Interparticle Gap Bridging to Boost Through-Plane Thermal Transport

To address the interfacial through-plane thermal resistance that limits $\kappa_\perp$ in AB composites, the ABBA process was developed, in which LM was added to the pre-aligned Ni columns within the AB scaffold. Cross-sectional optical images (**Figure 3a**) showed that at low LM contents (0.31-0.8vol% of the total composite), the LM occupied narrow interparticle gaps, preserving the structural integrity of the vertical Ni columns. At medium loadings (1.66-2.6 vol%), LM formed a more continuous vertical network integrated into the aligned structure. At high loading of 3.62vol%, the LM extended laterally, marking the onset of undesired in-plane connectivity. Elemental mapping via SEM–EDS (**Figure 3b)** showed that Ga and In signals overlapped with Ni-rich regions, consistent with LM preferentially localizing at Ni interparticle junction regions. Cross-sectional FIB-SEM with local EDS analysis and nanofocus XCT further support local LM enrichment of Ni-rich scaffold regions (**Figures S21-S22**), consistent with LM reducing effective interparticle thermal resistance within the pre-aligned scaffold. These observations were supported by 2D slices and 3D μ-CT renderings (**Figure 3c**), where LM appeared as bright, discrete inclusions integrated within the aligned columns. The retention of vertical order in the ABBA composites, despite the introduction of a secondary conductive phase, highlights the mechanical robustness of the AB scaffold and the compatibility of LM infiltration with pre-aligned structures.

Following LM infiltration, ABBA samples showed a modest reduction in anisotropy (**Figure 3d**), although the overall vertical alignment remained well preserved.

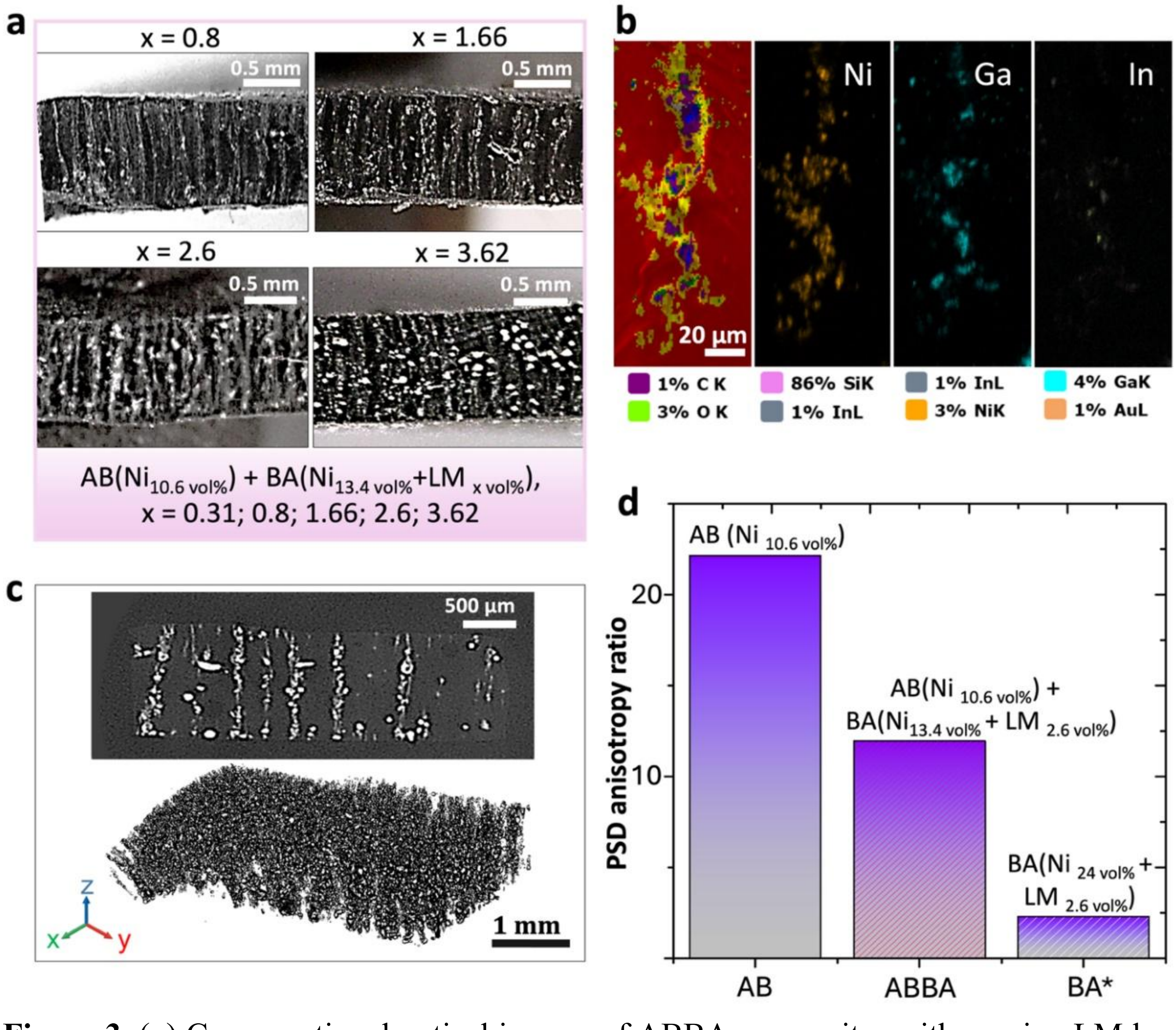

**Figure 3.** (**a**) Cross-sectional optical images of ABBA composites with varying LM loadings. (**b**) SEM–EDS elemental maps showing Ga and In infiltration along Ni-rich regions. (**c**) μ-CT slice (top) and 3D reconstruction (below) of the ABBA architecture with 6.05 vol% LM. (**d**) PSD anisotropy ratio of AB, ABBA, and bi-filler B composites.

The ABBA architecture yielded a substantial improvement in thermal transport. As shown in **Figure 4a**, $\kappa_\perp$ increased from ~1 W/(m·K) at low LM content (equivalent to the best AB composite) to 9.0 ± 0.2 W/(m·K) at 3.62 vol% LM, an order-of-magnitude enhancement. Quantitatively, this corresponds to a 15-fold increase relative to maximally loaded B composites, an eightfold increase relative to the maximally loaded BA composites, and a fivefold increase compared to the maximally loaded AB composites of the same particle size. Thermal IR imaging comparing the surface temperature response of B, BA, AB, and ABBA composites under identical hot-stage conditions further confirms the superior through-plane heat transport of the ABBA architecture across all four processing variants (**Figure S23**, **Video S6**). The improvement of ABBA is attributed to reduced interparticle thermal resistance enabled by LM-mediated junctions, while the pre-aligned Ni framework remains crucial to maintain directional heat flow. To prove the latter point, a Ni/LM bi-filler BA composite (called BA* to differentiate it from the Ni BA material) was prepared with the same overall Ni, LM (2.6 vol%), and Ecoflex content ratio as the optimized ABBA composite. This comparison isolated the effect of the processing sequence (pre-alignment versus direct blending and alignment). The BA* control exhibited a through-plane thermal conductivity of 3.7 ± 0.8 W m$^{-1}$ K$^{-1}$, approximately 2.5-fold lower than that of the optimized ABBA composite, and an in-plane electrical conductivity of $\sigma_{||}$ = 1 S/cm, twelve orders of magnitude higher than that of ABBA. Cross-sectional optical microscopy revealed a less continuous and more agglomerated morphology in the BA* control compared to ABBA, in which a more organized vertical scaffold was retained (**Figure S24**). This observation was bolstered by PSD analysis, which showed that direct BA processing of Ni/LM (BA*) produced only weak structural anisotropy, with a PSD anisotropy ratio of approximately 2–3, compared with approximately 12 for ABBA (**Figure 3d**). These observations confirm that the pre-alignment step played an important role in achieving effective through-plane thermal transport in this system.

A physical explanation at the microscopic level of the benefits of placing LM between Ni fillers to boost thermal conductivity must distinguish between potential contributions to heat transport coming from electrons and phonons (since $\kappa = \kappa_e + \kappa_p$). According to the Wiedemann-Franz law, the electronic contribution to the through-plane thermal conductivity, $\kappa_{\perp,e}$, depends on the through-plane electrical conductivity, $\sigma_\perp$, and the absolute temperature $T$ as $\kappa_{\perp,e} = L \cdot \sigma_\perp \cdot T$, where $L$ is the Lorenz factor, which for metals approaches the Sommerfeld value $2.44 \times 10^{-8}$ W Ω K$^{-2}$ [64]. Considering the ABBA $\sigma_\perp$ < 11 S/cm for the highest LM vol% tested of 3.62, $\kappa_{\perp,e} < 8 \times 10^{-2}$ W/(m·K) at room temperature. This value is orders of magnitude smaller than the $\kappa_\perp$ measured for ABBA, indicating that thermal transport is phonon mediated. Dispersed fillers contribute to thermal resistance because strong phonon scattering takes place at filler–matrix interfaces due to a large acoustic impedance mismatch between the stiff metal particles and soft elastomer. Within the acoustic mismatch approximation, the interfacial phonon intensity $I$ reflection coefficient is:

$$R = \frac{I_{reflected}}{I_{incident}} = \left(\frac{Z_1/Z_2 - 1}{Z_1/Z_2 + 1}\right)^2, \quad (1)$$

where $Z = \rho v_L$, is the material acoustic impedance, $\rho$ is the density and $v_L$ is the speed of sound in the material. Representative values illustrate the severity of this mismatch: Ecoflex has $Z \approx$ MRayl ( $v_L$= 1050 m/s, $\rho$ = 1050 kg/m³)[65], Ni $Z \approx 54$ MRayl ( $v_L$ = 6040 m/s, $\rho$ = 8900 kg/m³)[66], and EGaIn $Z \approx 18$ MRayl ( $v_L$= 2730 m/s, $\rho$ = 6440 kg/m³)[67]. The expression above predicts severe phonon reflection at Ni/Ecoflex interfaces ($Z_1/Z_2 \approx 54, R \approx$ 93%), compared with Ni/LM interfaces ($Z_1/Z_2 \approx 3$, $R \approx 25$).

To quantify the reduction in interfacial thermal resistance introduced by LM placed between particles, an analytical model was developed that decomposes $\kappa_\perp$ into matrix and aligned-scaffold contributions using a parallel heat-flow approximation and models the scaffold as a series of filler particles connected by thermal contacts (**Figure S17-S19**). The model confirms that the ABBA composite has a markedly lower effective contact-resistivity term ($R_c/D$ ~ 0.015 m K/W for 5 μm particles, **eq. S21** ) than the AB ($R_c/D$ ~ 0.17 m K/W, 5 μm particles), BA ($R_c/D$ ~ 0.4 m K/W, 5 μm particles), and LM/Ni bi-filler BA* ($R_c/D$ ~ 0.17 m K/W, 5 μm particles) controls, supporting the interpretation that LM incorporation reduces effective interparticle junction resistance within the pre-aligned scaffold.

Electrical transport behavior further highlights the vertical selectivity of the ABBA design (**Figure 4b**). In-plane electrical conductivity ($\sigma_{||}$) remains below $10^{-12}$ S /cm for LM contents up to 2.6 vol%, while through-plane conductivity ($\sigma_\perp$) ~ 3 S/cm, indicating that LM selectively reinforces vertical pathways without initiating lateral percolation. DC in-plane dielectric breakdown strength testing confirmed that the optimized ABBA composite retains substantial lateral insulation under high electric fields, with a breakdown strength of 6.6 ± 0.2 kV mm$^{-1}$ (171 ± 5 V/mil), compared to 8.7 ± 0.9 kV mm$^{-1}$ (230 ± 20 V/mil) for neat Ecoflex (**Figure S25**). At the highest LM content (3.62 vol%), $\sigma_{||}$ experienced a dramatic increase, consistent with the in-plane LM domain interconnection discussed in **Figure 4a**. Since ensuring in-plane electrical insulation is key for TIM, for the follow-up demos, we selected the composite with 6.05 vol% LM (marked with a star), which displayed a $\kappa_\perp$ = 7.5 ± 0.4 W/(m·K) for a $\sigma_{||} <$ $10^{-12}$ S /cm.

Importantly, the ABBA composite retains its enhanced $\kappa_\perp$ under tensile strains of at least 25%, indicating a mechanically stable structural morphology that preserves thermal transport performance during deformation (**Figure S26a**). To examine larger deformations, post-strain LFA measurements were performed after single tensile cycles at 50% and 100% strain (**Figure 4c**). ABBA retained approximately 94% and 81% of its initial $\kappa_\perp$ after 50% and 100% strain, respectively, indicating progressive disruption of the aligned scaffold only under larger deformation exceeding the wearable application window. Consistently, IR imaging of the surface temperature during uniaxial tensile stretching of the AB composite reveals a stable thermal response (**Video S6**). The strain stability of $\kappa_\perp$ under tensile deformation is rationalized using a scaffold resistance model in which LM-associated junctions act as deformable thermal bridges that preserve Ni–Ni contact during strain (**eq. S24**).

The mechanical durability of ABBA under repeated deformation was evaluated by tensile cycling followed by LFA measurement of $\kappa_\perp$. After 1000 stretching cycles at 25% strain, the ABBA composite retained approximately 94% of its initial conductivity, indicating good thermal stability under repeated wearable-relevant deformation (**Figure 4c**). The corresponding strain–time profile showed a stable cyclic response throughout the test (**Figure S26b**). Compressive loading at 10 psi (relevant for the intended application) over 1000 cycles showed stable $\kappa_\perp$ retention with no further degradation beyond the initial accommodation period, while higher pressures (50–100 psi) also showed stable $\kappa_\perp$ retention but produced permanent deformation of the composite (**Figure S27**). Surface inspection by microfiber wipe test before and after all tested deformation and cycling conditions, including 1000 tensile cycles at 25% strain, and 1000 compressive cycles at 10 psi revealed no visible LM residue, suggesting that the LM phase remains mechanically retained within the Ecoflex matrix under application-relevant operating conditions.

Long-term thermal stability was further evaluated by subjecting three ABBA samples to approximately 700 heating–cooling cycles over 24 hours of continuous operation on a hotplate under fan-assisted cooling, monitored by an IR camera with ambient temperature correction (**Figure S28**). The average standard error of the peak surface temperature across the three samples was 0.38°C (~0.5% of peak temperature) throughout the full test, confirming stable and reproducible thermal performance under prolonged cyclic thermal loading.

**Figure 4d** benchmarks ABBA against reported thermally conductive polymer systems, with detailed information on filler loading, alignment, $\kappa_\perp$, in-plane electrical conductivity, Young's modulus, stretchability, and processing route provided in **Table S2**. Although several boron nitride (BN)- and ceramic-based composites achieve comparable or higher $\kappa_\perp$, they generally do not combine high thermal transport with tensile stretchability and ultrasoft compliance. Carbon-fiber- and graphene-based composites can also reach high κ⊥ but commonly rely on in-plane electrically conductive networks. Among the 18 surveyed systems with κ⊥ ≥ 7.5 W/(m·K), only four are reported as stretchable, and all four are electrically conducting in-plane. Thus, none combines κ⊥ ≥ 7.5 W/(m·K) with both tensile stretchability and in-plane electrical insulation. The closest stretchable and electrically insulating directional composite identified is the aligned-BN system in reference S13 in the Supplementary Information, which reaches κ⊥ = 6.94 W/(m·K) using 43.75 vol% BN but has a reported Young's modulus of 1.3 GPa, far above the recommended range for skin electronics (<0.6 MPa [17]). ABBA is not intended to replace thermal greases in conventional rigid chip-cooling applications, where flowability, minimum bond-line thickness, and maximum filler loading are the principal considerations. Instead, it addresses applications requiring the simultaneous combination of through-plane heat transport, in-plane electrical insulation, tensile elasticity, ultrasoft compliance, dimensional integrity, and reusability, including wearable thermoelectric devices, flexible power electronics, and skin-conformal thermal-management systems.

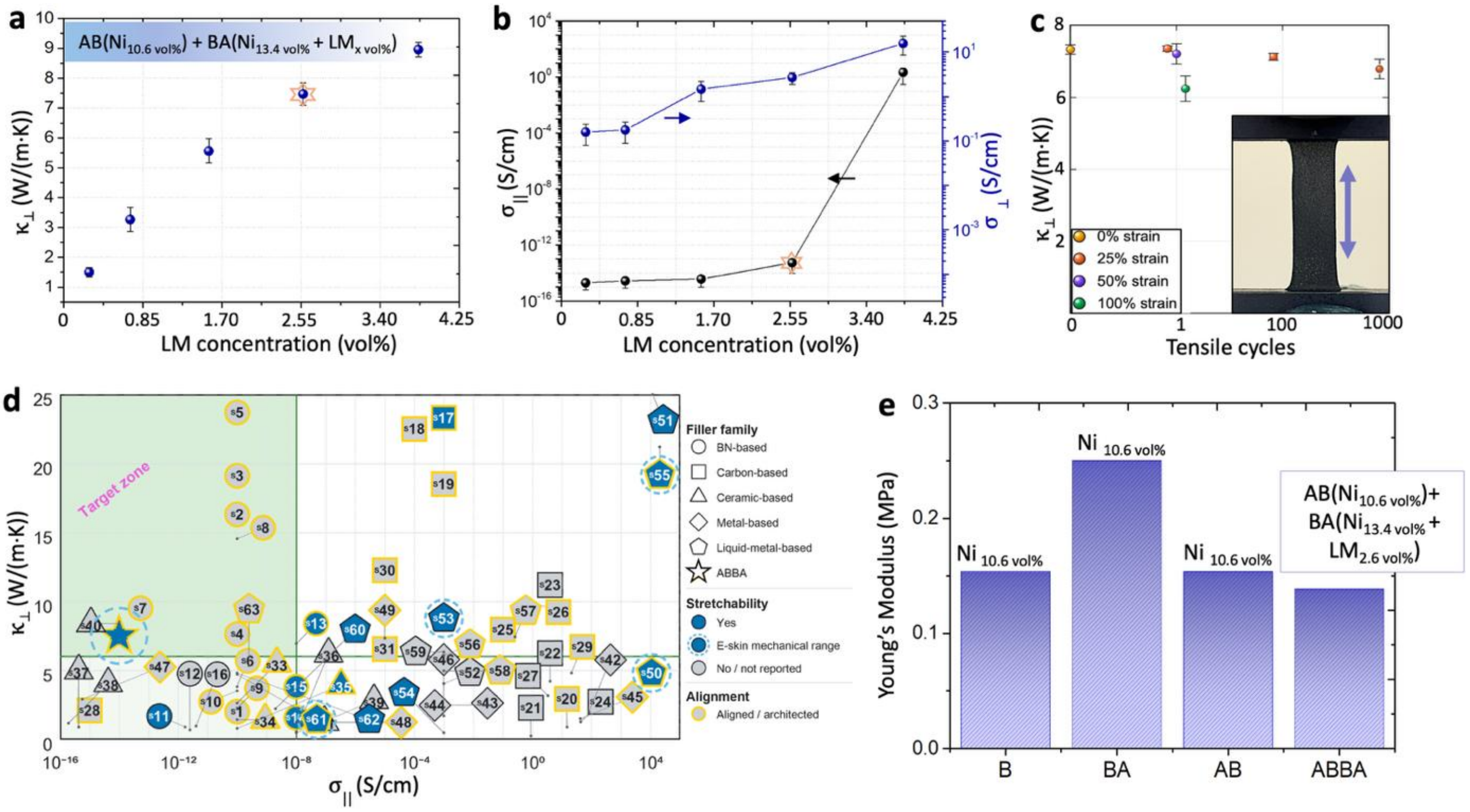

**Figure 4. Transport optimization and performance benchmarking of ABBA composites.** Effect of liquid metal (LM) loading on (**a**) through-plane thermal conductivity (κ⊥), and (**b**) in-plane (σ||) and through-plane (σ⊥) electrical conductivity of ABBA composites. The star represents the point providing the best trade-off between a high κ⊥ and a low σ||. (**c**) κ⊥ measured after 0, 1, 100, and 1000 stretching cycles at 25% strain (f = 0.25 Hz), and after one stretching cycle at 50% and 100% strain (f = 0.01 Hz). n = 3 per condition; error bars represent

standard error. (**d**) Thermal–electrical performance map of reported thermally-conductive composites classified by filler material family, stretchability, and alignment. The e-skin mechanical range is defined by an in-plane Young's modulus ≤ 0.6 MPa[17].The numbers inside the symbols correspond to references in the Supplementary Information. (**e**) In-plane Young's modulus of Ni-based composites processed differently.

### 2.4 Mechanical performance of composites

**Figure 2d** presents the in-plane Young's modulus as a function of Ni loading for B, BA, and AB composites with varying particle sizes. As expected for particle-reinforced elastomers, stiffness increased with filler concentration; however, the degree of mechanical reinforcement varied across processing routes due to differences in filler organization and network morphology. The particle-size-dependent onset of in-plane stiffening across 1, 5, and 15 μm systems is discussed in detail in the Supporting Information (**eq. S25**).

Interestingly, for composites containing 1 μm and 15 μm Ni particles, the AB route led to a larger increase in modulus compared to the B route. Particulate reinforcement models (e.g., Nielsen model) describe stiffness enhancement under the assumption of discrete fillers with limited interaction effects, where reinforcement arises primarily from micromechanical load transfer to the stiff filler phase. B composites exhibited a similar but systematically higher stiffness than those models, which could arise from indirect filler–matrix interactions that restrict polymer chain mobility and locally stiffen the elastomer matrix phase (29). AB composites form large column filler structures that could enable strong reinforcement approaching continuous-fiber limits. However, the reinforcing efficiency of continuous fibers remains limited when the columns are aligned perpendicular to the in-plane loading direction (corresponding with the Reuss model). Only at higher filler content would stiffness increase sharply, coinciding with the onset of in-plane column connectivity. This connectivity was evidenced by the simultaneous onset of electrical percolation (**Figure 2d, S16**). The mechanical reinforcement of in-plane connectivity can be modelled by the Voigt model for the limit case of perfect longitudinal fiber alignment (**Figure S29**). Our AB composites, with limited in-plane connectivity, would lie somewhere between Reuss and Voigt behavior, but closer to the Reuss limit. Interestingly, AB composites incorporating 5 μm Ni remained nearly as compliant as their B counterparts across the entire loading range (**Figure 4e**), indicating minimal reinforcement arising from in-plane connectivity due to superior vertical filler alignment as indicated by the PSD study in **Figure 2a.**

Among all the composites with 5 μm-diameter particles and 10.6 vol% loading, the BA composite exhibited the highest in-plane stiffness, which was attributed to poor vertical alignment. Introduction of LM via the ABBA process did not result in additional stiffening relative to the AB baseline. This is attributed to the ABBA architecture: LM improves thermal coupling within vertically aligned columns, while the columns remain laterally separated. Therefore, ABBA reduces effective interparticle thermal resistance without creating a laterally connected rigid network, allowing the composite to retain soft mechanical compliance.

**2.5 Demonstration of ABBA Substrates for Thermoelectric TIM Applications**

To assess the practical utility of enhanced through-plane thermal transport and in-plane electrical insulation, ABBA composites were evaluated as soft TIM and compared against pristine Ecoflex and a commercial reference thermal pad ($\kappa_{\perp}$ = 6 W/(m·K), soft but non-stretchable); RSPRO, 707-4764) of the same thicknesses. For practical TIM benchmarking, soft commercial pads were also evaluated mechanically. The 6 W/(m·K) pad showed permanent deformation after tensile loading, while the 12 W/(m·K) pad fractured at ~31% strain (Figure S32), highlighting the limited reversible stretchability of conventional high-κ TIMs. Under controlled heating conditions at the bottom of the film, infrared thermography from the top surface (**Figure 5a**) showed that the ABBA composite substrate consistently achieved the highest surface temperatures across the 30–60 °C range, indicating superior through-plane heat conduction. The commercial thermal pad demonstrated intermediate performance, while Ecoflex exhibited the poorest heat transfer efficiency. Time-resolved temperature profiles (**Figure 5b**) further confirmed these results (**Video S7**): the ABBA substrate top surface reached higher steady-state temperatures more rapidly.

To evaluate the device-level relevance of lateral electrical insulation, the optimized ABBA pad was compared against three control TIM acting as substrates for a commercial flexible multi-leg TEG module (**Figure 5c**, Asahi Rubber, FTEM series). The native TEG substrate was first peeled away to expose the copper electrodes for the thermoelectric legs, and each TIM was placed directly onto these electrodes. During testing, the TEG was heated using a Peltier element operated at constant input power, while the same heat-sink configuration and device geometry were maintained across all TIM tested (**Figure 5c**). The tested TIM were: (1) the optimized ABBA pad (~1 mm thick); (2) a in-plane conductive ABBA composite with a high LM vol% of 3.62 (~1 mm thick); (3) a commercial reference TIM with thickness matched to ABBA (~1 mm, achieved by stacking two 0.5 mm sheets, 7074764, RS PRO); and (4) a commercial reference TIM sheet at its original thickness (0.5 mm, 7074764, RS PRO). The optimized ABBA pad produced an open-circuit voltage $V_{OC}$ = 458.0 mV and maximum output power $P_{max}$ = 10.25 mW. In comparison, the conductive ABBA composite produced a $V_{OC}$ =

417.1 mV and $P_{max}$ = 8.31 mW, despite having a higher $\kappa_{\perp}$ (**Figure 5d**). This decrease is attributed to parasitic current leakage through the in-plane conductive network, which can short-circuit neighboring TE legs and reduce the effective device output voltage. This comparison confirms that maximizing $\kappa_{\perp}$ alone is insufficient for a performing TIM, as lateral electrical insulation is also essential for TEG integration.

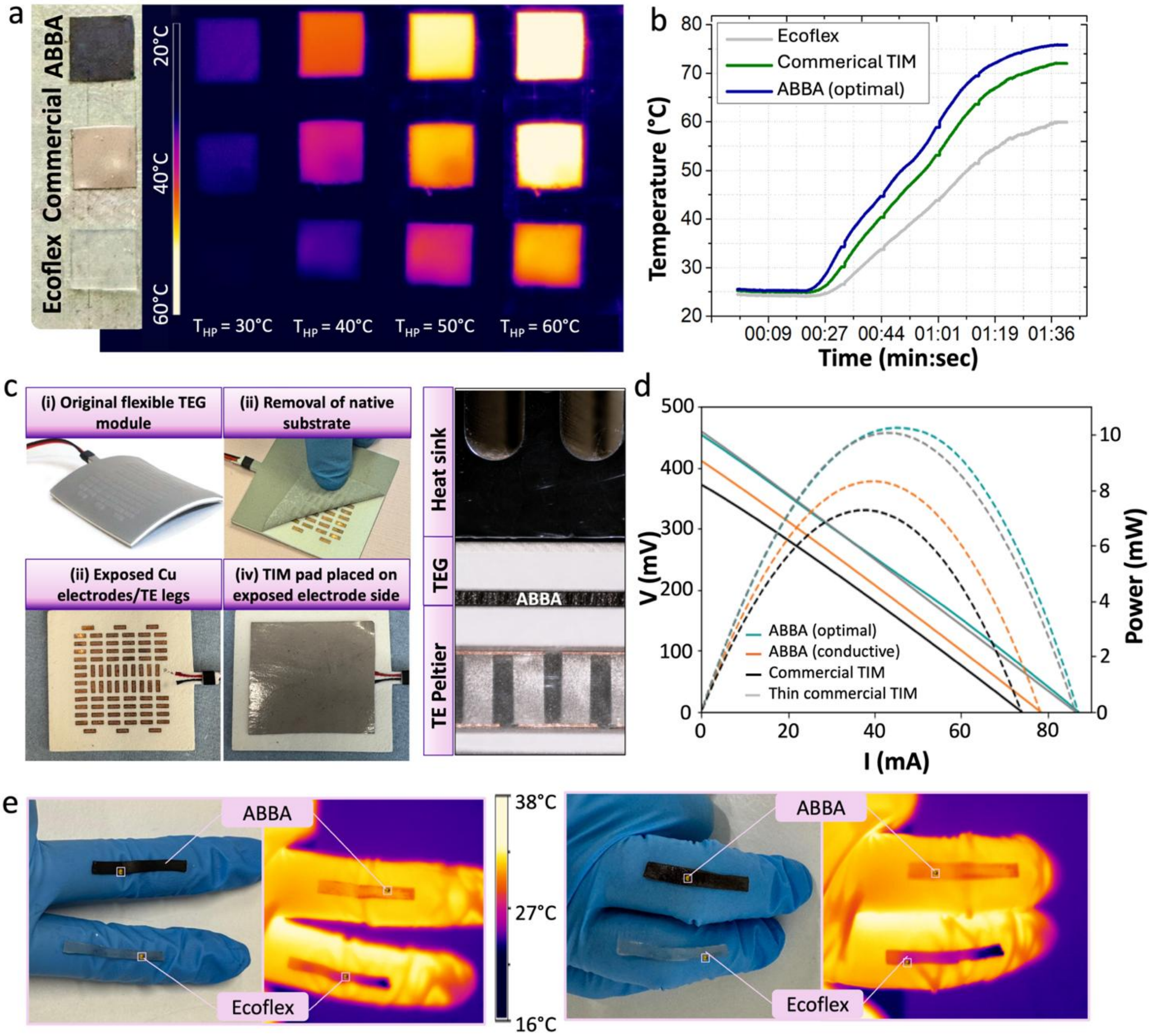

**Figure 5.** (**a**) Infrared thermographic images of the top surface of Ecoflex, the commercial thermal pad, and the ABBA substrates subjected to increasing hot-plate temperatures (30–60 °C). (**b**) Temperature–time profiles of the top surface of Ecoflex, the commercial thermal pad, and the ABBA substrates under identical heating conditions. (**c**) Flexible TEG demonstration setup. Images show the flexible multi-leg TEG module, removal of the original TEG substrate to expose the active electrode area, placement of alternative TIM pads, and the assembled heat sink/TEG/TIM/Peltier stack used for electrical output measurements. (**d**) Electrical output characteristics of the flexible multi-leg TEG. V–I and P–I curves measured using the optimized ABBA pad (~1 mm thick), conductive ABBA composite (~1 mm thick),

commercial TIM at matched thickness (~1 mm), and thin commercial TIM (~0.5 mm). (**e**) Demonstration comparing the top surface temperature of Ecoflex and ABBA substrates applied to gloved fingers in flat (left) and bent (right) configurations.

The commercial reference TIM at matched thickness (1 mm) produced $V_{OC}$ = 380.1 mV and $P_{max}$ = 7.27 mW, approximately 17% lower in $V_{OC}$ and 29% lower in $P_{max}$ than the optimized ABBA. The thin commercial reference TIM (0.5mm, half the thickness of ABBA) produced $V_{OC}$ = 460.9 mV and $P_{max}$ = 10.06 mW, comparable to the optimized ABBA pad (**Figure 5d)**. However, unlike the ABBA pad, the commercial TIM is non-stretchable, does not provide mechanical compliance for curved or dynamically deforming surfaces, and is intended for single-use deployment,  properties that preclude its application in wearable or skin-conformal thermoelectric systems.
To further assess performance in conformal applications, pristine Ecoflex and stretchable ABBA substrates were applied directly to a gloved human hand under identical geometric and thermal conditions. As shown in **Figure 5e** (and **Video S8**), the ABBA material achieves higher surface temperatures compared to Ecoflex while remaining equally compliant. The consistent temperature advantage, even under bending, indicates that the Ni–LM network within the ABBA composite maintains continuous thermal pathways under mechanical strain, reinforcing its applicability for wearable and skin-integrated thermal management systems. To quantify the interfacial adhesion relevant to skin-conformal deployment, lap-shear adhesion tests were performed using an MTS Criterion Model 41 universal testing machine on both glass and porcine skin substrates. The ABBA composite showed an adhesion strength of 22.2 ± 1.2 kPa on glass and 4.8 ± 0.3 kPa on porcine skin, comparable to or higher than neat Ecoflex (14.7 ± 0.3 kPa on glass, 4.6 ± 0.8 kPa on porcine skin), confirming that the addition of the thermally conductive filler network does not compromise the skin-contact adhesion of the elastomeric matrix (**Figure S30**).

## 3. Conclusion

The methodological contribution of this work is not magnetic alignment or a new filler chemistry but a change in the sequence of alignment. Conventional BA processing orients magnetic fillers inside a viscous precursor, where interparticle friction and drag limit the packing density and column continuity. The proposed magnetic alignment of Ni microparticles in air, before polymer introduction (AB), removes viscous drag from the alignment step entirely: the resulting columns are set by particle-particle magnetic interaction alone, not by a compromise between that interaction and matrix rheology. The second liquid-metal delivery step is not used to build the anisotropic scaffold, only to wet the junctions already present within it to reduce interparticle thermal resistance. Separating scaffold formation from junction bridging is an innovation that enables a structuring capability that neither single-stage blending nor single-stage alignment provides, and it is the basis for the two-step ABBA sequence outperforming either operation alone.

Ni-only AB scaffolds reach $\kappa_\perp$ approaching 1 W $m^{-1}K^{-1}$ at practical loadings (10.6 vol%, 5 μm particles), but thermal conductivity plateaus there because the bottleneck becomes the thermal resistance between Ni particles. Adding LM magnetically guided by a small number of extra Ni particles in the second step does not introduce a parallel conduction path in ABBA composites. Instead, LM preferentially moves to the pre-existing Ni scaffold, as evidenced by optical microscopy, μ-CT, and SEM–EDS, bridging the gap between Ni particles. In this way, the AB columnar architecture and its in-plane insulation remain intact in ABBA materials, while $\kappa_\perp$ increases in the latter by an order of magnitude at fixed Ni content.

The resulting optimal ABBA composite reaches $\kappa_\perp = 7.5 \pm 0.4$ W $m^{-1}K^{-1}$ while remaining electrically insulating in-plane ($\sigma_\parallel < 10^{-12}$ S $cm^{-1}$). The material is elastic, soft (Young's modulus $\approx 160 \pm 20$ kPa), and mechanically and thermally robust, surviving cyclic strain to 25% deformation, 1000 compression cycles at 10 psi, single cycles at 50–100 psi, and ~700 thermal cycles without property degradation or LM leakage. Individually, none of these transport or mechanical values represents a record; the notable merit of ABBA is that it achieves all these properties concurrently, without the trade-offs typically observed among them.

This unique combination of transport and mechanical properties of ABBA confers it a versatility that positions it ahead of existing TIMs/TIEMs for emerging applications other than the traditional rigid chip-scale gap-filling, for which the ASTM-qualified TIMs alternatives remain the appropriate choice. Those emerging applications are wearables, flexible thermoelectrics, skin-conformal heat spreaders, and soft robots, for which thickness control, elasticity, reusability, and low-pressure integration are required on top of the conformability already provided by conventional TIMs/TIEMs. Indeed, aligned-filler TIEMs (produced by ice-templating, electrospinning, or BA magnetic alignment) match our ABBA materials $\kappa_\perp$, but only at loadings that substantially stiffen the matrix, compromising stretchability/elasticity. Among liquid-metal composites, isolated LM-droplet elastomers stay electrically insulating but reach a limited $\kappa_\perp$, while continuous LM-matrix pastes reach high $\kappa_\perp$ by percolating, which makes them conductive in-plane as well; neither presents anisotropic transport, and both are prone to LM leakage.One potential limitation of the material is its finite through-plane electrical conductivity. While this characteristic can be advantageous for applications requiring through-plane electrical interconnections, such as the routing or connection of bus lines in a manner similar to anisotropic conductive adhesives, it may become problematic in architectures where electrical insulation between adjacent components is required. In such cases, the incorporation of a dielectric interlayer might be required.

We note that standardized thermal resistance characterization following ASTM D5470, including direct separation of bulk and contact resistance contributions under controlled assembly pressure, bond-line thickness as a function of pressure, and corrosion compatibility with common device metallization, was not performed in this study and remains an important direction for future work targeting industrial-grade TIM qualification. In addition, time-domain or frequency-domain thermoreflectance (TDTR) measurements could shed new light on the mechanism leading to boosted $\kappa\perp$, and accelerated aging under simultaneous electrical bias and cyclic strain would establish whether the Ni–LM junctions remain stable over realistic device lifetimes, beyond the single-session, 25%-strain testing performed here.

Because our ABBA process is largely matrix-agnostic, it can be readily extended to matrices providing application-specific functionalities, such as selective adhesion, biocompatibility, chemical resistance, etc. Consequently, the presented approach provides a general platform for engineering anisotropic multifunctional composites for applications spanning electronics, energy devices, soft robotics, and biomedical technologies.

## 4. Experimental Section/Methods

*Materials*: EcoFlex 00–30 (Smooth-On) was used as the polymer matrix. This platinum-cured silicone elastomer has an intrinsic thermal conductivity of 0.2 W/(m·K). Nickel (Ni) powders with average particle sizes of 1 μm (Ni-01, Jingmen Gem New Material Co., Ltd.), 5 μm (Type 123, Vale Canada Limited), and 15 μm (Cat. No. AA006631, Thermo Scientific Chemicals) were employed as thermally conductive fillers. Ferrite permanent magnets (FE-Q-18-10-06, Supermagnete Inc.) generating a magnetic flux density of 0.4T were used to induce particle alignment. The liquid metal (LM) phase was a gallium–indium eutectic alloy (99.99% metals basis, Product No. 12478, Thermo Fisher Scientific).

*Preparation of AB and ABBA composites:* For AB composites, Ni powders of selected particle size were dry-loaded into a customized Teflon mold (**Figure S31**). The mold was placed between two vertically oriented permanent magnets (0.4 T), as illustrated in **Figure 1a**, aligning the particles along the magnetic flux lines to form vertically continuous columns. After stable alignment was achieved, the mold-magnet assembly was rotated by 90°, and equal masses of Ecoflex 00–30 precursors (parts A and B) were injected through a side port to infiltrate the aligned particle scaffold. The filled mold was placed under vacuum for 10 minutes to remove trapped air and then cured at 70 °C for 30 minutes. AB composites were fabricated over a range of Ni loadings from 5wt% to 88 wt%. Three independently prepared samples were fabricated for the optimal and highest-performing compositions, and reported values represent the mean of the three replicates.

For ABBA composites, the AB scaffold (10.6 vol%, 5 μm-diameter particles) was infiltrated with a secondary blend-then-align (BA) mixture (**Figure 1b**). To prepare the BA mixture, Ni (5 μm-diameter) particles and Ga-In LM were combined using a mortar and pestle under ambient conditions. Ecoflex precursor (Parts A+B) was then added and gently mixed with a glass rod to form a uniform Ni/LM/Ecoflex paste. This mixture was injected into the uncured AB structure while maintaining the magnetic field. The LM-coated Ni particles infiltrated selectively into particle-particle gaps along the pre-aligned Ni columns. For ABBA composites, three independently prepared samples were fabricated at each loading. All reported values correspond to the mean of the three replicates.

*Processing duration study.* Dry Ni particles assembled into vertically aligned microcolumns instantaneously upon placement of the mold on the permanent magnet. This alignment is immediate and repeatable, as documented in **Video S1,** where a sample is repeatedly inserted and removed from the magnet**.** The subsequent Ni–LM co-grinding step required 5–7 min, after which the Ecoflex precursor (parts A and B) was added and gently mixed for approximately 2 min to yield a homogeneous Ni/LM/Ecoflex paste. Infiltration of this paste into the pre-aligned scaffold required approximately 1 min, followed by 10 min of vacuum degassing to remove entrapped air prior to thermal curing at 70 °C for 30 min. The total elapsed time (excluding curing) was therefore approximately 20 min per sample, throughout which the pre-aligned scaffold was continuously maintained under the applied magnetic field.

*Characterization:* The composite microstructure was characterized by optical microscopy (AM7115MZT Dino-Lite Edge 3.0), scanning electron microscopy (SEM, XL30 ESEM FEG, Philips) coupled with energy-dispersive X-ray spectroscopy (EDS), and micro-computed tomography (μ-CT, TECAN UniTOM XL, TESCAN GROUP, Brno, Czech Republic). Thermal conductivity (κ) was determined using a Light Flash Apparatus (LFA HyperFlash 467, LFA46700B-0805-L, Netzsch, Germany ) under controlled ambient conditions. Mechanical properties were evaluated using a universal tensile testing machine (Instron 5943, Norwood,

USA) with a crosshead speed of 10 mm min$^{-1}$. For the heat-dissipation experiments, the surface temperature evolution was monitored with an infrared camera (Py640i, Optris GmbH) under steady-state heating.

The in-plane electrical conductivity of the composites was measured using a four-point probe setup connected to a digital multimeter (Keithley 2000). The method follows the Ossila protocol for sheet resistance measurements, which incorporates geometric correction factors to account for finite sample dimensions. Each sample was prepared as a flat film (~1 mm thick), and measurements were performed at room temperature under ambient conditions. Through-plane electrical conductivity was measured using a two-probe configuration, where the sample was sandwiched between electrodes and connected to a precision multimeter (Agilent 34401A). Probes were attached onto silver droplets deposited on the top and bottom surfaces of the composite to ensure good electrical contact.

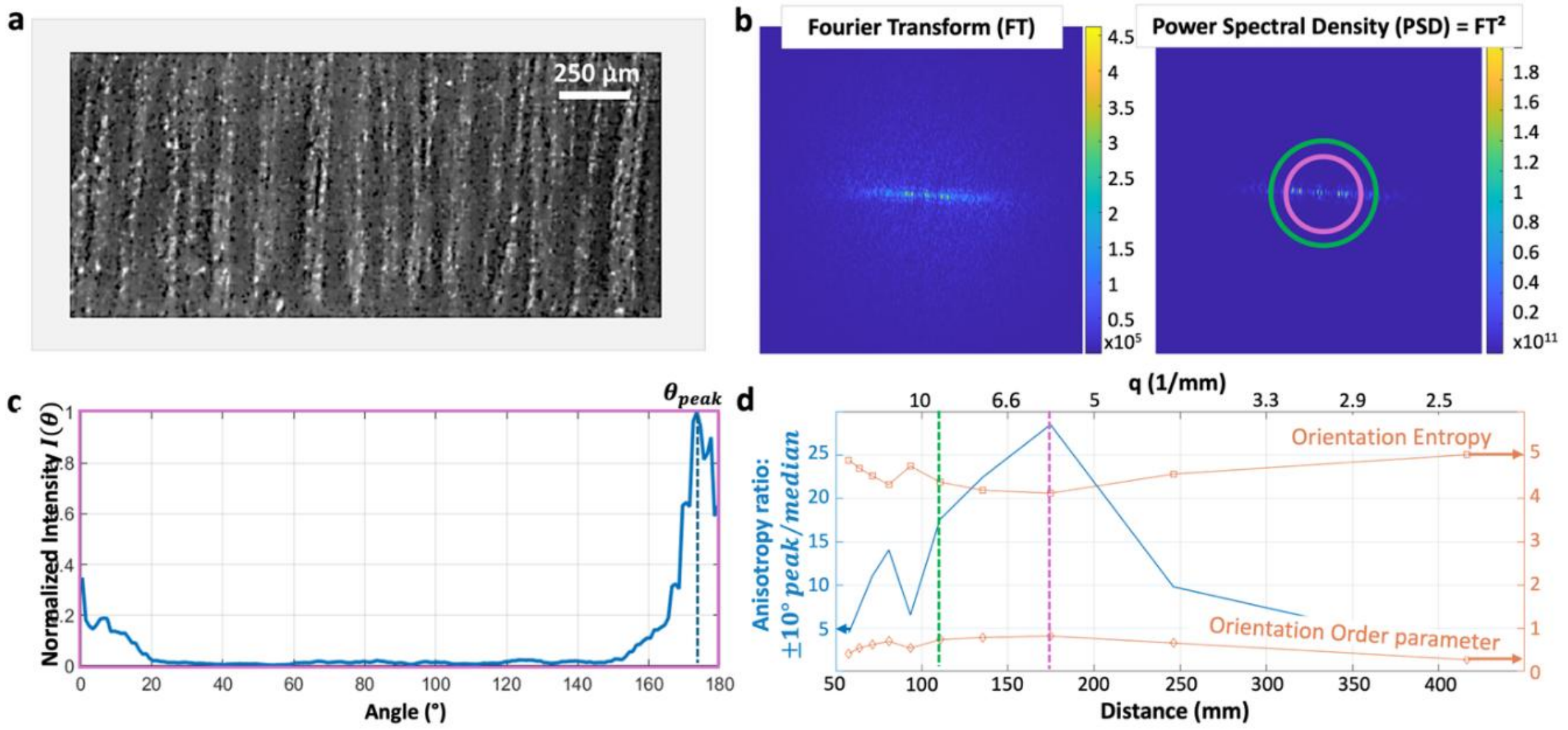

**Figure 6.** Quantitative characterization of structural anisotropy using Fourier and angular analysis. (**a**) Optical cross-section image. (**b**) Resulting Fourier Transform (FT) and PSD map. (**c**) Azimuthal integration of a discrete $q$-bin (example for the pink circle in panel b) provides orientation-dependent intensity profiles. (**d**) The anisotropy of each $q$-bin (pink and green lines correspond to the pink and green circles in panel b) was quantified using the PSD anisotropy ratio, orientation order parameter $R$, and Shannon orientation entropy $H$.

Microstructural anisotropy was quantified using Fourier Transform (FT) of cross-section images. Optical images were converted to grayscale, demeaned, edge-smoothed, and zero-padded prior to computing the 2D FT and its Power Spectral Density (PSD) (**Figure 6a,b**).

Azimuthal integration of the PSD was then performed within radial ($q$) bins, each corresponding to distinct length scales (pink and green circles as examples), yielding orientation-dependent intensity profiles I(θ) (**Figure 6c**).

For each q-bin, an anisotropy ratio was defined as the peak PSD intensity (within ±10° of the peak angle $\theta_{peak}$), divided by the median intensity. This ratio is particularly sensitive to single-alignment features such as magnetically aligned columns. The anisotropy ratio reported in **Figure 2a** corresponds to the $q_{peak}$ -bin with the strongest anisotropy (**Figure 6d** - pink line). This bin demonstrates the largest anisotropy ratio, the largest orientation order parameter *R*, and the smallest orientation entropy *H*; the characteristic column pitch *d* was obtained via $d=1/q_{peak}$.

*Contact-angle and dynamic deposition measurements.* Goniometry was performed using an OCA50 (Dataphysics, Germany) in a temperature- and humidity- controlled room at 23.5 ◦ C and 40 % RH, respectively. Two deposition methods are used to determine the apparent contact angle. The first quasi-static method gently deposits a 12 µl droplet on the sample surfaces and the contact angle is determined using the tangent fitting method due to the unusual shape of the droplet. The second dynamic method lets a droplet with a volume in the range of 20-24 µl fall from a height of ~7 cm to achieve a dynamic deposition of the droplet, breaking the oxide layer formed on the LM surface. The resulting contact angle is measured using the polynomial fitting method.

*Dielectric breakdown strength testing.* Four film specimens were tested for each material. Samples were cut into 14 × 10 mm² films with a thickness of 0.5 mm. Silver conductive ink was used to form 6 × 3 mm² in-plane electrodes separated by 1.1 ± 0.1 mm for Ecoflex and 1.3 ± 0.1 mm for the ABBA composite. A custom test fixture with steel electrodes and a 1 mm-thick microscope glass spacer was used to suppress air breakdown and ensure that failure occurred through the specimen/electrode gap. Film, electrode, and gap dimensions were measured using a TAGARNO TREND digital stereo microscope.

**Acknowledgements**

The authors thank the support from the KU Leuven Materials Engineering Department: Wout Veulemans for advice in powder processing, Iris Cuppens for support regarding chemical experiments, Danny Winant for the liquid nitrogen supply, Joop van Deursen, and the workshop personnel for helping with fabrication of molds. The KU Leuven XCT Core Facility is acknowledged for the 3D image acquisition and processing tools (https://xct.kuleuven.be/).

**Data Availability Statement**

The data supporting the findings of this study are publicly available in the KU Leuven Research Data Repository under DOI: https://doi.org/doi:10.48804/EWCNXH. The dataset includes processed experimental measurements and image-analysis data underlying Figures. 2–6 and Supplementary Information. Additional data is available from the corresponding author upon reasonable request.

**Supporting Information**

Supporting Information is available from the Wiley Online Library or from the author.

Magnetic fields organize nickel microparticles into vertical scaffolds before polymer infiltration, while liquid metal selectively bridges nanoscale gaps to enhance thermal coupling. This hierarchical strategy creates ultrasoft composites that transport heat efficiently through thickness while suppressing lateral electrical conduction, offering a new platform for safe and stretchable thermal management in wearable electronics.

**Magnetically Guided Sequential Multi-Filler Alignment Enables Anisotropic Heat Pathways in Stretchable Composites for Soft Electronics**

ToC figure:

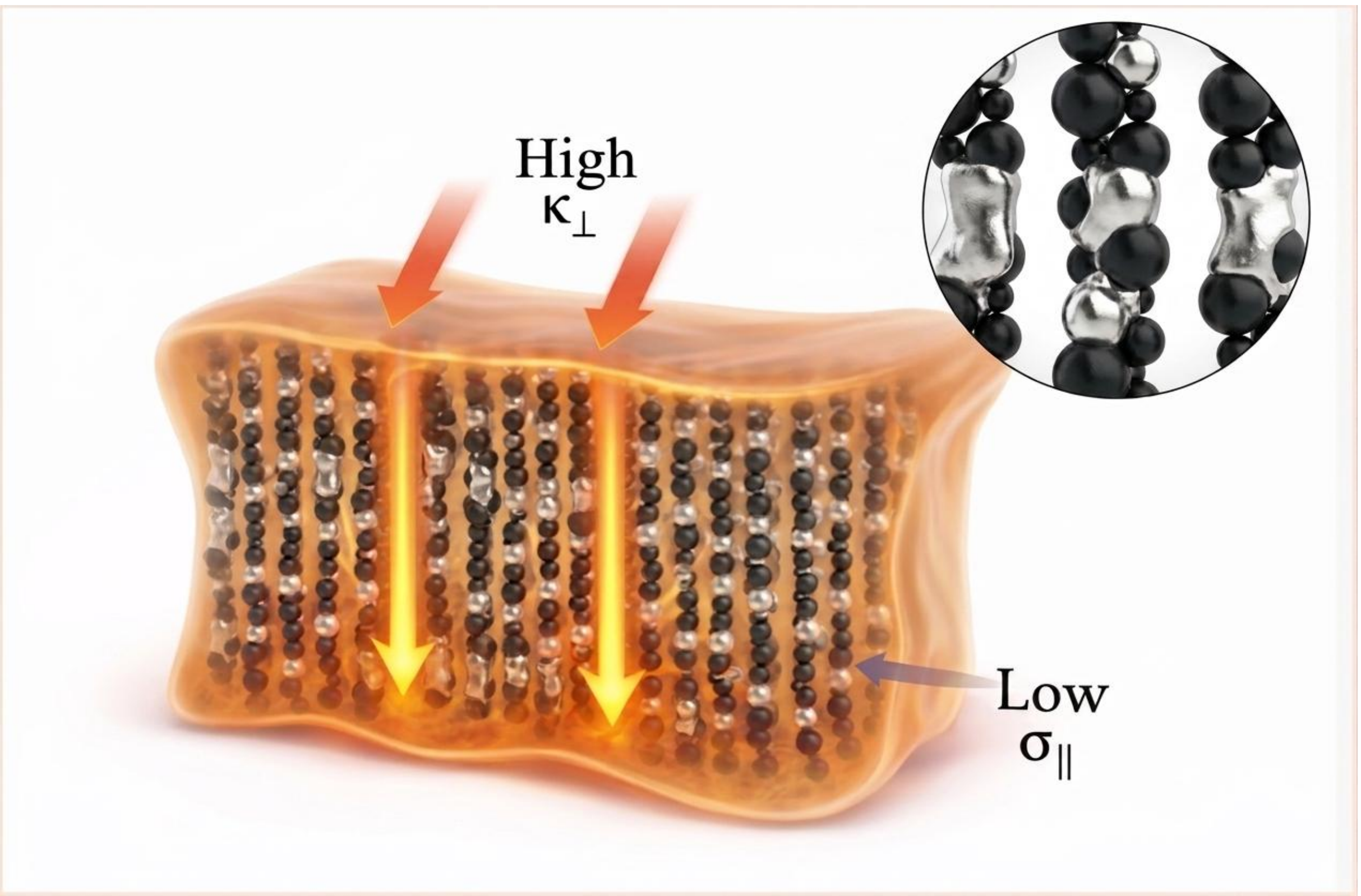

Supporting Information

**Magnetically Guided Sequential Multi-Filler Alignment Enables Anisotropic Heat Pathways in Stretchable Composites for Soft Electronics**

*Altynay Kaidarova, Viktor Naenen, Isidro Florenciano, Ruben Windey, Xiaoyan Qing, Laurens Snels, Juhyung Park, Tom Van der Donck, Jeroen Soete, David Seveno and Francisco Molina-Lopez**

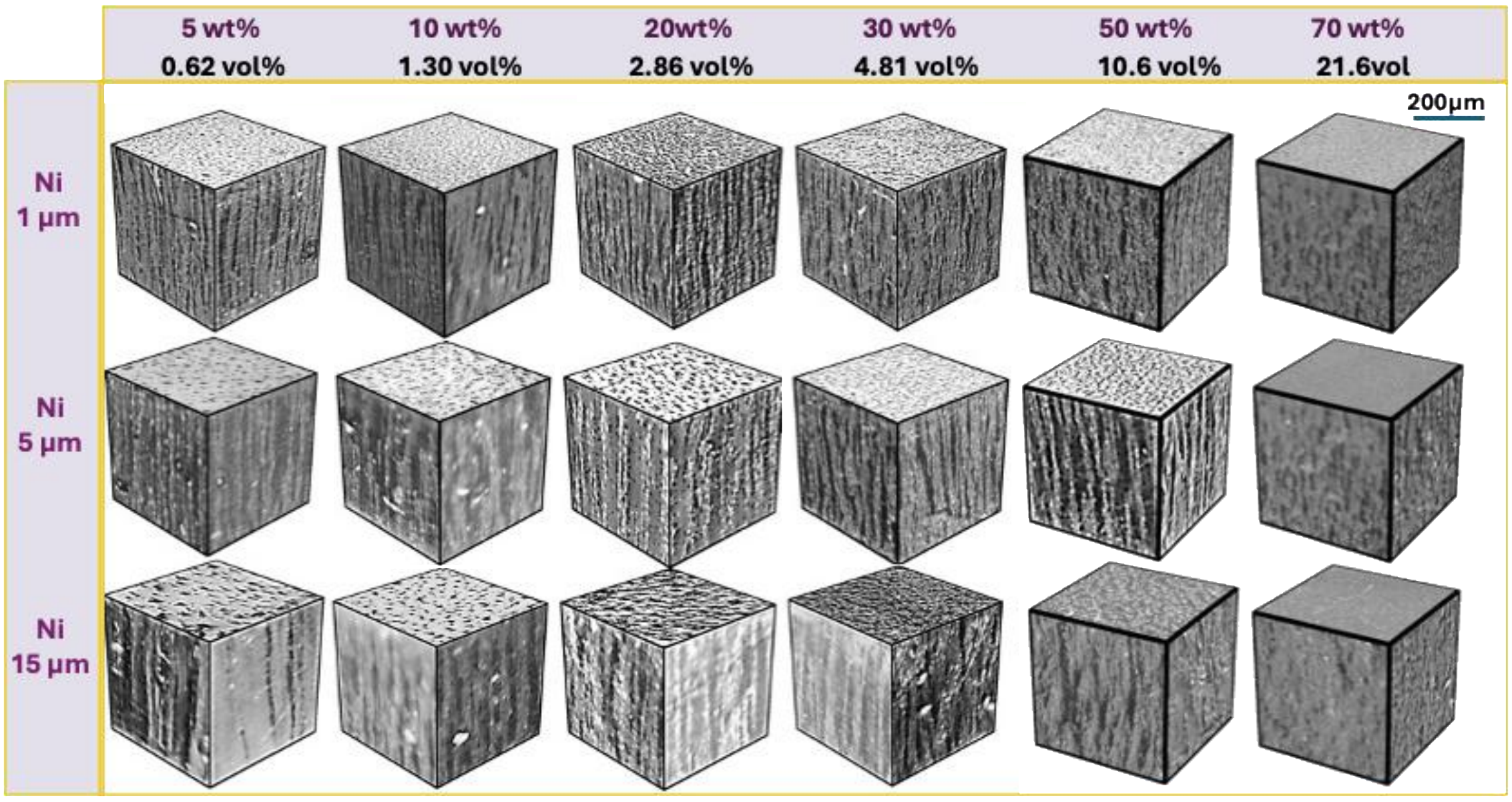

**Figure S1.** Optical 3D reconstructions of AB composites fabricated with 1 µm, 5 µm, and 15 µm Ni particles at increasing filler loadings (0.62–21.6 vol%). The reconstructions visualize the evolution of vertically aligned conductive pathways as a function of particle size and concentration, highlighting the transition from discrete columnar features to densely packed, continuous networks.

**Temporal stability of the dry Ni aligned scaffold.** To assess whether the columnar structure of the magnetically aligned Ni scaffold evolves over processing-relevant timescales and whether Ni particles undergo undesired agglomeration or gravitational settling, the dry aligned scaffold was monitored in air by side-view optical microscopy under continuous field application. Representative cross-sectional images were collected at t = 1 min, 5 min, 15 min, 30 min, 1 h, 6 h, and 24 h (**Figure S2a**). To test the robustness of the alignment process to any gravity-driven rearrangement, the mold was deliberately positioned on an inclined stage at approximately 15–20° relative to the horizontal. In addition to naked-eye visual inspection, temporal image-difference analysis was performed. A pixel-wise absolute intensity difference map between the first and last frame was generated to visualize the total accumulated change over the 24 h observation period (**Figure S2b**). In both analyses, the aligned Ni scaffold region showed negligible variation within the timescale of the experiment, with absolute intensity difference values close to those of the substrate/background.

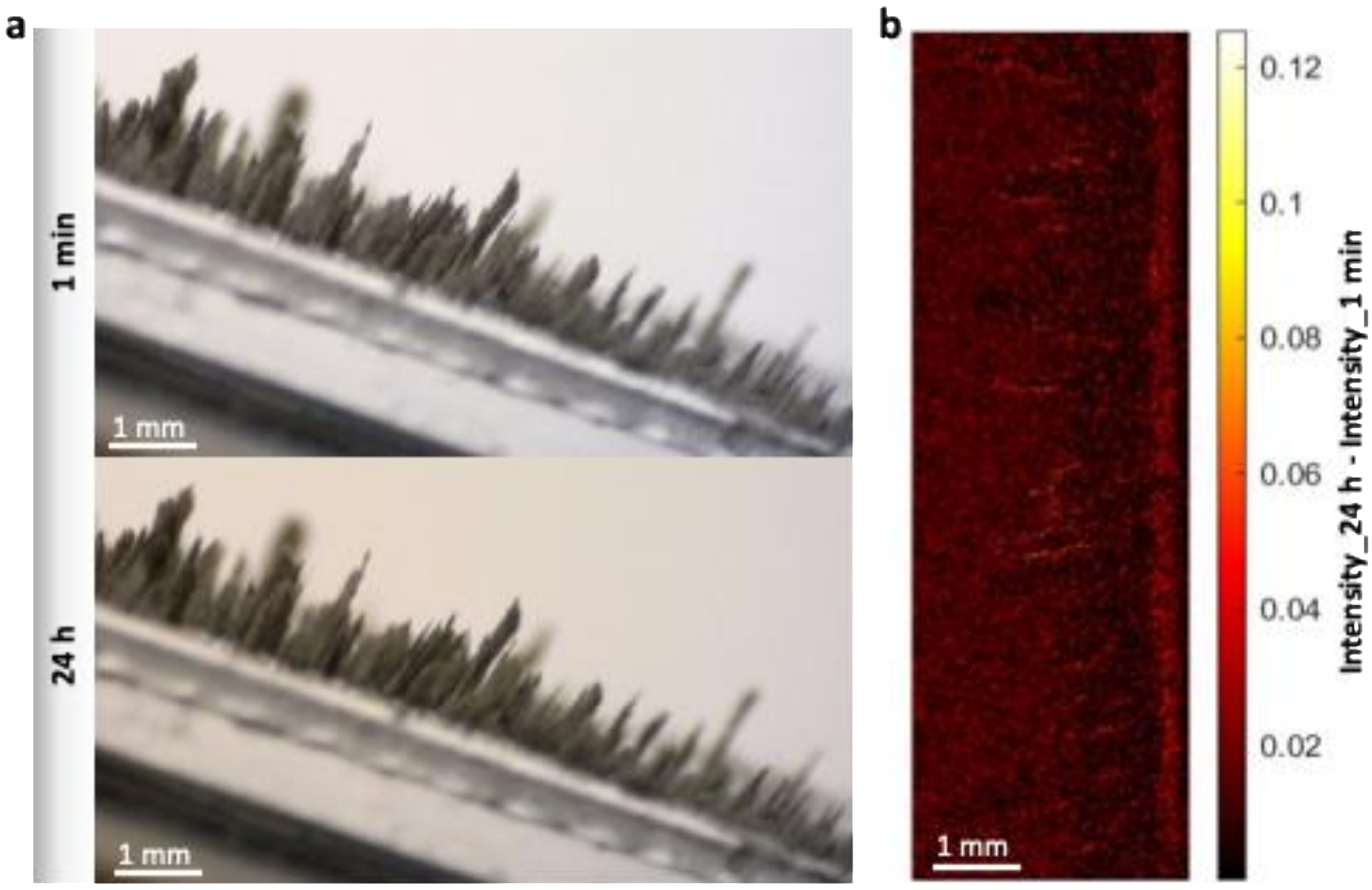

**Figure S2.** (a) Time-dependent side-view optical images of dry magnetically aligned Ni microcolumns in air, recorded under a continuously applied magnetic field at an angle of 15-20° from the horizontal. (b) Pixel-wise absolute intensity difference map between the first and last frame, visualizing the total image change over the 24 h observation period.

**Structural integrity during Ecoflex infiltration.** The response of the pre-aligned Ni scaffold to uncured Ecoflex infiltration was examined using time-resolved optical microscopy in both top-view and side-view geometries, including continuous video recordings (**Videos S2 and S3**) and representative timestamp images (**Figures S3-S6**).

In the top-view sequence (t=0–90 s; **Figure S3**), the Ni microcolumns cross-sections appear as dark dot-like against a bright background. During infiltration, the Ecoflex precursor advances across the field of view as a visible wetting front. The Ni columns appear to shift position and deform when the matrix first reaches them, but it is hard to determine whether this is an optical effect or an actual transient movement. To assess permanent structural changes during this process, the same region of interest was analyzed before and after infiltration (**Figure S4a**). The image was segmented to identify individual scaffold features. A red-blue overlay **(Figure S4b**) showed that the scaffolds underwent slight collective displacement consistent with limited field-constrained motion during infiltration. The overall column size and distribution were preserved, with the median filler chord length remaining approximately 37 μm (**Figure S4c**). The number of segmented scaffold objects remained similar, and no evidence of inter-column merging, fragmentation, or dissolution of the columnar pattern was observed.

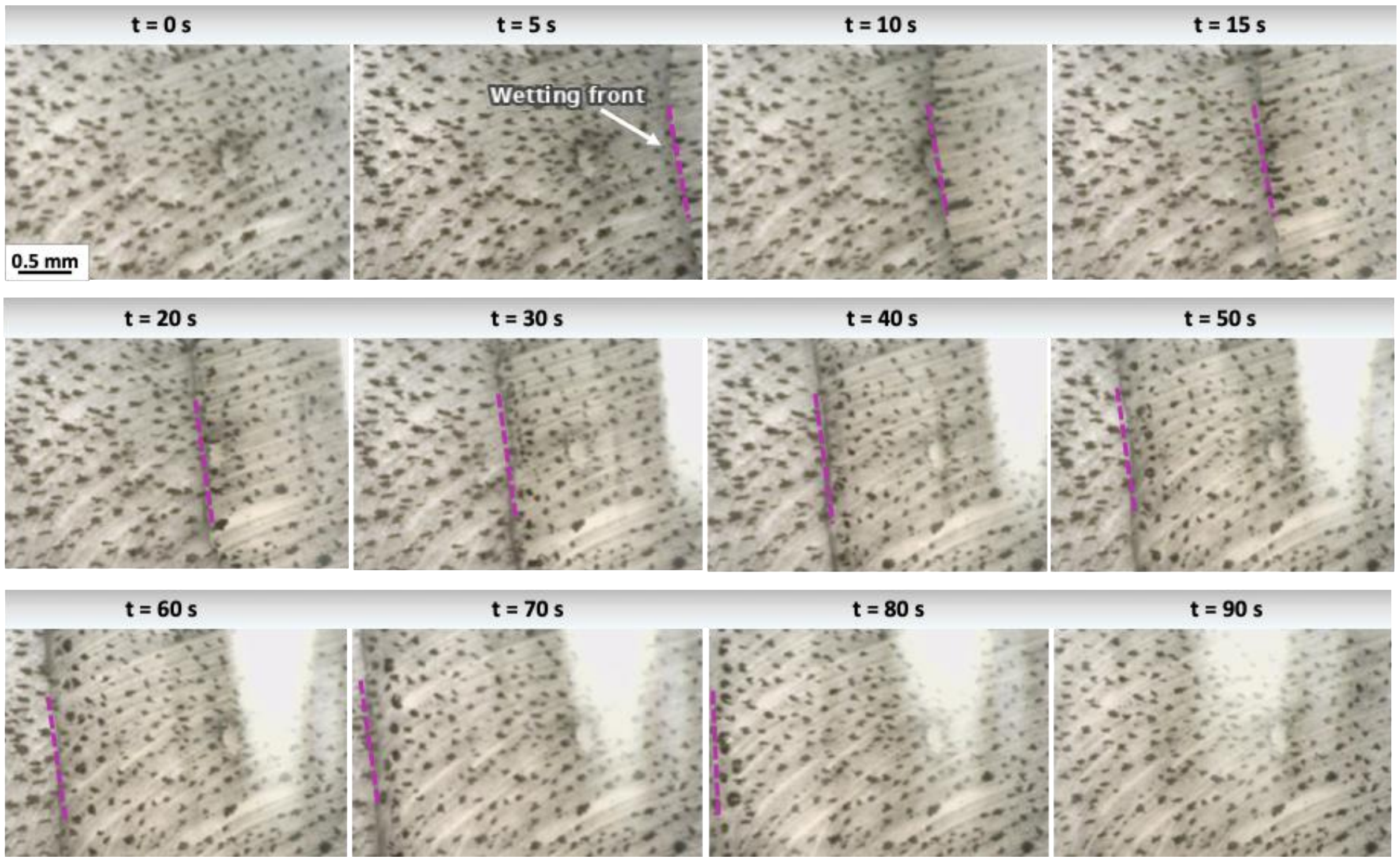

**Figure S3**. Time-resolved top-view optical microscopy of Ecoflex precursor infiltration into a preformed magnetically aligned Ni scaffold (the whole process takes place under the field).

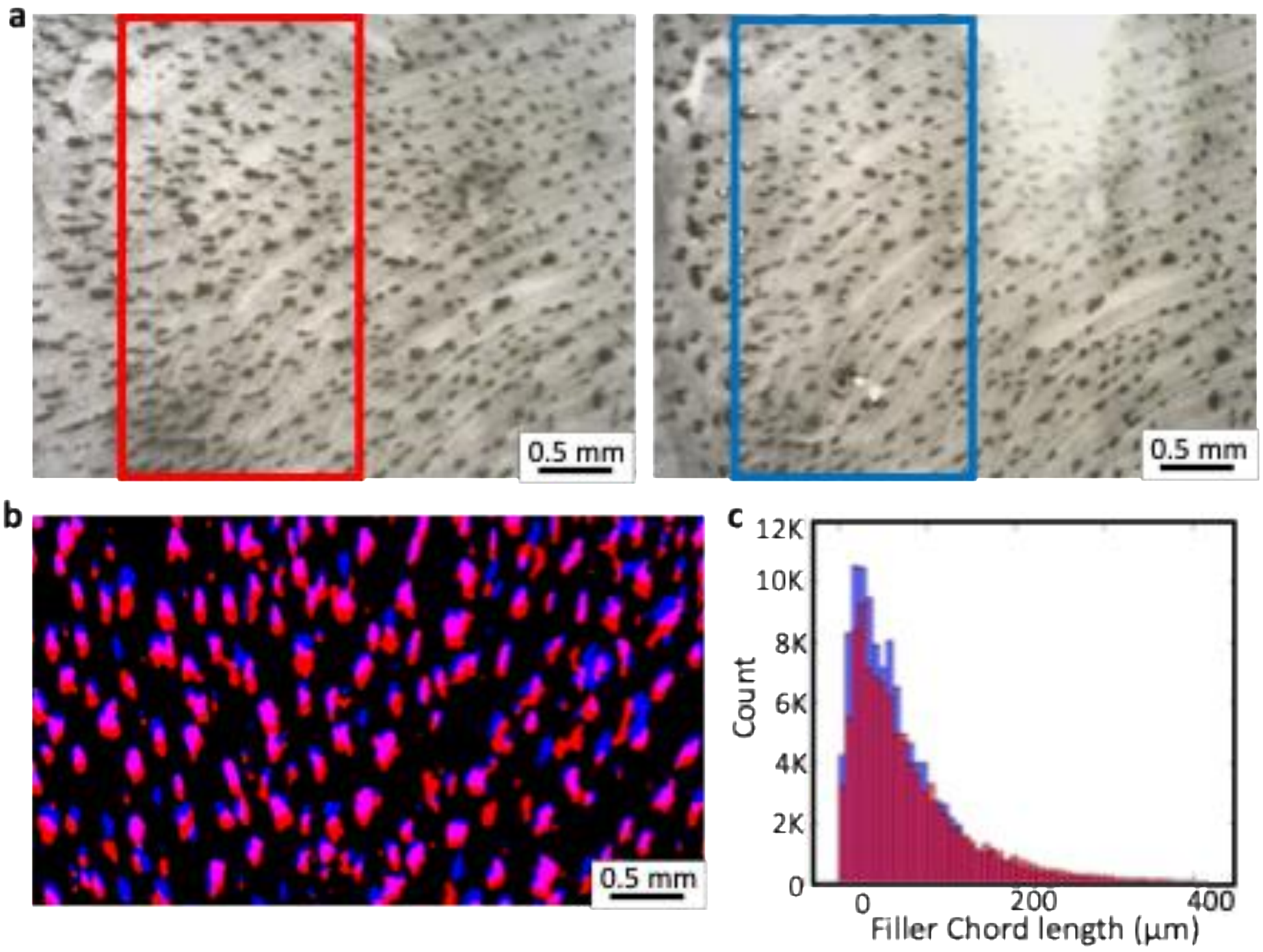

**Figure S4**. (a) Top-view optical images of aligned Ni columns before (t = 0 s) and after (t = 90 s) passage of the Ecoflex infiltration front, with analyzed regions marked in red and blue. (b) Overlay of segmented images before (red) and after (blue) infiltration, showing limited Ni column displacement and no fragmentation or merging. (c) Chord-length distributions before (red) and after (blue) infiltration, showing broadly preserved size distribution and the same median chord length after infiltration.

The side-view time-resolved sequence further supports this conclusion (**Figure S5**). As the Ecoflex precursor infiltrates the scaffold, the Ni microcolumns undergo limited positional adjustment. However, the columns remained continuous and vertically oriented throughout the t=4–20 s observation window, with no obvious fracture, collapse, or loss of columnar integrity.

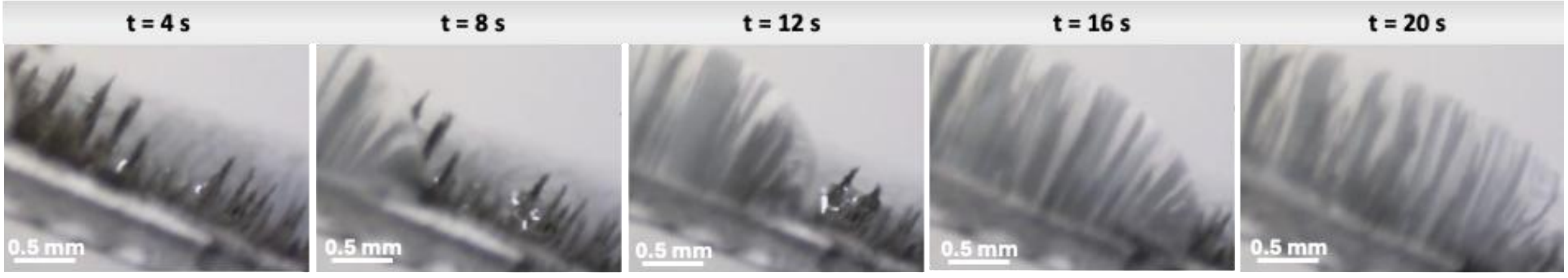

**Figure S5**. Time-resolved side-view optical microscopy of Ecoflex infiltration into a preformed magnetically aligned Ni scaffold.

Long-term post-infiltration stability under a constant magnetic field was further assessed using top-view optical images acquired after completion of Ecoflex infiltration at different time points (**Figure S6a**). Drift-corrected overlays of consecutive frames (**Figure S6b**) show that the spatial distribution of the Ni microcolumns remains unchanged. No pillar fracture, merging, collapse, or delayed redistribution was observed. A minor apparent mismatch in the earliest transition is likely attributable to an unintentional camera angle shift rather than true scaffold rearrangement, as the mismatch varies across the field of view and coincides with changes in illumination.

Taken together, this analysis indicates the micro-columnar architecture appears to be preserved during and after infiltration as long as the magnetic field is present, with no evidence of uncontrolled agglomeration, column breakup, collapse, or delayed post-infiltration restructuring.

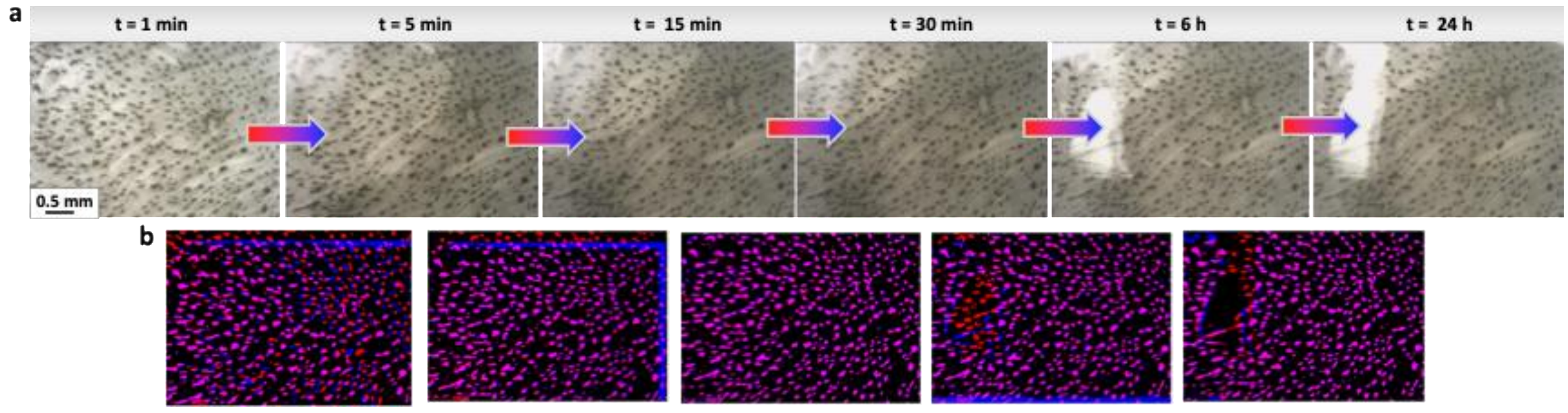

**Figure S6** (a) Top-view optical images acquired after completion of Ecoflex infiltration at different time points. (b) Drift-corrected overlays of consecutive frames. The earlier frame is shown in red and the later frame in blue; overlapping regions appear pink. The strong overlap between consecutive frames indicates that the spatial particle pattern remains stable, with no obvious pillar fracture, merging, collapse, or delayed redistribution after infiltration. Minor mismatch in the earliest transition is attributed to an unintentional camera angle drift rather than a structural change.

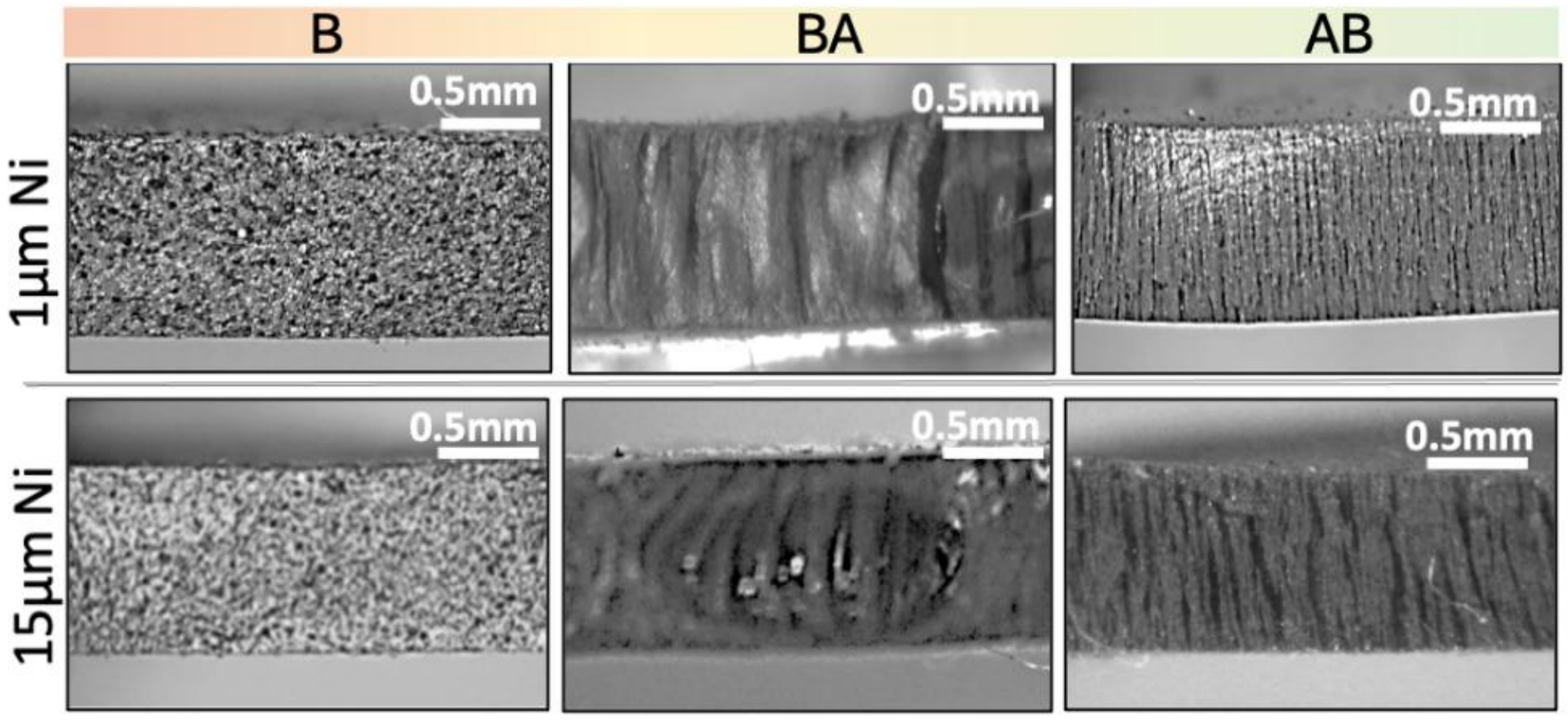

**Figure S7.** Cross-sectional optical micrographs of blend-only (B), blend-then-align (BA), and align-before-blend (AB) composites fabricated using 1 μm and 15 μm Ni particles. The images reveal how both processing sequence and particle size influence the formation and continuity of vertically aligned filler pathways.

**Morphological characterization of Ni–LM agglomerates.** SEM characterization of the mechanically ground Ni–LM (5 µm diameter) powder revealed agglomerates with heterogeneous surface morphology (**Figure S8 a–c**). The agglomerates contained rough granular regions, consistent with Ni-rich particle clusters, likely merged by LM (see also **Figure S20**), together with smoother and locally flattened regions consistent with surface-redistributed LM. Agglomerate-size analysis from calibrated SEM images (using ImageJ) yielded an apparent projected agglomerate size (equivalent circular diameter) of 49 ± 6 µm, with a corresponding Feret diameter of 59 ± 6 µm, indicating irregular Ni–LM agglomerates containing likely a few hundreds of Ni particles. BSE imaging exhibited contrast heterogeneity within the agglomerates, indicating local variation in composition and/or surface topography (**Figure S8b**). At higher magnification, the closely packed Ni particulate features exhibit a cauliflower-like texture (**Figure S8c**). The EDS overlay is consistent with the coexistence of Ni agglomerates and EGaIn interpenetrated within the agglomerates and around their surface (**Figure S8d**). These results are consistent with mechanical grinding producing mixed Ni–LM agglomerates in which LM is locally associated with Ni-rich particle clusters, rather than a simple mixture of isolated Ni particles and separate spherical LM droplets.

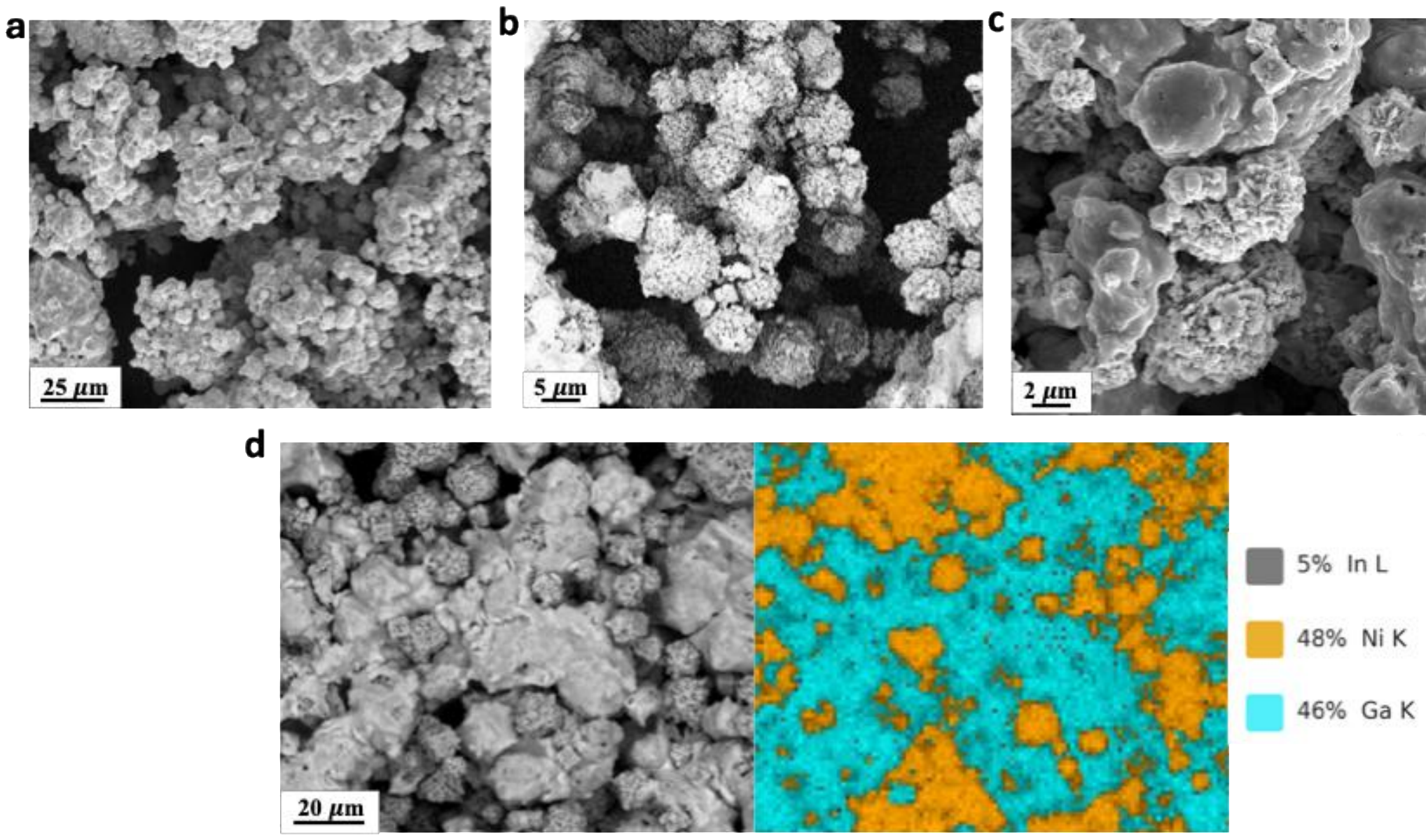

**Figure S8.** (a) SE-SEM overview (ETD detector) showing multi-particle Ni–LM agglomerates with characteristic cauliflower texture, formed by LM-mediated assembly of individual Ni particles during co-grinding. (b) Contrast-enhanced BSE image showing heterogeneous contrast within the agglomerates, consistent with compositional variation. (c) High-magnification SE image (ETD detector, PtPd-coated) showing two co-existing morphological motifs: smooth, rounded LM-coated entities and cauliflower-textured Ni clusters. (d) SEM image (left) and corresponding EDS elemental overlay map (right) of a representative agglomerated region, confirming the association of Ni and EGaIn within the mechanically ground powder.

**Elemental verification by EDS point analysis.** To support the morphological interpretation, EDS point/area analysis was performed on representative smooth and cauliflower-textured regions of the mechanically ground Ni–LM powder (**Figure S9**). Three representative measurements spanning the observed compositional range are summarized in **Table S1**. A region assigned morphologically as LM-rich showed 12.8 wt% Ni, 59.5 wt% Ga, and 7.8 wt% In, consistent with an EGaIn-dominated phase with minor Ni. A cauliflower-textured region assigned as Ni-rich contained 92.5 wt% Ni and 4.0 wt% Ga (In in low amounts was undetectable), consistent with a Ni-rich particle cluster with only minor LM-derived signal. A region positioned at the geometric boundary between a smoother LM-rich area and an adjacent Ni-rich cluster showed an intermediate composition of 62.3 wt% Ni and 36.1 wt% Ga, which sits between the two phases and is assigned to a mixed junction. These three measurements span a compositional gradient from Ni-rich to LM-rich, with the boundary region showing intermediate character consistent with local Ni–LM association at the contact zone. We note that EDS point analysis does not resolve interfacial geometry at the nanoscale, and the measured compositions reflect the average over the EDS excitation volume rather than a sharp interface.

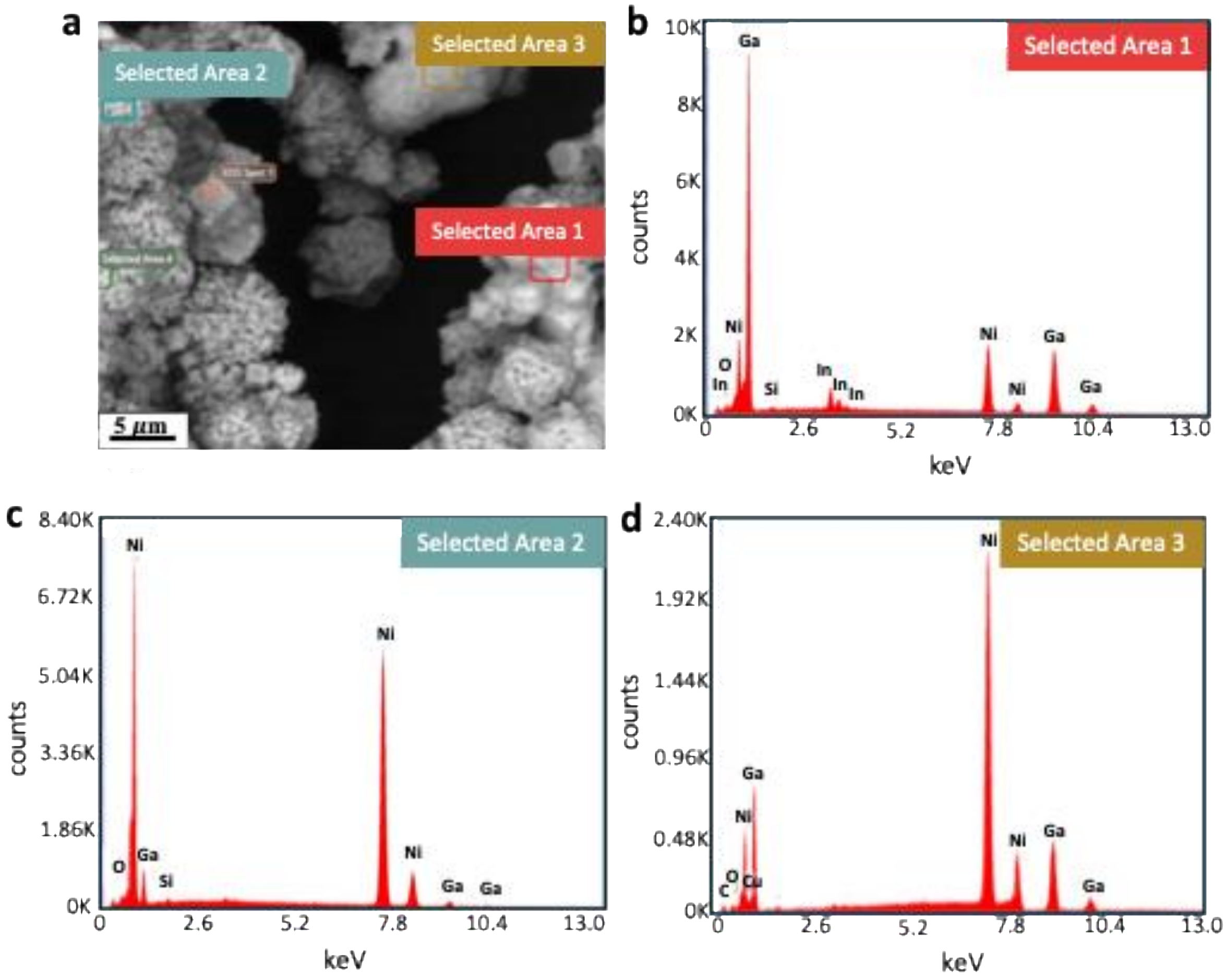

**Figure S9.** (a) SEM image showing selected regions used for EDS analysis. (b) EDS spectrum of an LM-rich region. (c) EDS spectrum of a Ni-rich cauliflower-like granular region. (d) EDS spectrum of a mixed Ni–EGaIn region showing dominant Ni peaks together with Ga contribution.

**Table S1. EDS point analysis of Ni–LM powder agglomerates.**

| Measurement | Location description | Ni (wt%) | Ga (wt%) | In (wt%) | O (wt%) | Phase assignment |
|---|---|---|---|---|---|---|
| Selected Area 1 | Smooth LM region | 12.82 | 59.5 | 7.81 | 1.49 | LM-rich |
| Selected Area 2 | Granular Ni cluster | 92.52 | 3.97 | n.d. | 2.86 | Ni-rich |
| Selected Area 3 | LM–Ni contact zone | 62.26 | 36.14 | n.d. | 0.74 | Mixed junction |
| *n.d. = not detected. All measurements at 20 kV, eZAF Smart Quant* | | | | | | |

**Large-area phase mapping.** EDS mapping over a large field of view shows that Ni and Ga signals are broadly co-distributed across the agglomerated powder, rather than appearing as completely separated Ni-only particles and large isolated LM droplets (**Figure S10**). As expected for the eutectic alloy, Ga and In overlap. This spatial co-distribution is consistent with mechanical grinding promoting local Ni–LM association and redistribution of LM over/around Ni-rich particle clusters.

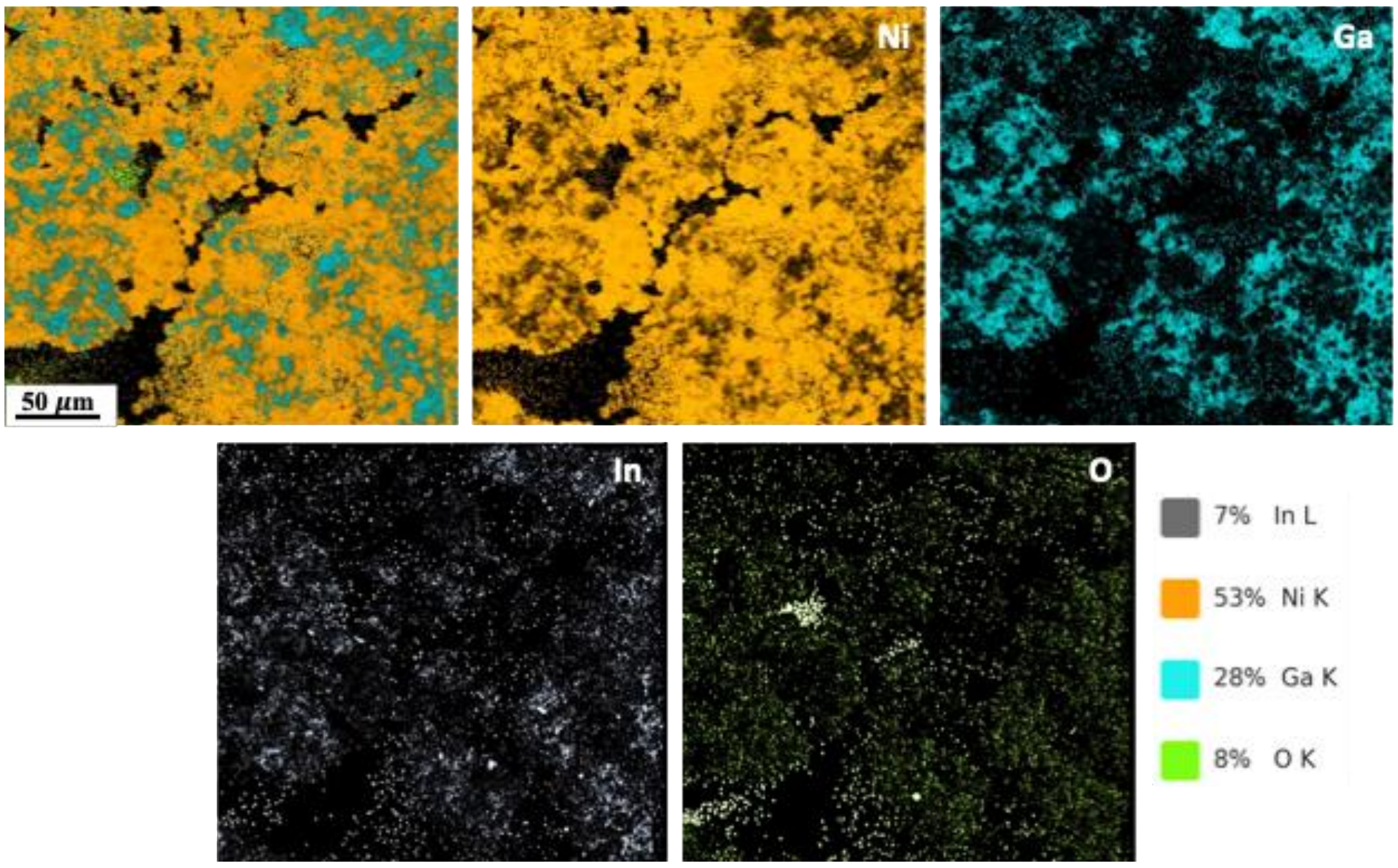

**Figure S10.** Large-area EDS elemental overlay showing the spatial distribution of Ni, Ga, In, and O within the ground Ni–LM powder.

**Apparent contact of EGaIn.** Our claim that LM preferentially associates with Ni particles to form bridges rather than mixing with the Ecoflex matrix stems from differences in processing rather than equilibrium wettability between materials. This was evaluated using two contact-angle/deposition conditions. In the quasi-static deposition condition, EGaIn was deposited gently from the needle onto an Ecoflex or Ni substrate using standard sessile-drop contact-angle procedure (Dataphysics OCA50). The oxide-skinned droplet exhibited a high contact angle and characteristic tip on both substrates, indicating limited static spreading regardless of the substrate material **(Figure S11).** The dynamics of this test are representative of the processing steps of Ni/LM gently mixing with Ecoflex, and magnetic alignment, where the oxide skin remains largely intact, suppressing the mixture of LM into the polymer matrix. To explain the observed association of LM and Ni, we resort then to dynamic impact deposition test, in which a 20–24 µL EGaIn droplet was released from the needle from a ~ 7 cm height onto either Ni or Ecoflex substrate to mimic the rapid mechanical contact during mortar-and-pestle grinding. Dynamic impact likely breaks the native LM oxide skin, considerably reducing the apparent contact angle of LM on both substrates. Under dynamic deposition, the apparent contact angle decreased to **50 ± 10° on Ni** and **29± 6° on Ecoflex**. Therefore, during mortar-and-pestle grinding, repeated oxide rupture and forced mechanical contact promoted local Ni–LM association and formation of Ni–LM agglomerates. Then, a new oxide skin grew on those agglomerates, preventing further association with any material in subsequent process steps, unless mechanical force was employed. During the subsequent gentle mixing with Ecoflex precursor and magnetic alignment steps, the LM oxide skin in the Ni-rich agglomerates remained intact, preventing LM from spreading into the Ecoflex matrix. At the same time, the compliant and morphing nature of LM (even with the native oxide) presumably contributed to close Ni interparticle distance, explaining the observed morphology and the dramatic decrease in thermal contact resistance of ABBA composites (see **Figure S19**).

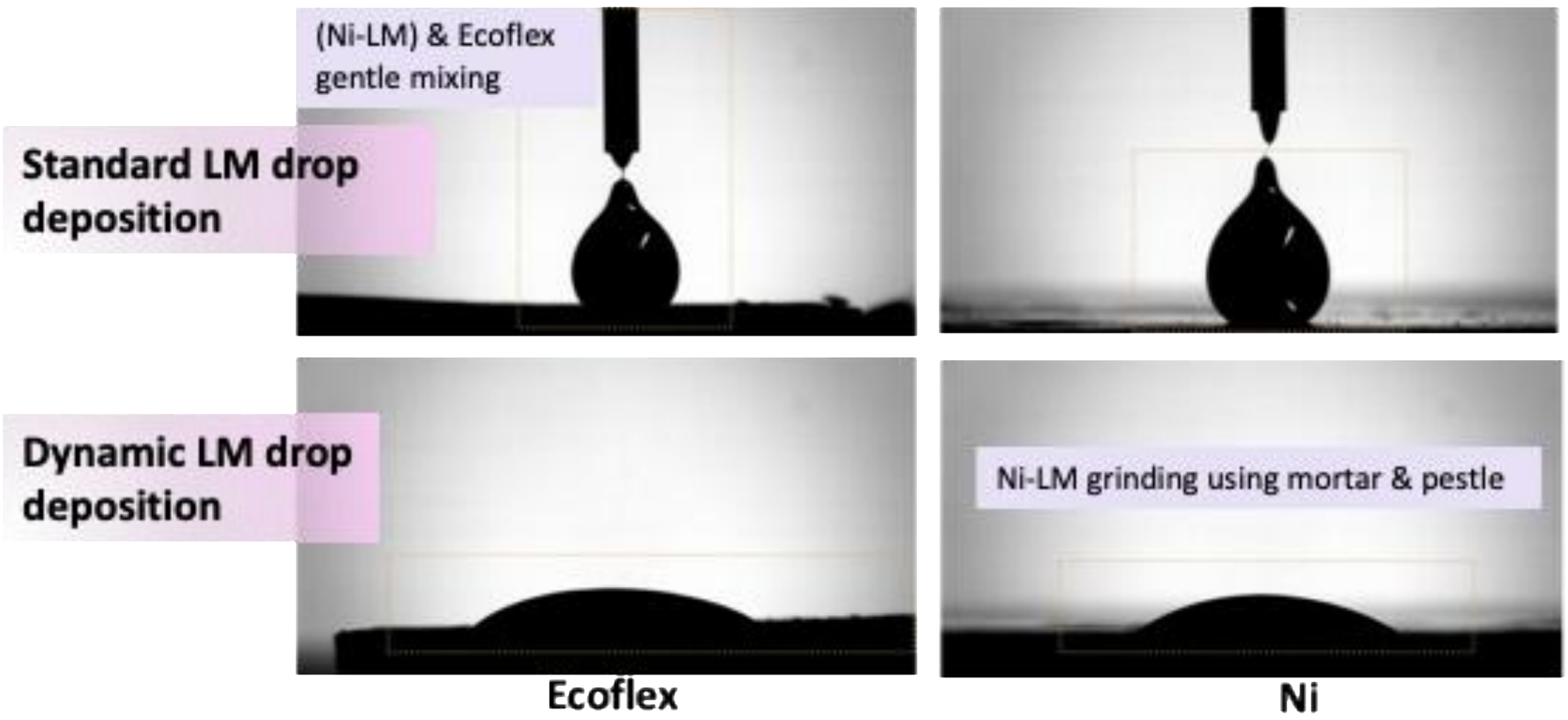

**Figure S11. Effect of deposition dynamics on apparent EGaIn wetting behavior on Ecoflex and Ni.** Under standard drop deposition, oxide-skinned EGaIn droplets showed roundness on both Ecoflex and Ni, corresponding to the gentle-mixing and magnetic alignment steps where the LM oxide skin remains largely intact. Under dynamic drop deposition, which mimics the high-shear mortar–pestle grinding step, oxide-skin disruption promoted spreading on both substrates.

**Quantitative alignment analysis.** To quantify the structural organization of the aligned fillers, we analyzed optical microscopy images using both chord-length (line-intercept) and Fourier transform-based methods.

First, top-view optical microscopy images were converted to grayscale and binarized using global thresholding to distinguish between the filler and matrix phases (**Figure S12a**). 10800 line probes were overlaid at multiple orientations. The lengths of consecutive intercepts in filler and matrix phase were measured to generate chord-length distributions **(Figure S12b)**. The median length of the filler-phase chords was defined as the effective column width, while the median matrix-phase chord length was used to estimate inter-column spacing **(Figure S12c).** The characteristic column pitch, $d$, was calculated as the sum of these two medians.

To validate this measurement, we performed Fourier transform (FT) analysis on the same images. The magnitude of the FT and its logarithmic scale are shown in **Figure S13a**. The power spectral density (PSD) was computed as the squared FT magnitude and integrated over the polar angle at each radial ($q$) value, yielding intensity as a function of spatial frequency $q$ (**Figure S13b**). The position of the dominant peak $q_{\text{peak}}$ was then converted to real-space periodicity using $d = 1/q_{\text{peak}}$. Both techniques show a relatively good agreement and reflect the same trend in column width and spacing as filler loading increases. Note that chord length-based spacings are systematically larger than FT spacings, with FT spacings being a lowest limit of chord-length spacings because FT registers only the nearest neighbours, while the arbitrary character of chord-length spacing include also non-nearest neighbours.

The column pitch for AB composites fabricated with different particle sizes and filler loading was calculated by the chord-length (**Figure S14**) and FT methods (from top-view and cross-sectional images, as shown in **Figure S15a** and **S15b**, respectively). Both methods show good agreement. The pitch for the AB composite with 10.6 vol% 5 μm-diameter Ni particles selected in this work was ~ 160 μm, dictating the minimum size of hot spots that could be efficiently mitigated by the proposed material.

The column pitch *d* decreases with increasing loading for all particle sizes (**Figure S14** and **Figure S15**), indicating tighter packing with reduced inter-column spacing. For small particles (1 μm), the pitch *d* stabilizes at relatively low filler loadings, suggesting an early formation of the final scaffold framework. In contrast, larger particles (5 μm and 15 μm) exhibit a more gradual decrease in *d*, consistent with a later formation of the final framework. Larger particles also lead to wider columns and larger pitches. Consistent with these observations, Figure S14 shows that the column width increases only slightly with filler loading, whereas inter-column spacing decreases sharply at low concentrations and then plateaus. This observation indicates that densification of the aligned scaffold is primarily driven by introducing new columns in the framework (evidenced by the reduction of matrix gaps) rather than by significant column thickening.

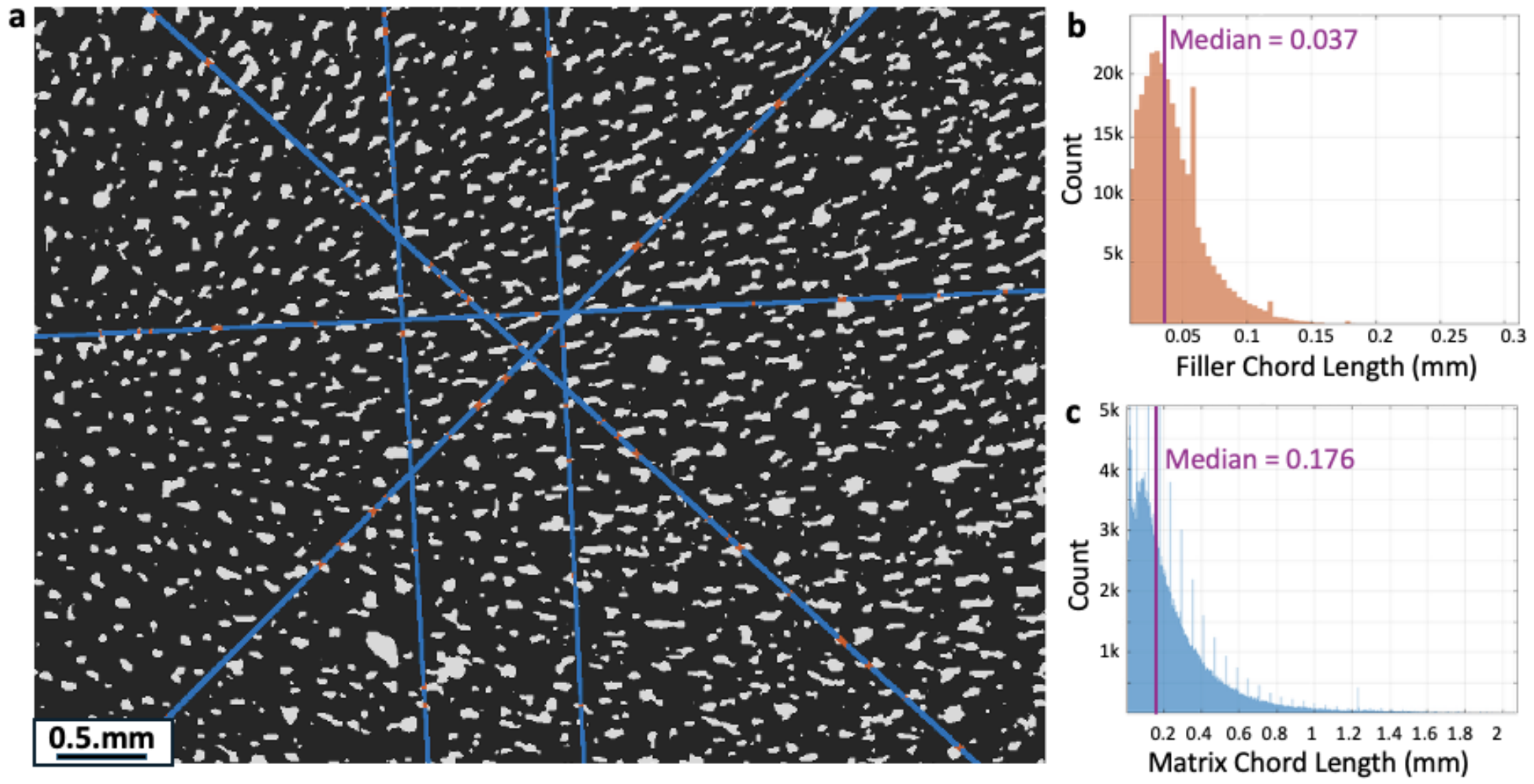

**Figure S12.** (a) Top-view optical picture of an aligned composite (AB, 0.62vol%, 1 µm-diameter Ni particles) to illustrate the chord-length analysis for 5 randomly selected line probes. The chord-length distributions of the filler (b, orange) and matrix (c, blue) phases were used to estimate the effective column width and inter-column spacing, respectively.

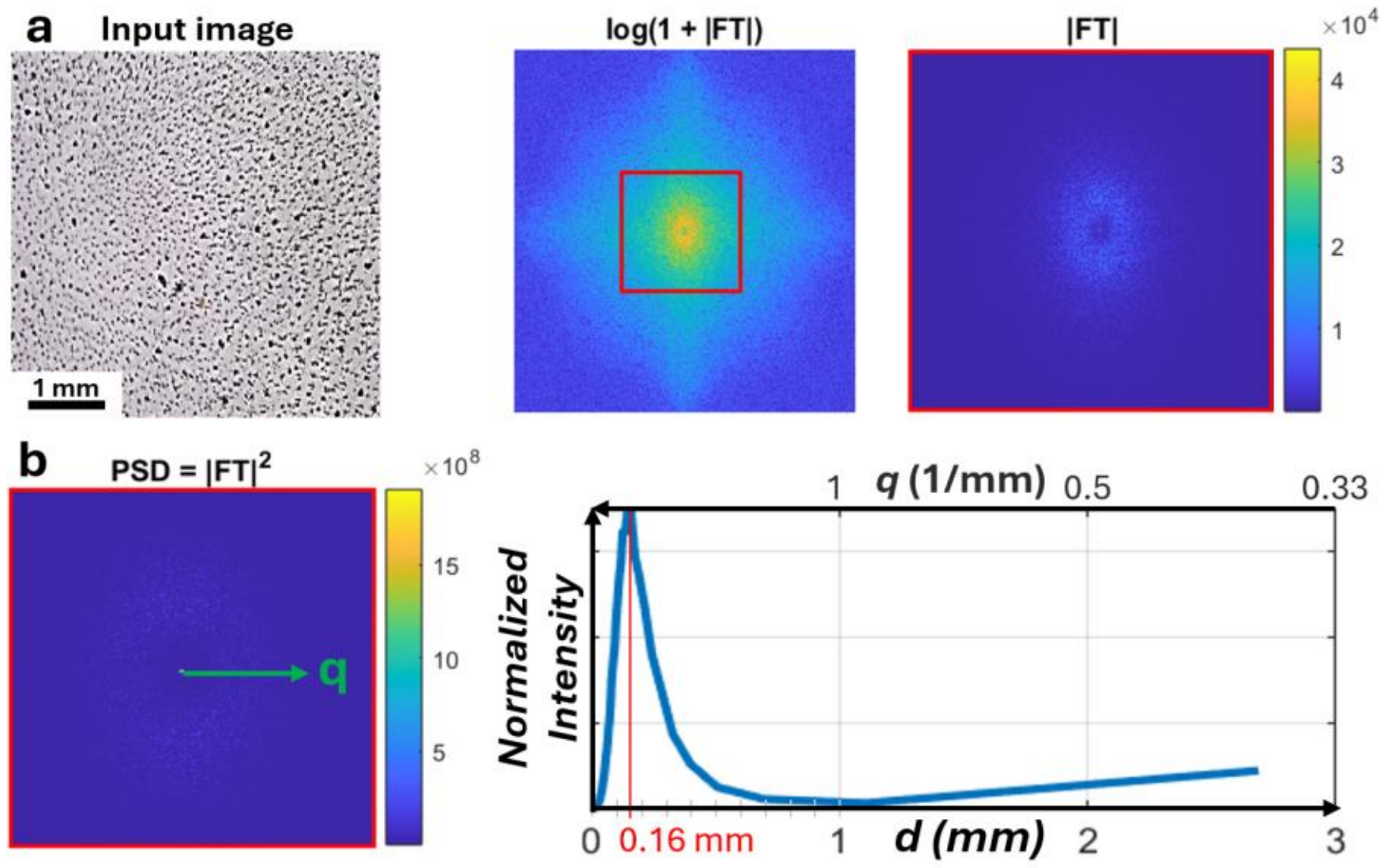

**Figure S13.** (a) Representative (logarithmic map of) Fourier transform for top-view microscopy image (AB, 0.62 vol%, 1 µm-diameter Ni particles). (b) The PSD is radially integrated to quantify the texture intensity versus spatial frequency ($q$) and corresponding real-space periodicity ($d = 1/q$).

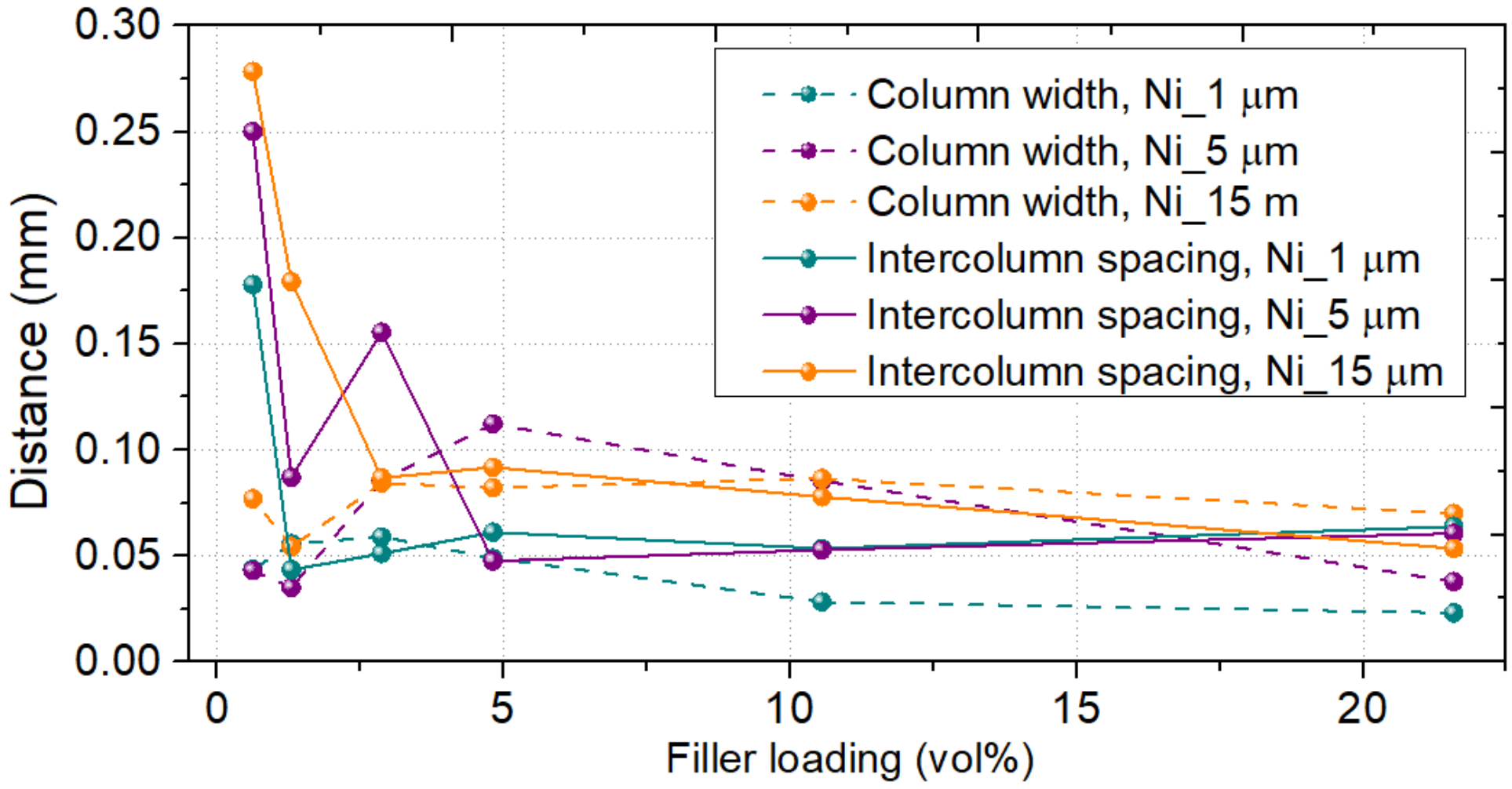

**Figure S14**. Chord-length analysis for AB composites: Evolution of effective column width and inter-column spacing with filler loading for different Ni particle sizes, extracted from top-view images. Values represent the mean of three independent samples (n = 3).

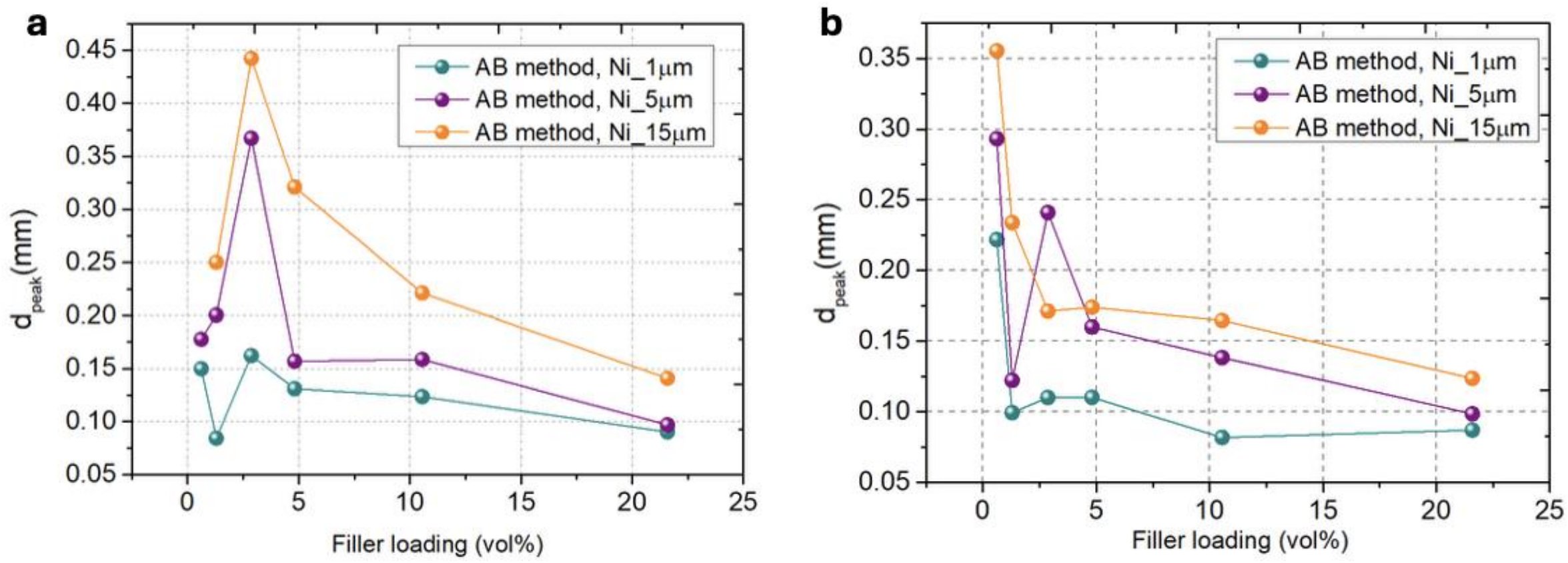

**Figure S15.** Fourier Transform analysis for AB composites: Pitch ($d_{peak}$) as a function of filler loading for different Ni particle sizes, extracted from (a) top-view, and (b) cross-section images. Values represent the mean of three independent samples (n = 3).

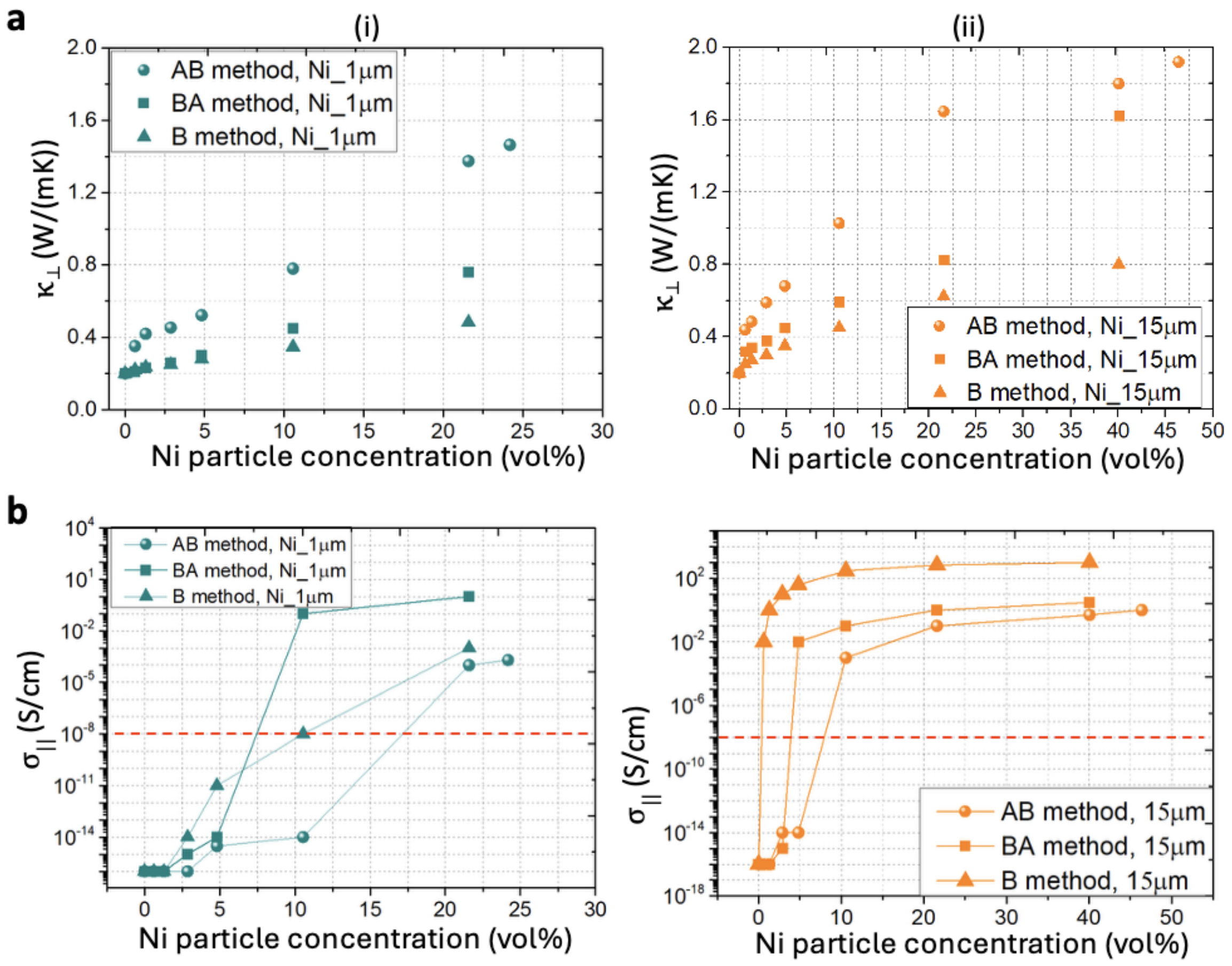

**Figure S16.** (a)Through-plane thermal conductivity and b) in-plane electrical conductivity of composites fabricated by B, BA, and AB routes for (i)1 μm and (ii) 15 μm Ni particles.

## Analytical model of thermal contact resistivity

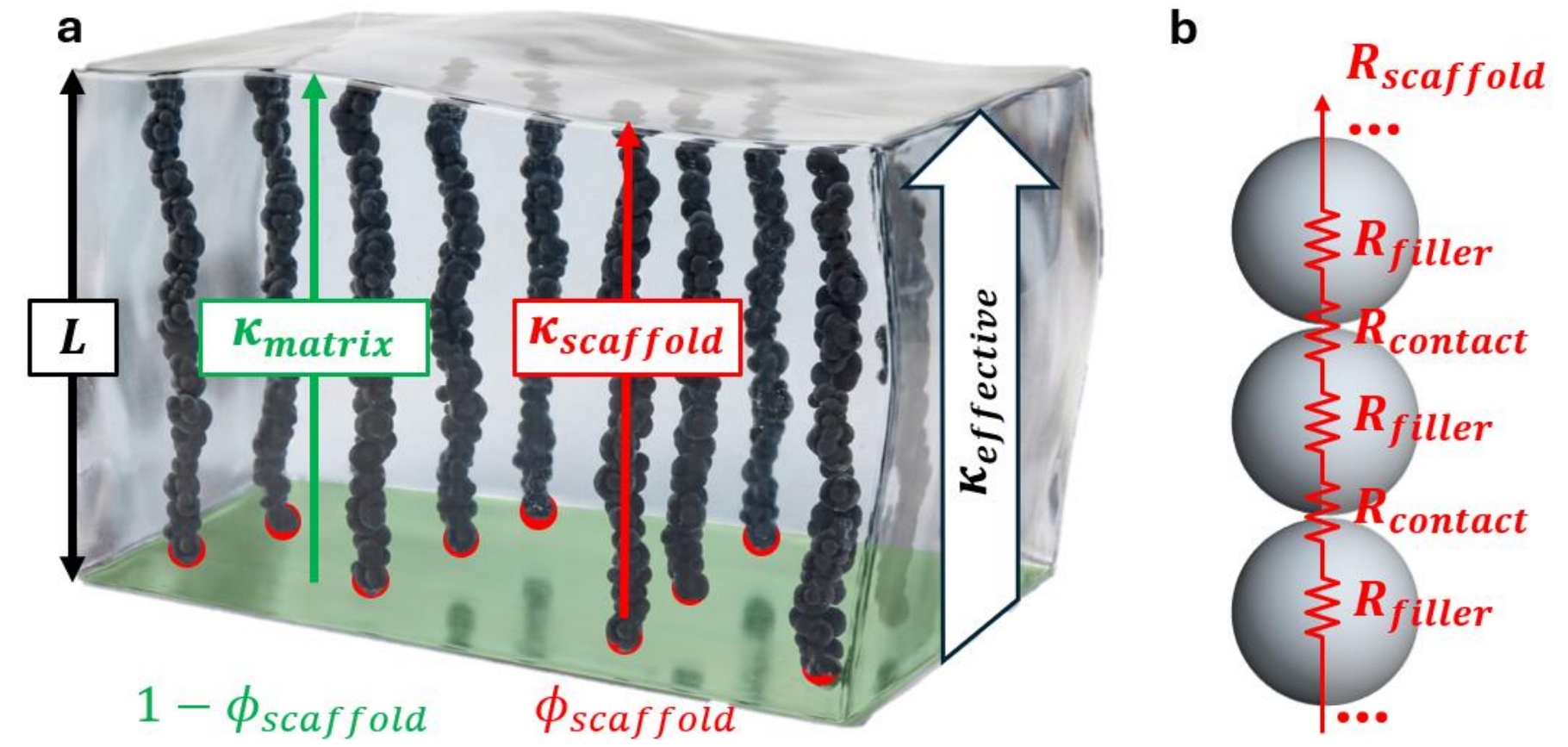

**Figure S17**. Effective thermal conductivity and network resistance model. (a) Schematic of heat transport in anisotropic composites comprising a matrix phase (green) and an aligned filler scaffold (red) with volume fraction $\phi_{scaffold}$. The effective thermal conductivity, $\kappa_{\mathrm{eff}}$, reflects contributions from both phases for a particular $\phi$. Independent measurement of $\kappa_{\mathrm{eff}}$ and $\kappa_{\mathrm{matrix}}$ enables the extraction of the scaffold conductivity $\kappa_{\mathrm{scaffold}}$. (b) The scaffold is modeled as a series of individual fillers ($R_{\mathrm{filler}}$), connected by inter-particle thermal contacts ($R_{contact}$). Estimating $R_{filler}$ allows the determination of the effective contact resistance within the scaffold.

### 1. Effective thermal conductivity and scaffold conductivity

The experimentally measured composite conductivity is denoted $\kappa_{\mathrm{eff}}$. This conductivity results from the heat flow through 2 distinct phases, as shown in **Figure S17**:

- a filler scaffold with conductivity $\kappa_{\mathrm{scaffold}}$, which is modeled as a cylinder
- a matrix phase with conductivity $\kappa_{\mathrm{matrix}}$.

The scaffold occupies volume fraction $\phi_{\mathrm{scaffold}}$, the matrix $1-\phi_{\mathrm{scaffold}}$.

For through-plane alignment, the two phases conduct in parallel:

$$R_{\mathrm{tot}} = R_{\mathrm{scaffold}} \parallel R_{\mathrm{matrix}} \tag{S1}$$

For a sample of thickness $L$and area $A$:

$$\frac{L}{A_{composite}\kappa_{\mathrm{eff}}} = \frac{L}{A_{\mathrm{scaffold}}\kappa_{\mathrm{scaffold}}} \parallel \frac{L}{A_{\mathrm{matrix}}\kappa_{\mathrm{matrix}}} \tag{S2}$$

With

$$A_{\text{scaffold}} = \phi_{\text{scaffold}} A_{composite}, \qquad A_{\text{matrix}} = (1 - \phi_{\text{scaffold}}) A_{composite}, \tag{S3}$$

we obtain the arithmetic mixing rule:

$$\kappa_{\text{eff}} = \phi_{\text{scaffold}} \kappa_{\text{scaffold}} + (1 - \phi_{\text{scaffold}}) \kappa_{\text{matrix}} \tag{S4}$$

Solving for the scaffold conductivity:

$$\boxed{\kappa_{\text{scaffold}} = \frac{\kappa_{\text{eff}} - (1 - \phi_{\text{scaffold}}) \kappa_{\text{matrix}}}{\phi_{\text{scaffold}}}} \tag{S5}$$

**2. Volume fraction correction for spherical fillers**

A spherical filler of diameter *D* occupies a volume

$$V_{\text{filler}} = \frac{\pi D^3}{6}, \tag{S6}$$

while a scaffold column, which is modeled as a cylinder of the same diameter *D*, presents a cell volume of

$$V_{\text{scaffold}} = \frac{\pi D^3}{4}. \tag{S7}$$

Thus the scaffold volume fraction is

$$\phi_{\text{scaffold}} = \frac{V_{\text{scaffold}}}{V_{\text{filler}}} \phi_{\text{filler}} = \frac{3}{2} \phi_{\text{filler}} \tag{S8}$$

Substituting into (S5):

$$\boxed{\kappa_{\text{scaffold}} = \frac{\kappa_{\text{eff}} - \left(1 - \frac{3}{2} \phi_{\text{filler}}\right) \kappa_{\text{matrix}}}{\frac{3}{2} \phi_{\text{filler}}}} \tag{S9}$$

This completes the determination of $\kappa_{\text{scaffold}}$ directly from measurable quantities.

## 3. Scaffold resistance and the role of inter-particle contacts

At the **macroscopic** level, the scaffold columns behave as a homogeneous conductor of conductivity $\kappa_{\text{scaffold}}$. Its thermal resistance is therefore

$$R_{\text{scaffold column}} = \frac{L}{A_{column}\,\kappa_{\text{scaffold}}} \tag{S10a}$$

At the **microscopic** level, however, the picture is more complicated. Each column in the scaffold consists of $N = L/D$ spherical fillers of diameter $D$, connected in series by thermal contacts, as shown in **Figure S17b**. Each filler contributes an axial resistance $R_{\text{filler}}$, and each inter-particle junction contributes a contact resistance $R_c/A_{column}$, with $R_c$ being the contact resistivity in [K $W^{-1}$ $m^2$]. Approximating $N - 1 \approx N$:

$$R_{\text{scaffold column}} \approx NR_{\text{filler}} + N\frac{R_c}{A_{column}} \tag{S10b}$$

Equating (S6a) and (S6b), dividing by $L$, and using $N = L/D$:

$$\frac{1}{A_{column}\kappa_{\text{scaffold}}} = \frac{R_{\text{filler}}}{D} + \frac{R_c}{A_{column}D} \tag{S11}$$

Thus:

$$\boxed{\frac{R_c}{D} = \frac{1}{\kappa_{\text{scaffold}}} - \frac{A_{column}R_{\text{filler}}}{D}} \tag{S12}$$

The only remaining task is to compute the axial filler resistance $R_{\text{filler}}$.

## 4. Axial resistance of a spherical filler embedded in matrix

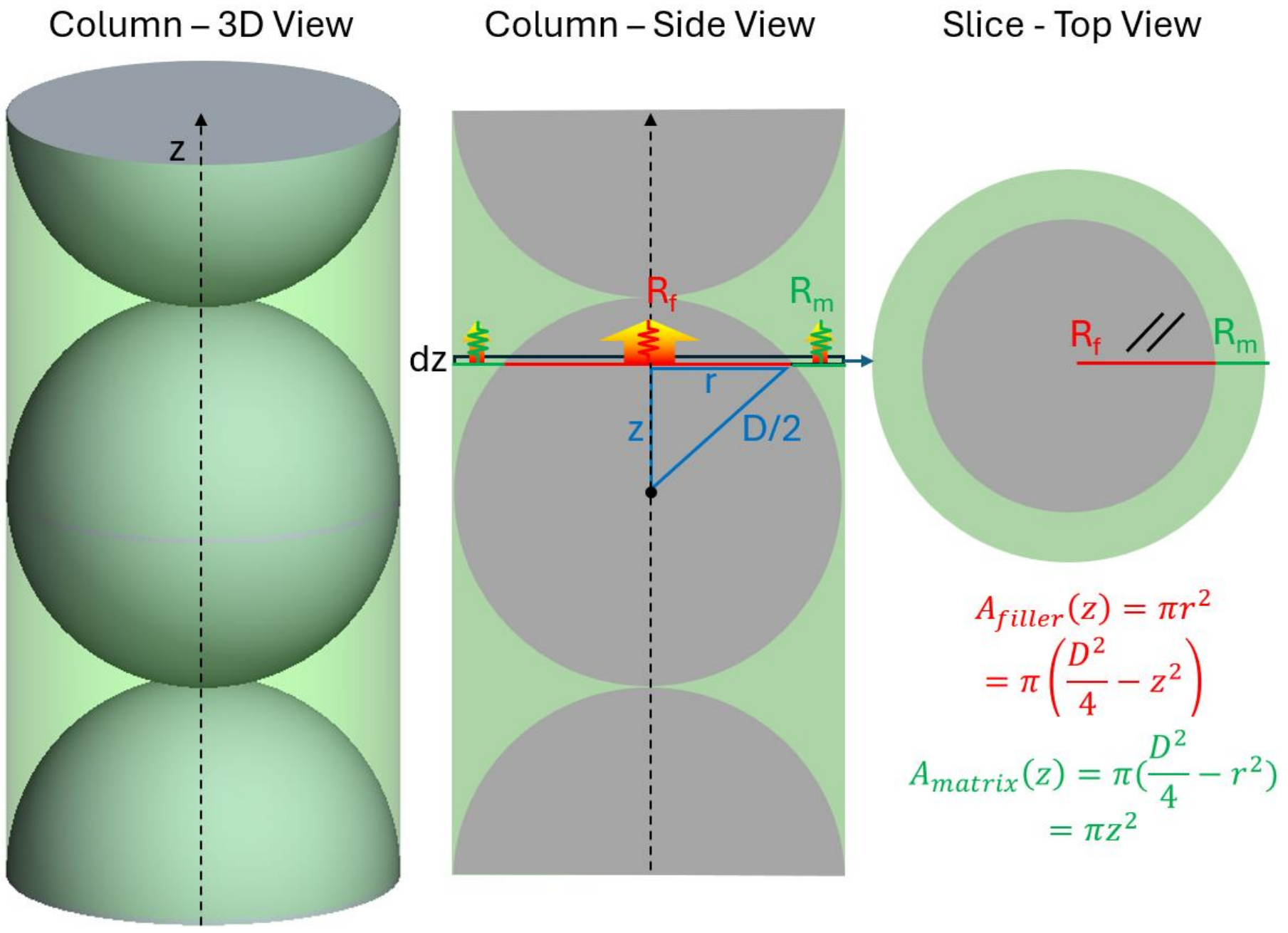

**Figure S18**. Schematic representation of spherical fillers embedded in a scaffold column. A column of spheres is aligned along the $z$-axis, and sliced into infinitesimal disks. At each axial position, the slice consists of a circular filler core surrounded by a matrix, which act as parallel heat-flow paths. This geometric construction underlies the integral used to compute the axial thermal resistance of a single spherical filler in the scaffold column.

To determine the axial thermal resistance of a spherical filler embedded within the scaffold column, we analyze heat flow through the filler and its surrounding matrix by slicing the column into infinitesimal disks of thickness $dz$, as shown in **Figure S18**. At each axial position, heat can pass through the filler and its surrounding matrix, which act as parallel conduction pathways.

At axial position $z$, the spherical filler has cross-section

$$A_f(z) = \pi\left(\frac{D^2}{4} - z^2\right) \tag{S13}$$

and the surrounding matrix occupies

$$A_m(z) = \pi z^2. \tag{S14}$$

The differential resistances of the two parallel heat paths are therefore

$$R_f(dz) = \frac{dz}{A_f(z)k_f}, \qquad R_m(dz) = \frac{dz}{A_m(z)k_m} \tag{S15}$$

Since these two paths conduct simultaneously, the total differential resistance is the parallel combination

$$R_{\text{par}}(dz) = \frac{dz}{A_f(z)k_f + A_m(z)k_m} \tag{S16}$$

Substituting the spherical cross-section areas:

$$R_{\text{par}}(dz) = \frac{dz}{\pi[(D^2/4 - z^2)k_f + z^2 k_m]} \tag{S17}$$

Simplifying the expression:

$$R_{\text{par}}(dz) = \frac{1}{\pi}\frac{dz}{b + az^2}, \qquad a = k_m - k_f < 0, b = \frac{D^2}{4}k_f > 0 \tag{S18}$$

The integrand has the form $1/(az^2 + b)$with $ab < 0$, which yields an **arctanh**. Carrying out the integral from $-D/2$ to $+D/2$:

$$\boxed{R_{\text{filler}} = \frac{4}{\pi D\sqrt{(k_f - k_m)k_f}}\text{arctanh}\left(\sqrt{\frac{k_f - k_m}{k_f}}\right)} \tag{S19}$$

We find the filler-resistance term in (S12) by multiplying with $A_{column}$ and dividing by $D$:

$$\boxed{\frac{A_{column}R_{\text{filler}}}{D} = \frac{1}{k_f}\frac{\text{arctanh}\left(\sqrt{1 - \frac{k_m}{k_f}}\right)}{\sqrt{1 - \frac{k_m}{k_f}}} = \frac{f}{k_f}} \tag{S20}$$

where the contrast factor $f$ is defined as $\frac{\text{arctanh}\left(\sqrt{1-\frac{k_m}{k_f}}\right)}{\sqrt{1-\frac{k_m}{k_f}}}$.

## 5. Final expression for the inter-particle contact resistance

Substituting (S20) and (S9) into (S12), we find the final expression for $R_c/D$: the total contact resistivity of a scaffold, normalized by film thickness. Note that all $N$ junctions add up to the total contact resistivity as $R_c{\cdot}N = R_c{\cdot}L/D$:

$$\frac{R_c}{D} = \frac{\frac{3}{2}\phi_{\text{filler}}}{\kappa_{\text{eff}} - \left(1-\frac{3}{2}\phi_{\text{filler}}\right)\kappa_{\text{matrix}}} - \frac{1}{k_f}\frac{\text{arctanh}\left(\sqrt{1-\frac{k_m}{k_f}}\right)}{\sqrt{1-\frac{k_m}{k_f}}} = \frac{\frac{3}{2}\phi_{\text{filler}}}{\kappa_{\text{eff}} - \left(1-\frac{3}{2}\phi_{\text{filler}}\right)\kappa_{\text{matrix}}} - \frac{f}{k_f} \tag{S21}$$

$R_c/D$ has units of bulk thermal resistivity, [K W$^{-1}$ m]. While $R_c$ gathers both insight into the interparticle material thermal properties and its geometry (poorer alignment and larger gap between particles will translate into a larger $R_c$), $R_c/D$ also takes into account the fact that smaller particles will introduce more junctions within a scaffold.

## 6. Application to Ni–Scaffolds

For a Ni scaffold ($k_f = 90\,\text{W}\,\text{m}^{-1}\text{K}^{-1}$) embedded in Ecoflex ($k_m = 0.16\,\text{W}\,\text{m}^{-1}\text{K}^{-1}$), the contrast factor $f$ defined in Eq. (S20), and the axial filler resistance term in eq. (S21) become

$$f = 3.86;\ \frac{f}{k_f} = 3.86/90 \approx 0.043\,\text{K}\,\text{W}^{-1}\text{m} \tag{S22}$$

This yields physically meaningful $R_c/D$ values for the AB and BA scaffolds, where Ni particles are surrounded by Ecoflex. This behavior is shown in **Figure S19a**. In contrast, ABBA composites produce negative $R_c/D$ values, which is unphysical. This occurs because the filler-resistance term exceeds the scaffold resistance $1/\kappa_{\text{scaffold}}$, indicating that Ecoflex is not the material occupying the inter-particle regions in ABBA scaffolds.

A consistent interpretation emerges when liquid metal is taken as the matrix phase. For $k_m = k_{\text{LM}} = 33\,\text{W}\,\text{m}^{-1}\text{K}^{-1}$, the contrast factor reduces to

$$f = 1.24; \approx 0.0138\ \mathrm{K\,W^{-1}m}. \qquad \text{(S23)}$$

With this correction, the extracted $R_c/D$ values for ABBA become positive and physically meaningful, confirming that liquid metal bridges dominate the inter-particle regions. Still, the extracted $R_c$ remains far smaller than in AB and BA scaffolds (**Figure S19a**), reflecting the much lower interfacial resistance of liquid metal compared to Ecoflex. This improvement directly follows from the much lower acoustic impedance contrast. and thus far weaker phonon reflection at the Ni/LM interface (≈25%) compared with Ni/Ecoflex (≈93%) (eq. 1).

**Figure S19a** summarizes the evolution of $R_c/D$ across processing routes (AB, BA, ABBA, BA**). The more aligned AB scaffolds outperformed BA, as expected. Liquid-metal bridging provided a pronounced benefit: BA* exhibited much lower interfacial resistivity than any BA, and the two-step ABBA composites outperformed any other material, as expected. Notably, **Figure S19b** showed that even though the average filler conductivity decreased when liquid metal is introduced (Ni: 90 W/(m·K), LM: 33 W/(m·K)), the measured ABBA $\kappa_{\text{scaffold}}$ increased by a factor of seven for only 2.5 vol% more LM content (BA* composite was added as a reference). This result demonstrates that liquid metal eliminated interparticle resistance, the dominant thermal transport bottleneck, thereby enabling highly efficient thermal pathways

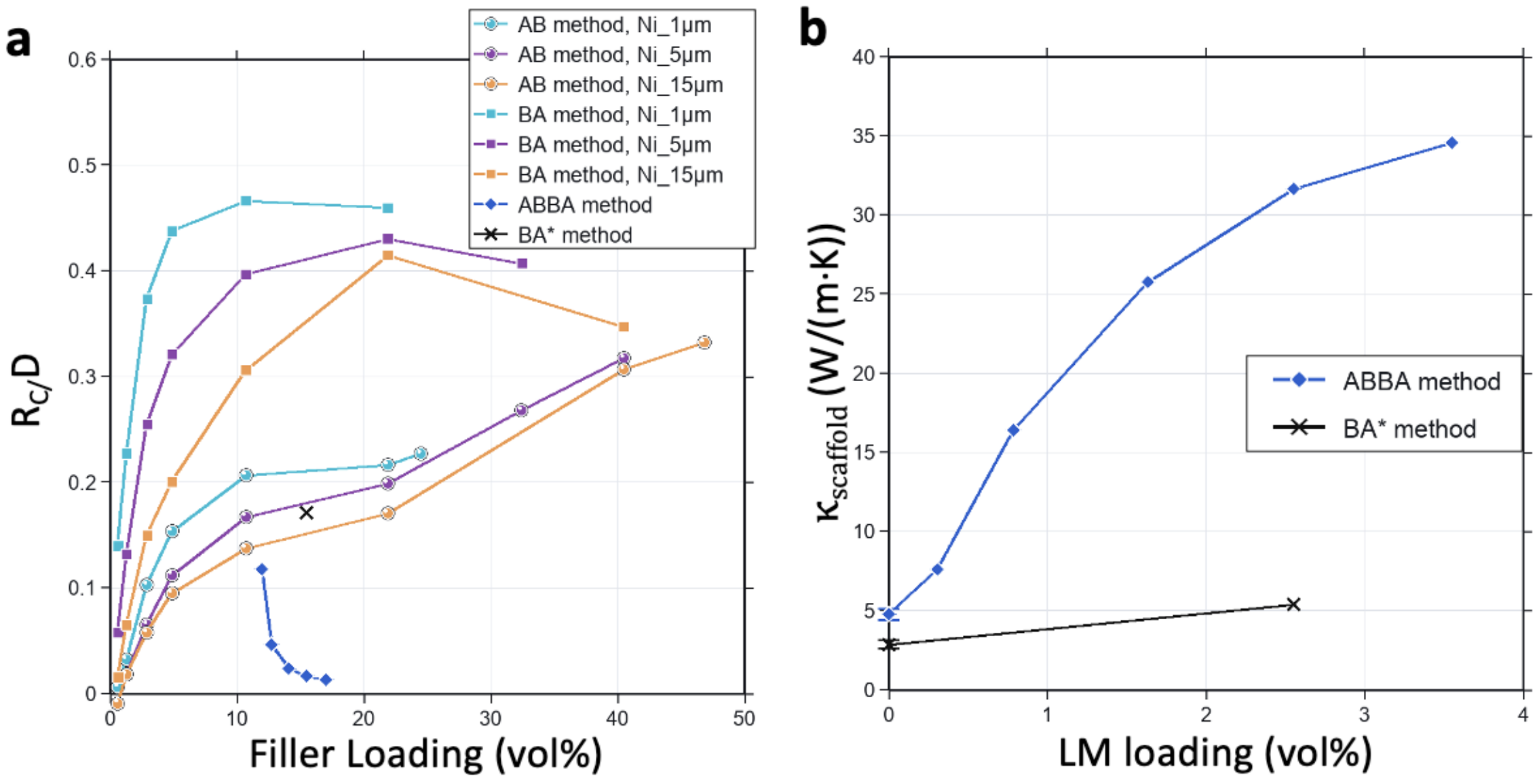

within the scaffold.

**Figure S19.** Evolution of the thickness-normalized contact resistivity $\mathrm{R_c/D}$ as a function of Ni filler loading for different processing routes (AB, BA, ABBA, BA*). **(b)** Improvement of the scaffold conductivity $\kappa_{\text{scaffold}}$ with increasing liquid-metal loading for ABBA and BA* composites (all data at 12 vol% Ni). Liquid-metal bridging significantly reduced inter-particle resistance, resulting in a strong enhancement of the scaffold conductivity.

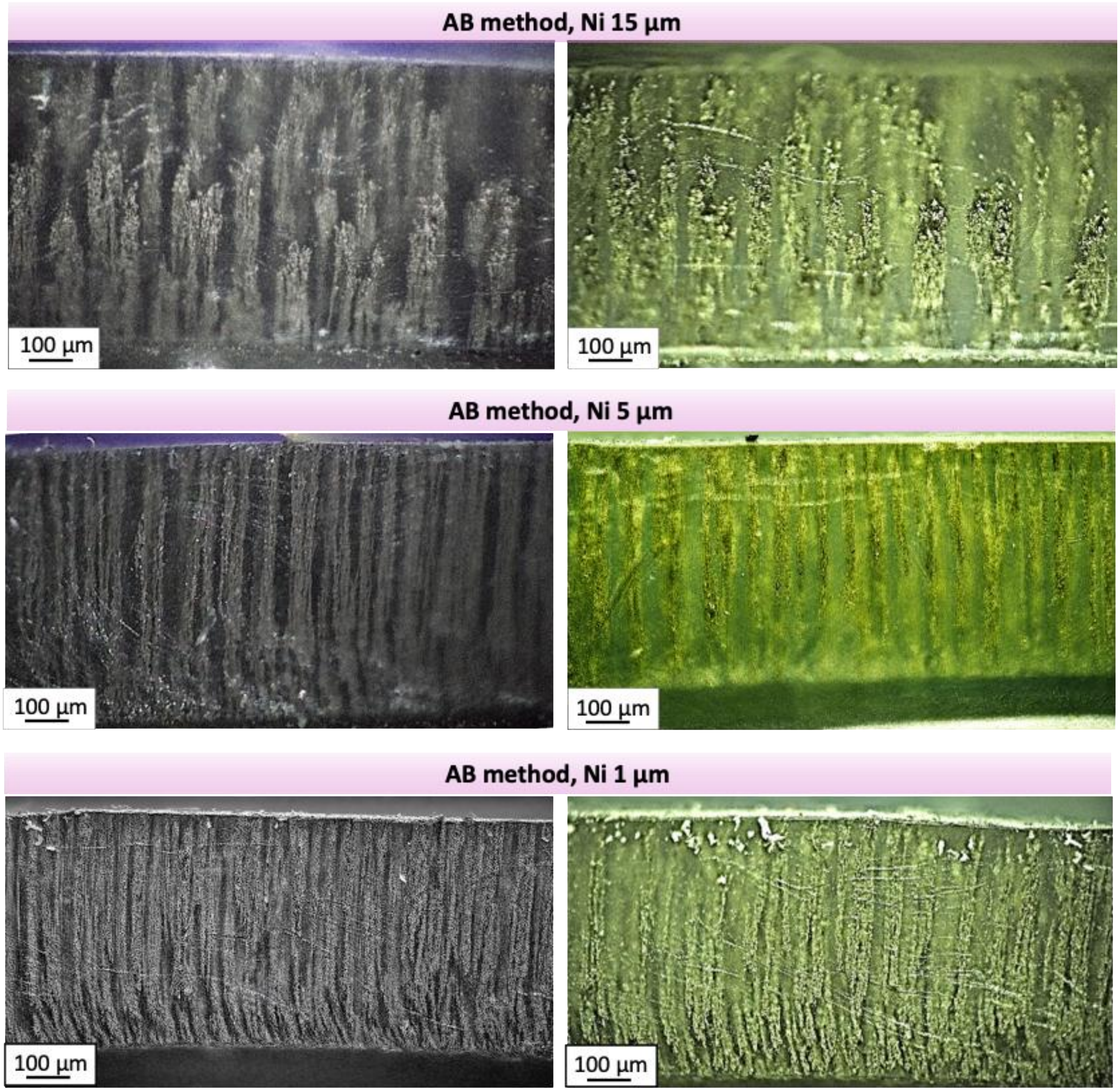

**Figure S20.** Particle-size-dependent scaffold morphology in AB composites at 10.6 vol%. Cross-sectional stereo optical microscopy images of AB scaffolds prepared with 15 µm and 5 µm and 1 µm Ni particles, acquired under standard (right) and green-light (left) illumination (TAGARNO A/S, Horsens, Denmark).

**Cross-sectional SEM–EDS images evidencing LM localization at Ni interparticle regions.** To examine the internal distribution of the LM phase within the ABBA composite, high-magnification cross-sectional microstructural and elemental analysis was performed. First, BSE-SEM imaging of the composite cross-section shows that Ni particles (~ 5 μm in diameter) are organized into vertically oriented scaffold chains (**Figure S21a**). These vertically aligned Ni-rich regions define the primary through-plane conduction pathways. Although a few poorly connected particles are observed in the cross-sectional plane, they connect at different depths (F**igure 3a-c**), minimizing the occurrence of interparticle insulating gaps where thermal contact resistance is expected to be significant.

To further investigate the internal structure of the scaffolds, FIB-prepared cross-sections were taken from Ni-rich scaffold regions and analyzed by SEM–EDS (**Figure S21b–c**). The FIB cross-sectional images reveal a heterogeneous morphology within the scaffold, with contrast variations at the particle/interparticle scale. Local EDS analysis shed light on the nature of the different phases. Selected Area 1, positioned within a bright particle-rich region, shows a dominant Ni signal (**Figure S21b, c**). Selected Area 2, positioned in the darker interfacial region adjacent to the Ni-rich particle, shows a dominant Si signal together with detectable Ga, indicating the presence of silicone matrix and LM-derived species near the Ni interface (**Figure S21c**). **Figure S21d** presents a larger-area EDS elemental map of a representative region within the vertically aligned Ni-rich column/scaffold. The maps show that Ga and In signals are distributed around Ni particles, connecting them physically. These results provide evidence that the scaffold consists of Ni particles effectively bridged by LM. This observation supports the proposed role of the LM phase in reducing local interparticle thermal junction resistance within the vertically aligned Ni scaffold.

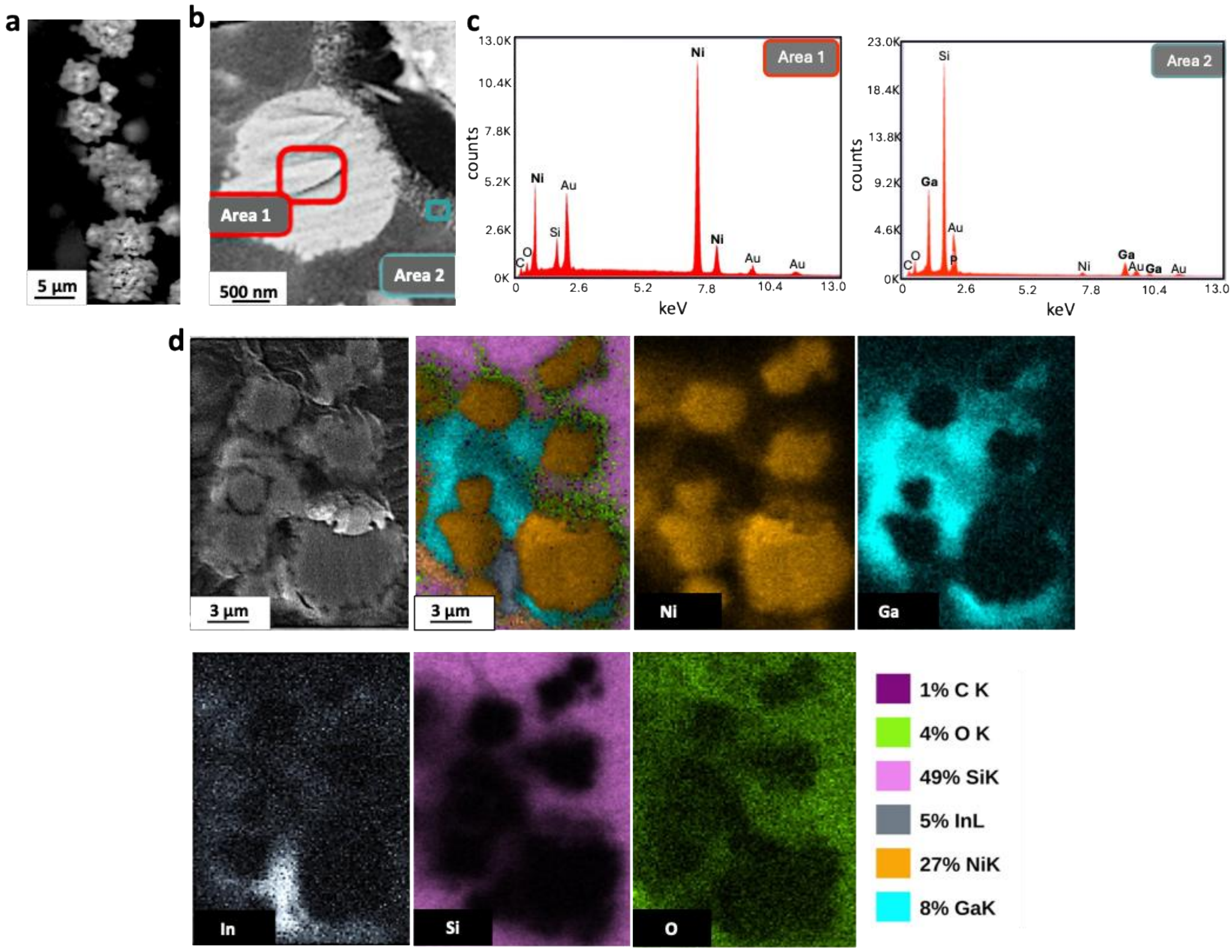

**Figure S21.** High-magnification cross-sectional microstructural and elemental characterization of the ABBA composite. (a) BSE-SEM image of the vertically organized Ni-rich scaffold taken from a cross-section produced manually with a blade cutter. (b) BSE-SEM image of the FIB-prepared cross-section of a Ni-rich scaffold region acquired with the CBS detector. (c) Corresponding EDS spectra collected from Area 1 and Area 2, respectively. (d) Zoomed-out SEM image and corresponding EDS elemental maps of the FIB-prepared cross-section.

**Nano-XCT visualization of 3D Ni–LM spatial association.** Because 2D cross-sectional SEM–EDS cannot fully describe three-dimensional connectivity and might be subject to cross-sectioning defects, the 3D structure of the ABBA composite was studied in a non-destructive way by a nanofocus XCT using a UniTOM HR system (voxel size 500 nm, 120 kV, 2998 projections over 360°). The resulting 3D reconstructions of scaffold segments are shown in **Figure S22a-b**, revealing that the scaffold is composed of irregular domains composed of connected Ni-LM agglomerates rather than isolated phases. In the segmented reconstruction of a representative agglomerate, two morphologically distinct phases can be identified: ~5 µm-diameter spherical particles and irregularly shaped material in between (**Figure S22c**), consistent with the proposed morphology of Ni particles bridged by LM. We note that nano-XCT contrast alone cannot chemically assign every voxel because both Ni and Ga/In are X-ray attenuating metals; the CT data are therefore interpreted in conjunction with the cross-sectional SEM–EDS mapping in **Figure S21**.

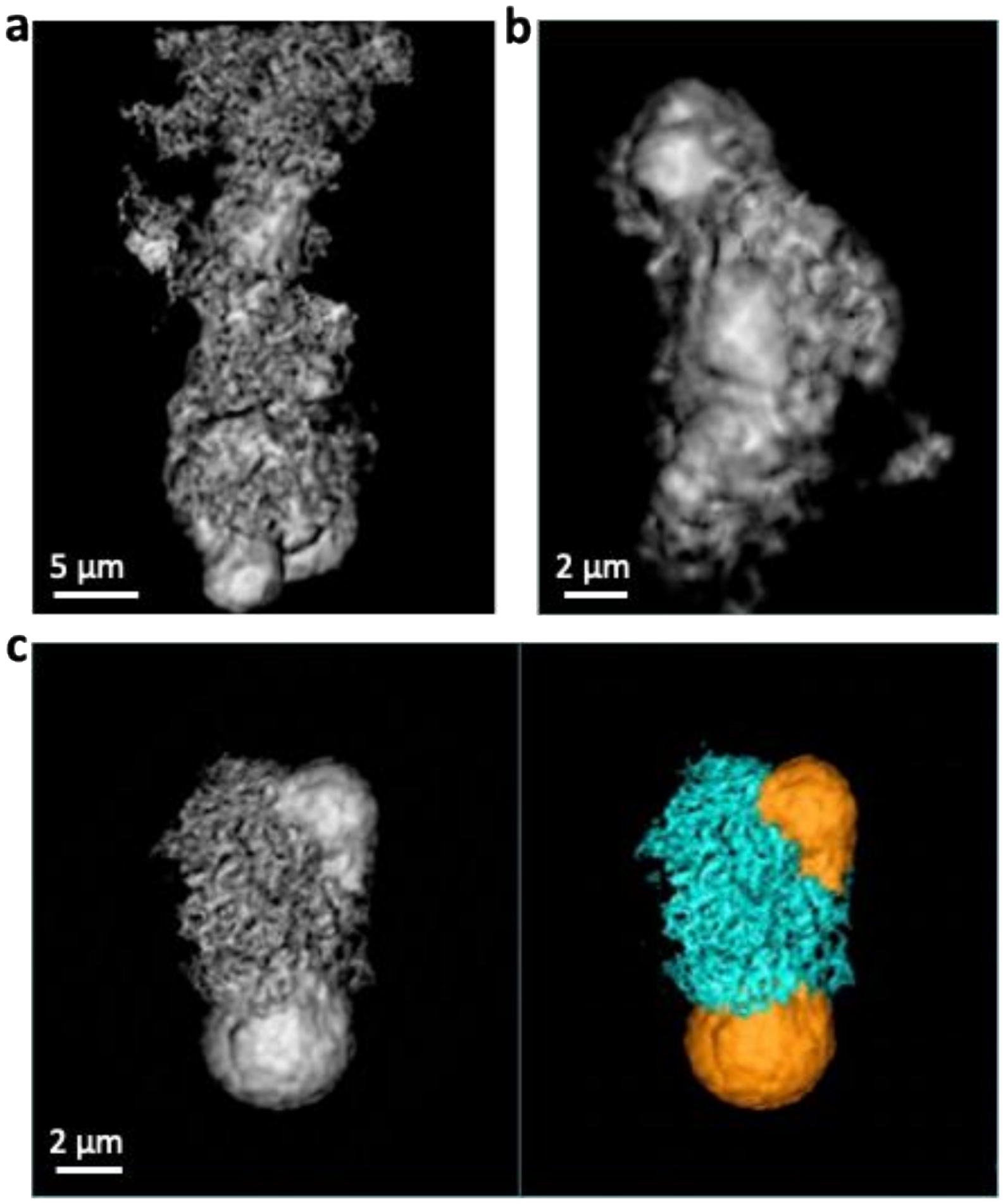

**Figure S22.** Nano-XCT visualization of Ni–LM agglomerates in the ABBA composite. (a) 3D nano-XCT reconstruction showing a vertically elongated Ni–LM agglomerate within a scaffold segment. (b) Magnified reconstruction of an individual agglomerated region. (c) Grayscale reconstruction (left) and corresponding false-color segmentation (right) of a representative agglomerate.

To visually evaluate the advantage of the ABBA architecture for thermal management, relative to other composites, thermal IR imaging was performed on B, BA, AB, and ABBA samples under identical thermostatic hot-stage conditions. Samples were placed on the same hot stage and imaged at set temperatures of 30, 40, 50, 60, and 70 °C. Quantitative temperature–time curves confirmed that ABBA exhibited the fastest heating rate and highest final surface temperature, followed by AB, BA, and B (see also **Video S6**). These results complement the results in **Figure 2e** and **Figure 4a**, i.e., that the ABBA strategy provides superior through-plane heat transport compared with the corresponding B, BA, and AB counterparts.

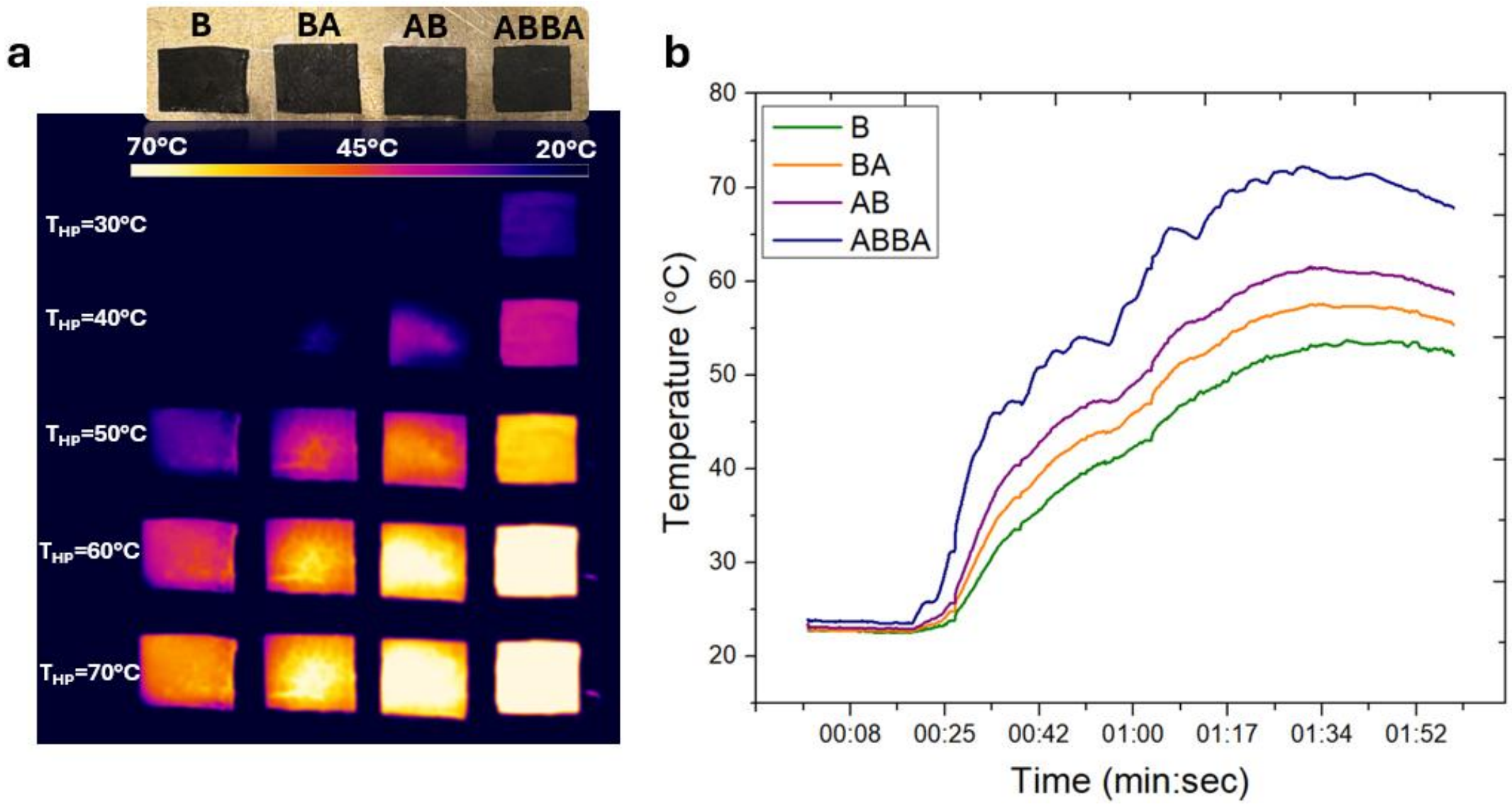

**Figure S23.** (a) Optical image and corresponding IR thermal images of B, BA, AB, and ABBA samples placed on a thermostatic hot stage at set temperatures up to 75 °C. (b) Quantitative surface-temperature time evolution during heating under identical hot-stage conditions of 75 °C.

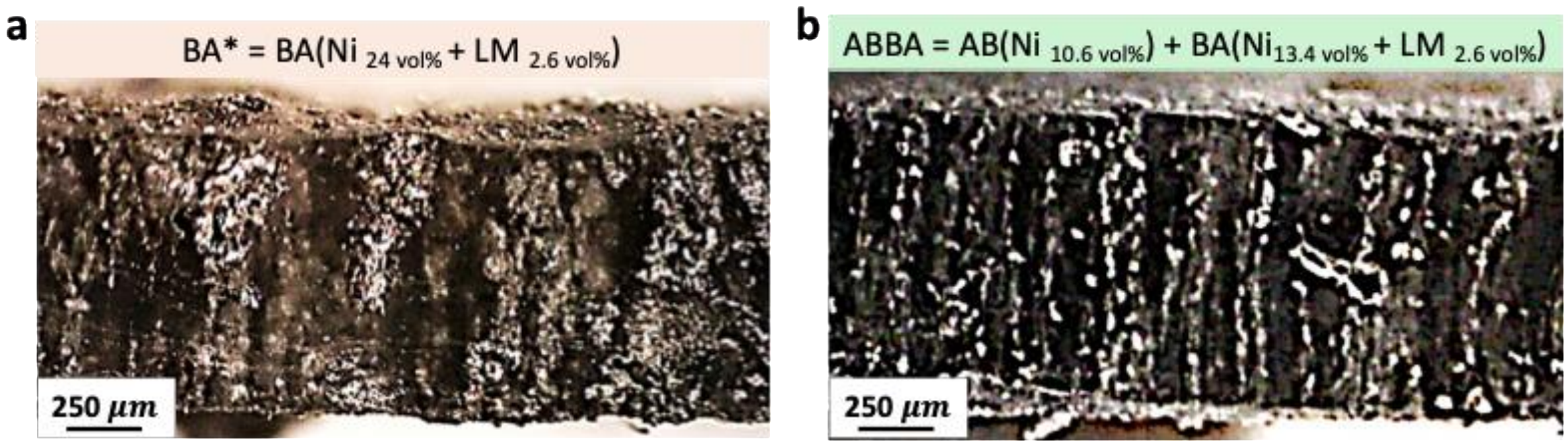

**Figure S24.** Optical cross-section images comparison between (a) direct BA processing of Ni/LM agglomerates (BA*) and (b) optimized ABBA.

**Dielectric breakdown strength testing.** Dielectric breakdown strength (DBS) testing was performed using a custom DC high-voltage setup with a maximum output of 20 kV. The test protocol was guided by ASTM D149 principles for dielectric breakdown testing of solid insulating materials, but adapted to an in-plane electrode geometry to reflect the lateral electrical-insulation requirement of the intended TIM/TEG application. The applied voltage was increased in steps of 500 V, with a hold time of 20 s at each step. Breakdown was identified by a sudden collapse of the applied voltage accompanied by a marked increase in leakage current across the sample. Neat Ecoflex showed a DBS of 8.7 ± 0.9 kV/mm (230 ± 20 V/mil), while the optimized ABBA composite showed 6.6 ± 0.2 kV/mm (171 ± 5 V/mil). The lower DBS of the composite is expected due to the presence of conductive Ni/LM inclusions, which can locally intensify the electric field. However, the measured DBS of the optimized ABBA composite falls within the ~100–300 V/mil range commonly used as a practical reference for insulating materials in electronic devices, further supporting its electrical safety under the tested in-plane configuration.

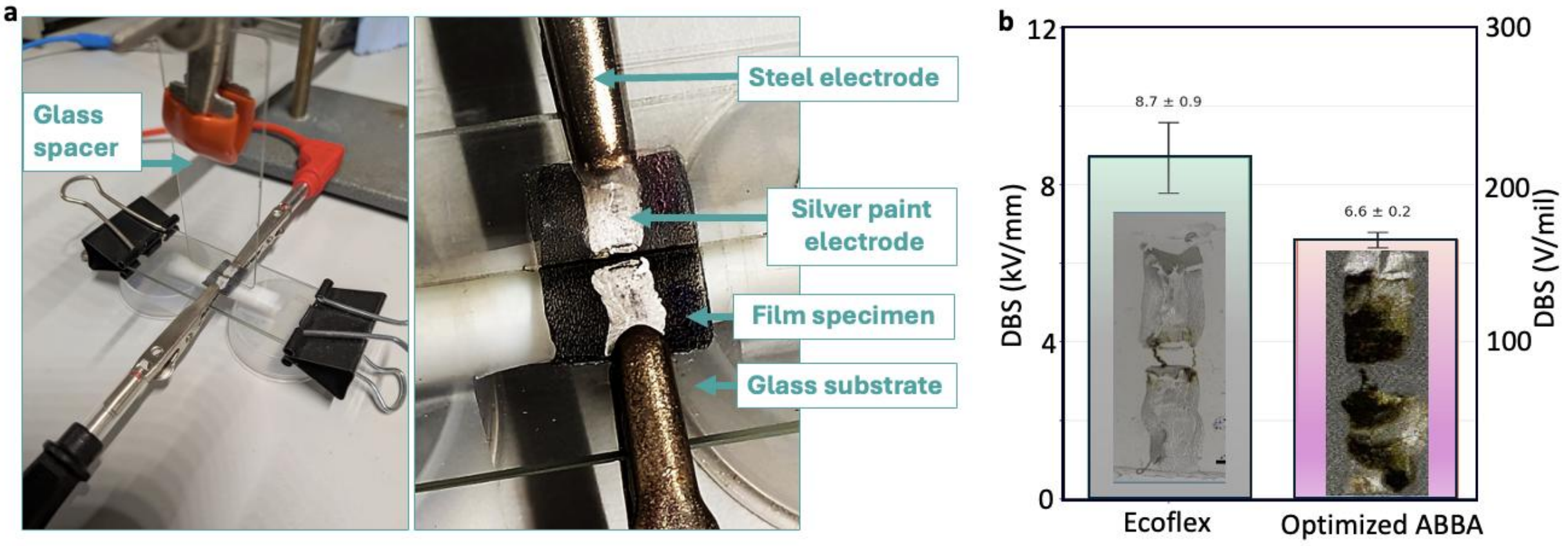

**Figure S25.** (a) Custom DC in-plane dielectric breakdown setup and electrode configuration. (b) Dielectric breakdown strength of neat Ecoflex and optimized ABBA measured under the same in-plane DC configuration. Insets show representative post-breakdown specimens. Error bars represent standard error (kV/mm), n = 4.

**The strain stability** of κ⊥ of can be understood using the scaffold resistance model, where the effective scaffold resistance contains an interparticle contact-resistance term, $R_c/D$ (**eq. S21**):

$$R_{\text{scaffold}} \sim \frac{1}{\kappa_{\text{scaffold}}} = \frac{f}{\kappa_{\text{filler}}} + \frac{R_c}{D}. \quad \text{(S24)}$$

Under in-plane tensile strain, the elastomer undergoes through-plane Poisson contraction, which can compress the vertically aligned scaffold and modify the effective interparticle contact resistance $R_c$. In a rigid particle–particle network, mechanical deformation can disrupt solid–solid contacts or change the local contact geometry, making $R_c$ highly sensitive to strain. In contrast, the ABBA scaffold contains LM-associated interparticle junctions that are both thermally conductive and mechanically compliant. The liquid-like LM phase can accommodate local particle displacement during stretching or compression while maintaining contact between neighboring Ni-rich domains. In this way, the LM acts as a deformable thermal bridge: it preserves the heat-transfer pathway even when the particles shift, compress, or slightly rearrange under mechanical loading. As a result, the effective $R_c$ remains comparatively stable under moderate deformation, and κ⊥ is preserved over the 0–25% strain range (**Figure 4c, Figure S26a**).

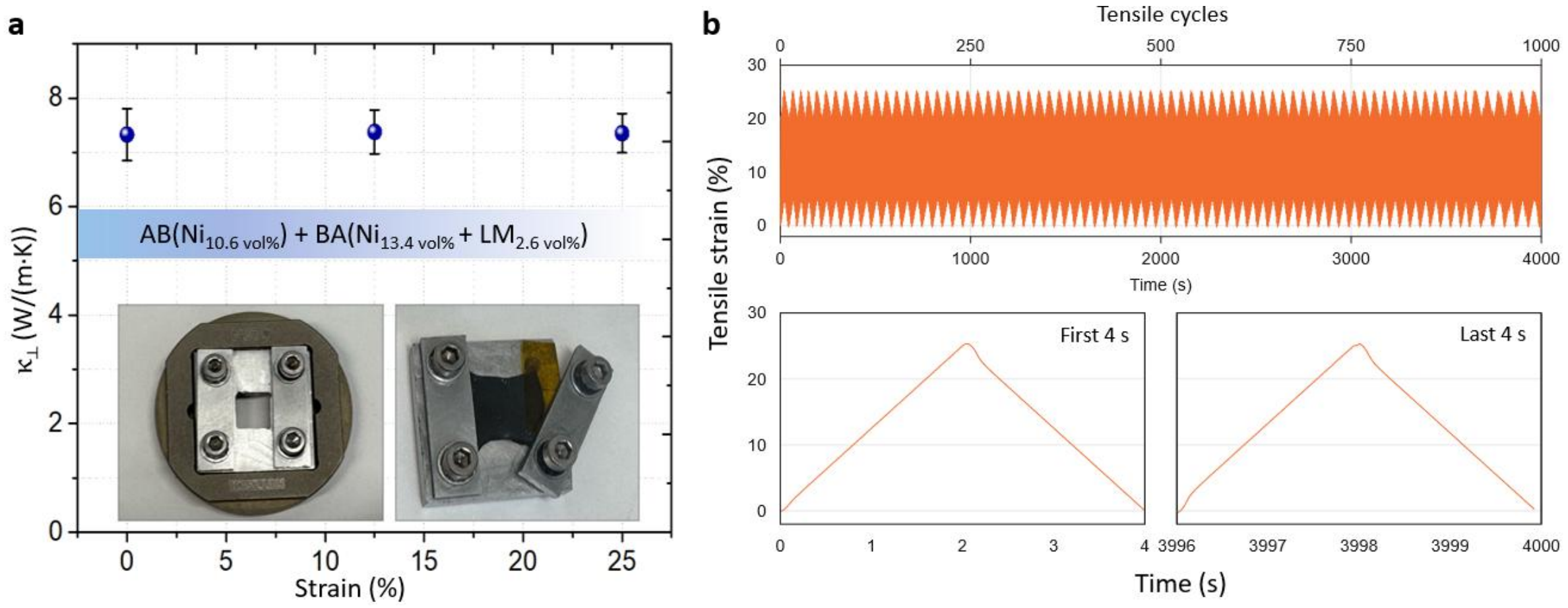

**Figure S26. Thermomechanical stability of ABBA.** (a) Strain-dependent through-plane thermal conductivity of the 2.6 vol% LM ABBA composite n = 3 per condition; error bars represent standard error. (b) Representative strain–time profile during the full 1000-cycle tensile test at 25% strain (f = 0.25 Hz) with expanded views of the first 4 s and last 4 s showing stable cyclic deformation throughout the test.

**Thermal conductivity retention under repeated compressive loading.** Cyclic compression was evaluated at 10, 50, and 100 psi using an Instron E3000 electropuls dynamic testing setup. κ⊥ was measured by LFA after 0 cycles (uncompressed baseline), and after 1, 100, and 1000 cycles at 10 psi, as well as after 1 cycle at 50 and 100 psi (n = 3 per condition). Residual strain and recovery rate were extracted from the corresponding force–displacement traces (**Figure S27 c,d**). Residual strain quantifies irreversible deformation, i.e., the thickness that does not recover after unloading: $\varepsilon_{pl} = \frac{t_0 - t_{res}}{t_0}$, with $t_0$ the initial thickness and $t_{res}$ the thickness after unloading. Recovery rate quantifies the fraction of the total deformation ($t_0 - t_{peak}$) that is recovered once the compressive load is removed ($t_{res} - t_{peak}$): $RR = \frac{t_{res} - t_{peak}}{t_0 - t_{peak}}$, with $t_{peak}$ the thickness at the peak compressive load.

At 10 psi, the pressure range most relevant for wearable and skin-conformal applications, where typical skin contact pressures are reported in the range of 1–10 psi, κ⊥ was largely retained across all tested cycle counts. After 1 cycle at 10 psi, κ⊥ remained close to the baseline value (7.5 W $m^{-1}$ $K^{-1}$). A modest reduction was observed after 100 cycles (~6.8 W $m^{-1}$ $K^{-1}$), with no further significant change between 100 and 1000 cycles, indicating that κ⊥ stabilizes after an initial accommodation period. Residual strain at 10 psi remained low across all cycle counts (2.39% after 1000 cycles). The recovery rate was broadly consistent across all cycle counts, ranging between approximately 63% and 70%, with a slight decrease from ~70% after 1 cycle to ~65% after 100 and 1000 cycles. This relatively stable recovery rate suggests that mechanical accommodation occurs primarily within the first compression cycle, with no evidence of progressive irreversible damage accumulating over subsequent cycles

At 50 and 100 psi, κ⊥ was largely retained after a single compression cycle despite the substantially high applied pressure. Compressive loading at these pressures could, in principle, induce competing effects: it may improve through-plane thermal contact by reducing interparticle spacing and increasing contact area, but it may also degrade thermal transport if the aligned scaffold is irreversibly damaged or if interparticle conduction pathways are disrupted. The observation that κ⊥ is largely preserved despite residual strains of approximately 25% and 40% at 50 and 100 psi, respectively, and recovery rates of only approximately 20%, suggests that the LM-mediated interparticle junctions remain thermally functional even under substantial compressive deformation. This behavior is consistent with the compliant nature of the LM-associated junctions: unlike purely rigid solid–solid contacts, the deformable LM phase can accommodate local particle displacement and maintain thermal coupling between neighboring Ni-rich domains during compression.

Nevertheless, the large residual deformation observed at 50 and 100 psi indicates a significant permanent set of the elastomeric pad. Therefore, cyclic fatigue testing at these pressures was not pursued. We note that 50–100 psi exceeds the pressure range most relevant to wearable and skin-conformal applications.

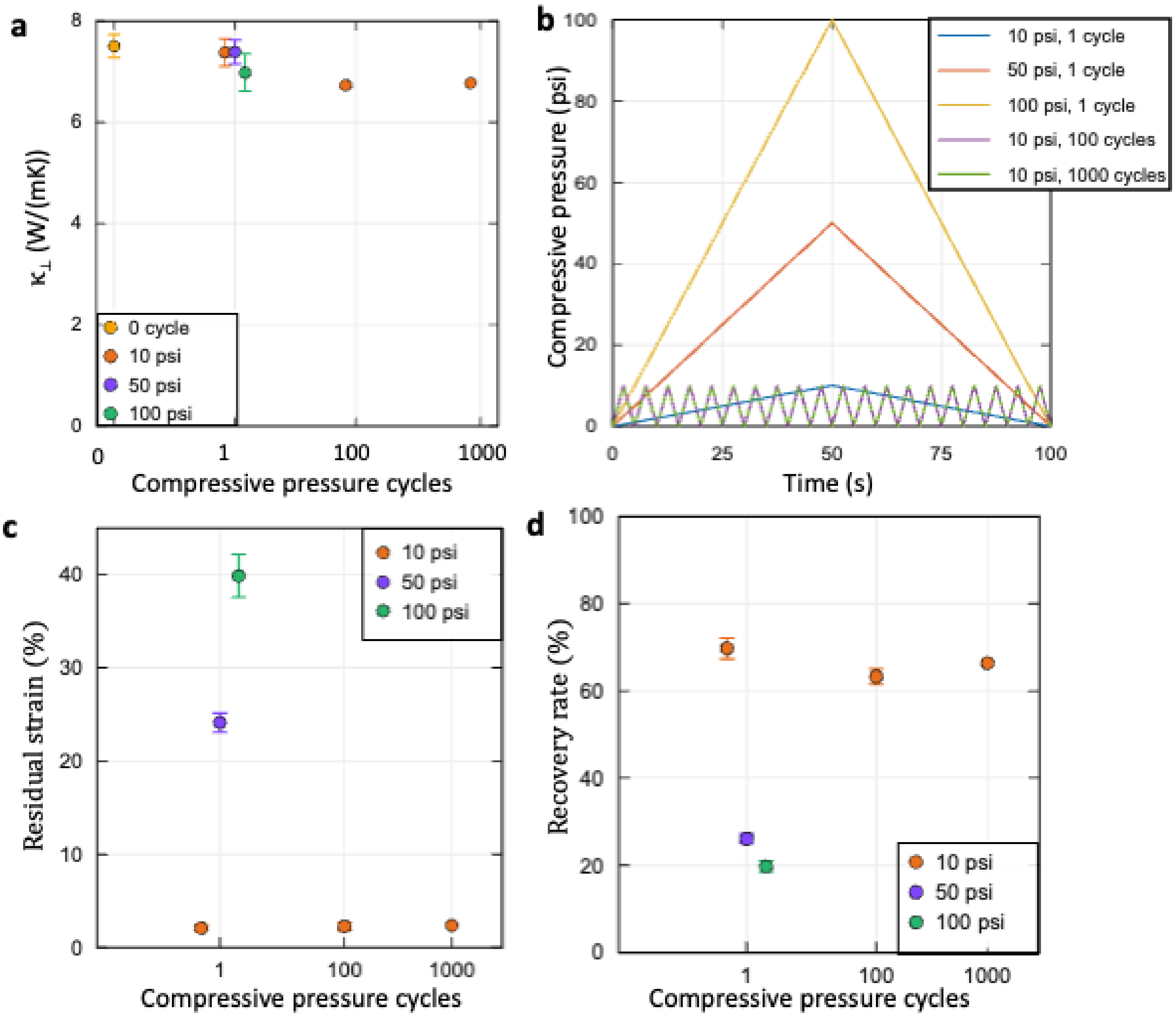

**Figure S27.** Thermomechanical response of the ABBA composite under repeated compressive loading. (a) Through-plane thermal conductivity (κ⊥) measured after 0, 1, 100, and 1000 compressive cycles at 10 psi, and after 1 cycle at 50 and 100 psi (n = 3 per condition; error bars = standard error). (b) Representative compressive pressure–time profiles at 10, 50, and 100 psi for 1 cycle, and at 10 psi for the first 20 cycles of 100 and 1000 cycles. (c) Residual strain and (d) recovery rate as a function of compressive pressure and number of cycles. The 1000-cycle condition is shown without error bars because only one valid replicate was retained.

LM leakage under practical operating conditions was assessed by visual inspection of sample surfaces before and after deformation/cycling using a clean microfiber tissue. No visible metallic LM residue was detected after 1000 tensile cycles at 25% strain, 1000 compressive cycles at 10 psi, single-cycle compression at 50 and 100 psi, or approximately 700 heating–cooling cycles over 24 h. These observations suggest that the LM phase remains mechanically retained within the Ecoflex matrix under the tested conditions. The absence of visible LM transfer is consistent with the stable retention after mechanical cycling and the stable thermal response during repeated heating–cooling operation. We note that this wipe test provides qualitative evidence of no visible LM squeeze-out, while quantitative leakage characterization under prolonged high-pressure or elevated-temperature conditions remains a subject for future work.

**Thermal cycling stability.** Three optimized ABBA composite samples were subjected to repeated heating–cooling cycles using a custom IR camera-based setup (**Figure S28a**). The samples were placed on a hotplate operated at a variable setpoint and cooled by a fan between cycles to accelerate thermal cycling. Surface temperature was recorded continuously using an IR camera. A reference sample was included in the field of view to correct for ambient temperature fluctuations during the measurement. Long-term cycling was performed for approximately 700 cycles over ~24 h (**Figure S28b**). The corrected surface-temperature response remained stable over the full measurement period, with no systematic drift or degradation. The average standard error of the peak temperature across the three samples was 0.38 °C, corresponding to approximately 0.5% of the peak temperature. An expanded time window in **Figure S28c** shows in detail how all three ABBA samples followed the hotplate temperature response reproducibly. These results confirm the stable and reproducible thermal response of ABBA under repeated thermal cycling

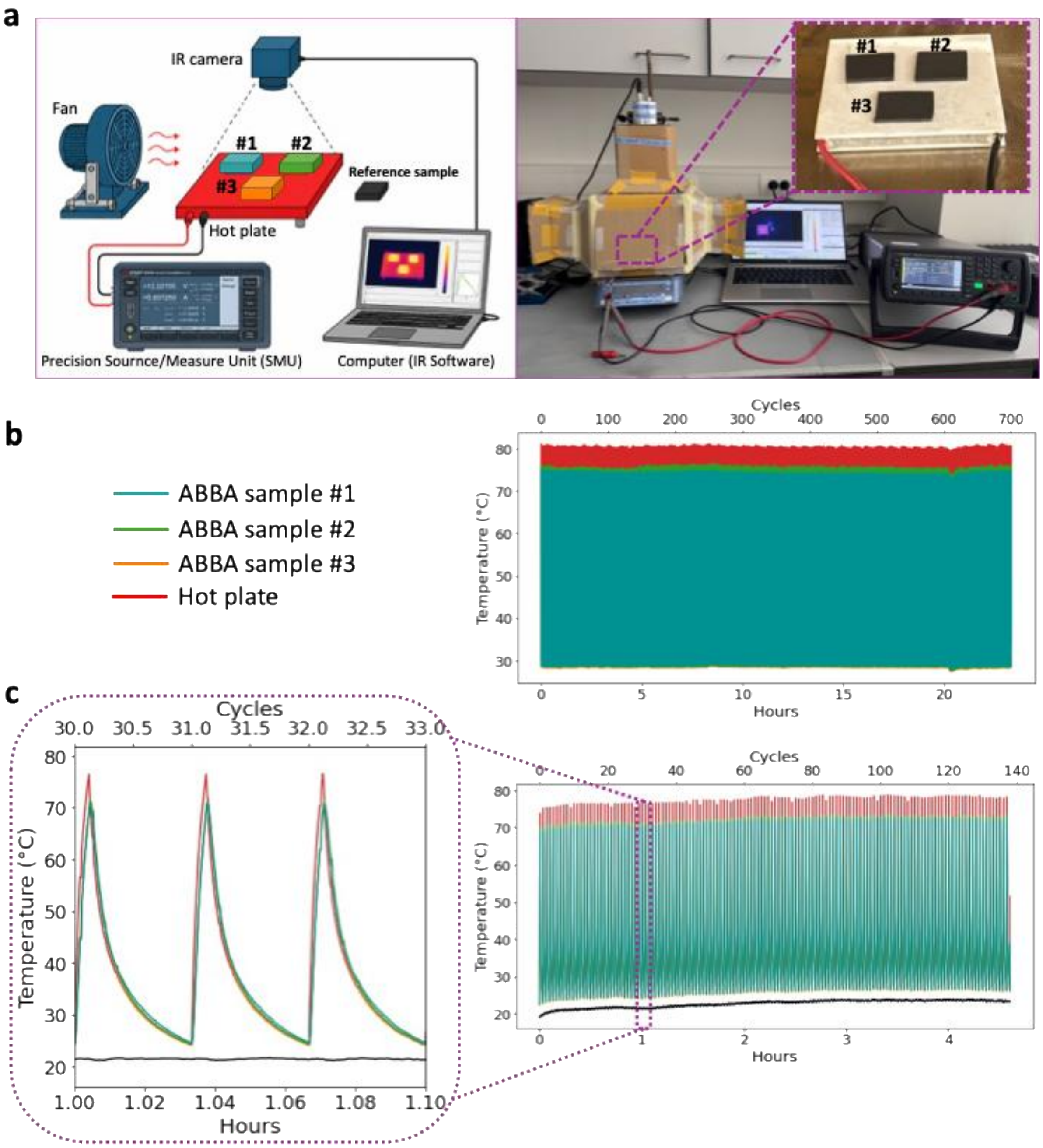

**Figure S28.** Thermal cycling stability of the optimized ABBA composite under repeated heating–cooling cycles. (a) Custom IR camera-based thermal cycling setup. Three ABBA samples and a reference sample were placed on a hot plate and monitored continuously during fan-assisted heating–cooling cycles. (b) Long-term thermal cycling response over approximately 700 cycles in ~24 h, showing stable corrected surface-temperature response without systematic drift or degradation. (c) Representative expanded time window of 140 cycles in ~4.5 h, showing reproducible temperature responses across the three ABBA samples.

**Mechanistic explanation of Young modulus**. The in-plane modulus is governed not only by filler loading but by the degree to which the vertically aligned scaffold develops lateral connectivity. Under through-plane alignment, vertically isolated columns contribute little to in-plane stiffness $E_{in\text{-}plane}$, as mechanical load transfer occurs predominantly in the vertical direction. This can be expressed using an orientation-modified rule of mixtures

$$E_{\text{in-plane}} = \eta_0\, E_{\parallel} + (1 - \eta_0)\, E_{\perp}, \tag{S25}$$

where $E_{\parallel}$ is the modulus in the Voigt limit (corresponding to fully in-plane connected columns loaded along their axis), $E_{\perp}$ is the composite modulus in the Reuss limit (corresponding to perfectly isolated vertical columns loaded in-plane), and $\eta_0$ is the Krenchel orientation factor, which quantifies how much of the filler network is effectively oriented towards the in-plane loading direction. Isolated vertical columns correspond to a low $\eta_0$ and weak in-plane reinforcement; laterally connected columns correspond to a rising $\eta_0$ and a sharp modulus increase. The two limiting cases are the Reuss lower bound (series, isolated columns, $\eta_0 \to 0$) and the Voigt upper bound (parallel, fully connected columns, $\eta_0 \to 1$). AB composites with limited lateral connectivity lie close to the Reuss limit, while composites with lateral bridging progressively approach Voigt behavior.

The onset of lateral connectivity, and therefore the stiffening transition, depends strongly on particle size, as evidenced by the cross-sectional optical images (**Figure S20**):

1 µm particles form very fine, densely packed columns with minimal inter-column spacing. The structure is nearly continuous in-plane. This near-continuous morphology promotes early lateral connectivity, driving an early rise in $\eta_0$ and consequently, early stiffening.

Likewise, the 15 µm particles form wide, coarse, and somewhat irregular columns with visible inter-column gaps and partial lateral bridging. This morphology promotes earlier onset of in-plane connectivity between neighboring columns, enabling mechanical load transfer across the scaffold and leading to an intermediate rise in stiffness.

5 µm particles show the most uniform, well-defined, and widely spaced columnar architecture. This geometry suppresses lateral connectivity over the widest loading range, keeping $\eta_0$ persistently low and the modulus near the Reuss limit across the practically relevant concentration window.

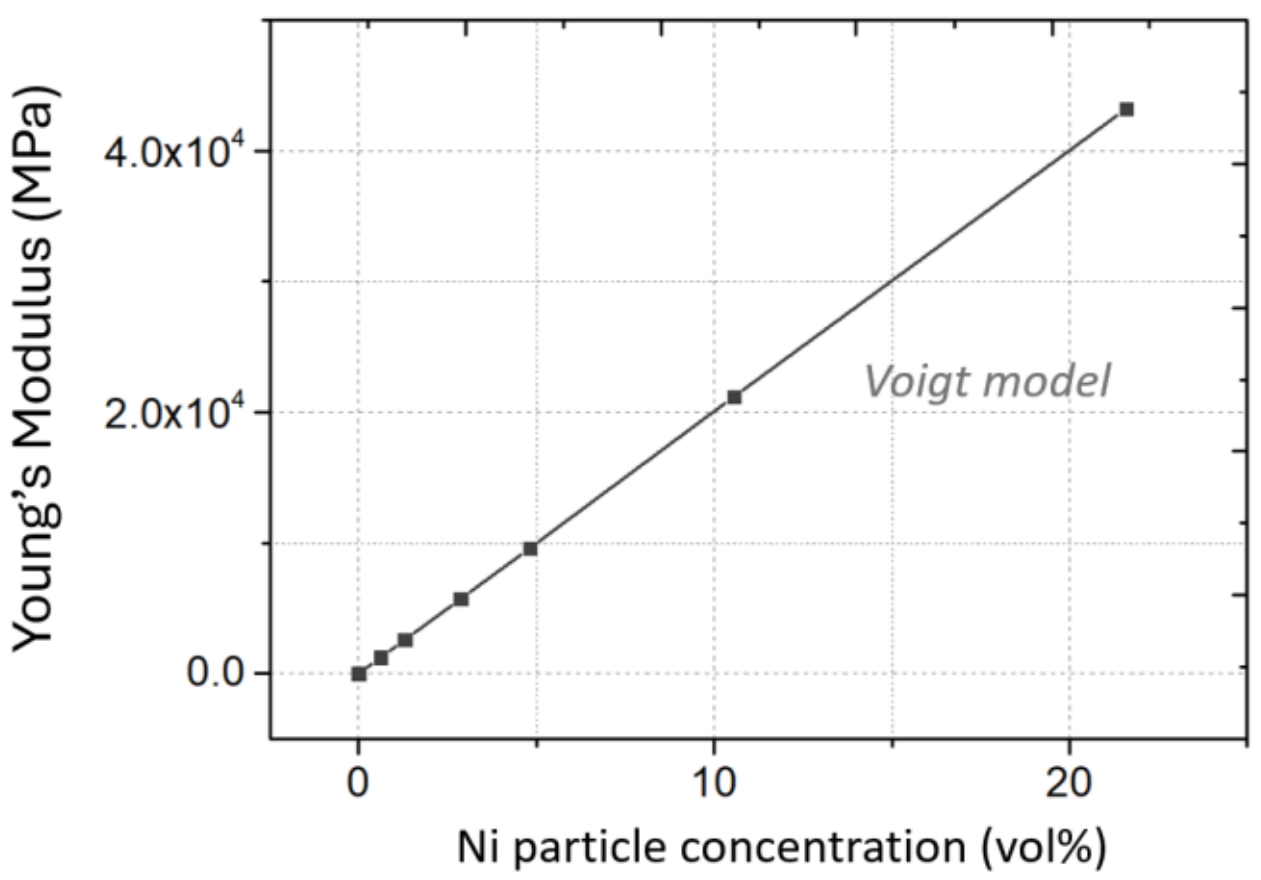

**Figure S29.** Voigt limit for mechanical reinforcement assuming continuous, longitudinally aligned, and fully strained Ni fibers. This reinforcement is orders of magnitude larger than what was observed for the AB composites, which is not surprising considering that our Ni columns are aligned mostly transversely to strain.

**Surface adhesion of Ecoflex and ABBA.** Lap-shear adhesion tests were performed on neat Ecoflex and the optimized ABBA composite on both glass and porcine skin substrates, without any surface modification or adhesion promoter (MTS Criterion Model 41 universal testing machine). On glass, neat Ecoflex showed an adhesion strength of 14.7 ± 0.2 kPa, while the ABBA composite showed a higher value of 22.2 ± 0.6 kPa, corresponding to an approximately 51% increase relative to neat Ecoflex. This increase indicates that incorporation of the Ni/LM scaffold does not compromise adhesion to a smooth, rigid substrate and may enhance interfacial shear resistance, possibly due to roughness-induced increased contact area..

On porcine skin, neat Ecoflex and ABBA showed comparable adhesion strengths of 4.6 ± 0.5 kPa and 4.8 ± 0.2 kPa, respectively. This indicates that the filler network does not significantly alter the skin-contact shear adhesion of the Ecoflex matrix. The lower adhesion values on porcine skin compared with glass are expected because skin is softer and more heterogeneous than glass.

Overall, these results show that the ABBA composite maintains or even increases shear adhesion compared to neat Ecoflex on application -relevant surfaces, such as skin or glass. Thus, incorporation of the Ni/LM scaffold does not compromise surface adhesion, supporting the use of ABBA as a conformal thermal interface layer for wearable applications. If stronger or application-specific adhesion were required, the Ecoflex surface could be further modified, for example, by oxygen plasma treatment or adhesion-promoting surface chemistry, without necessarily changing the bulk thermal or electrical properties of the ABBA composite.

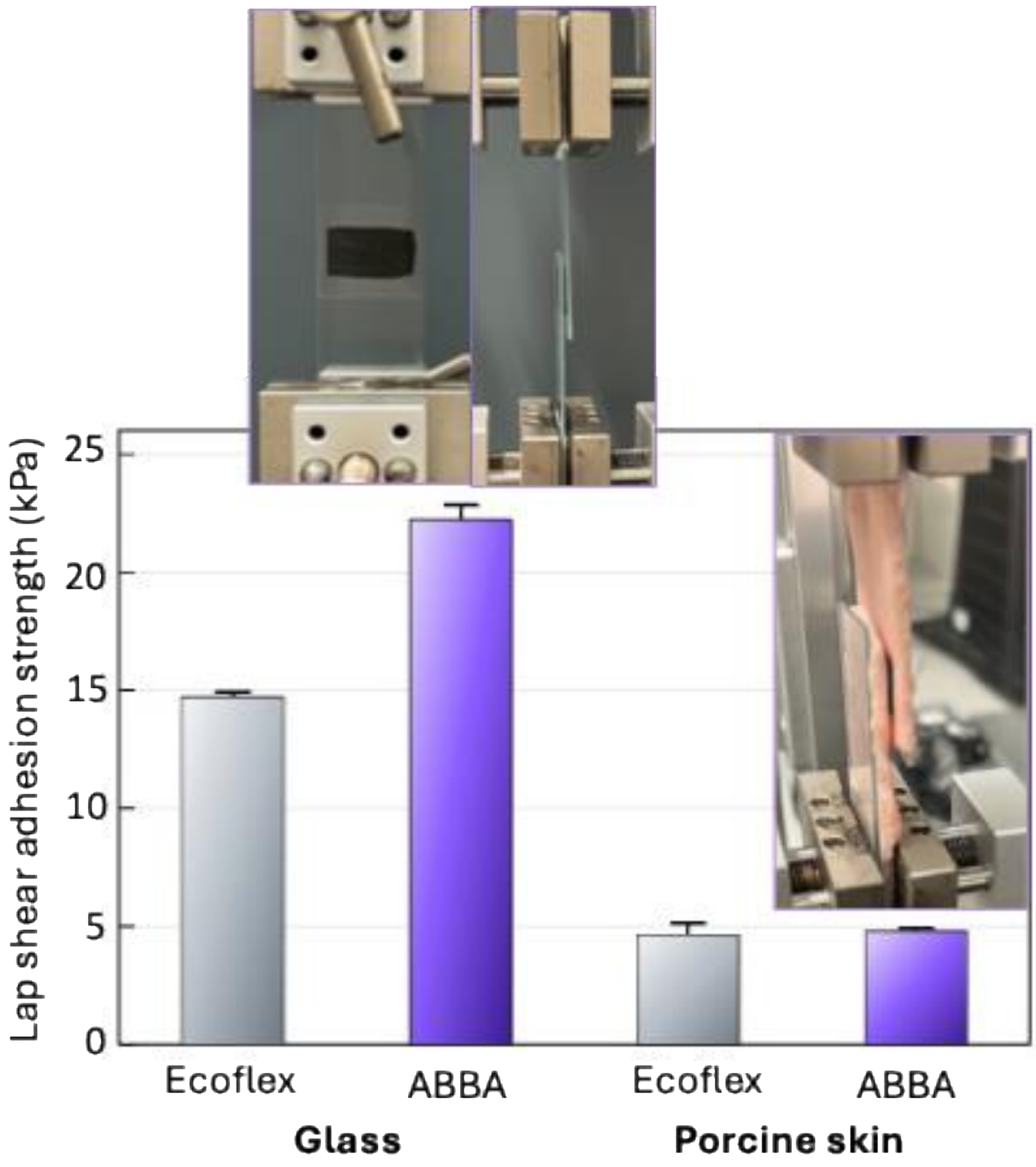

**Figure S30.** Surface adhesion characterization of the ABBA composite. Lap-shear adhesion strength of neat Ecoflex and optimized ABBA measured on glass and porcine skin substrates. Insets show representative photographs of the lap-shear test configuration on glass and porcine skin. Tests were performed using an MTS Criterion Model 41 universal testing machine. Error bars represent the standard error, n = 3.

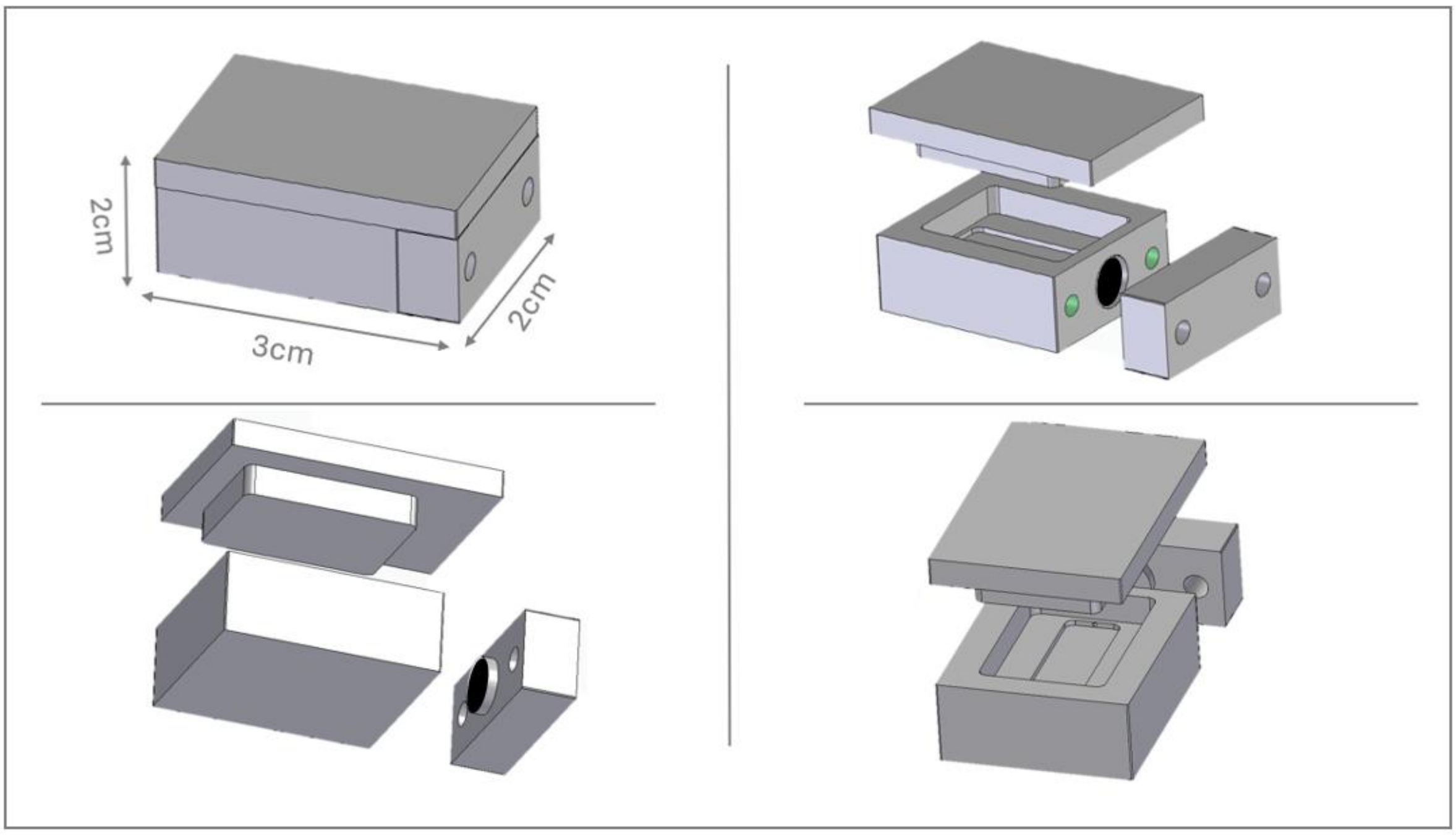

**Figure S31.** Assembled views of the modular mold showing the housing, insert, and lid components used to define sample geometry and enable controlled processing during composite fabrication.

**Tensile response of commercial thermal pads.** Two commercial gap-filler pads were evaluated: RS PRO 707-4764, rated at 6 W/(m·K), and RS PRO 707-4774, rated at 12 W/(m·K). For the 6 W/(m·K) pad, three specimens were loaded to approximately 50% engineering strain, released, remeasured, and loaded a second time. The second-loading strain was recalculated relative to each specimen's original gauge length. Accordingly, the second-loading curves begin at positive strain and do not retrace the first-loading response, demonstrating incomplete dimensional recovery and a changed mechanical response (Figure S32a). The specimens retained a permanent extension of 2.1 ± 0.4 mm, corresponding to a residual strain of 6.3 ± 1.1% (mean ± SE, n = 3; Figure S32b). The 12 W/(m·K) pad showed a more severe tensile limitation: all three specimens fractured during the first loading at approximately 31% engineering strain, preventing a second-loading measurement (Figure S32c). Although both products showed a low in-plane modulus below 0.5 MPa, they failed the elastic test in different ways. The 6 W/(m·K) pad survived the first loading but exhibited substantial plastic deformation, whereas the 12 W/(m·K) pad showed an early fracture.

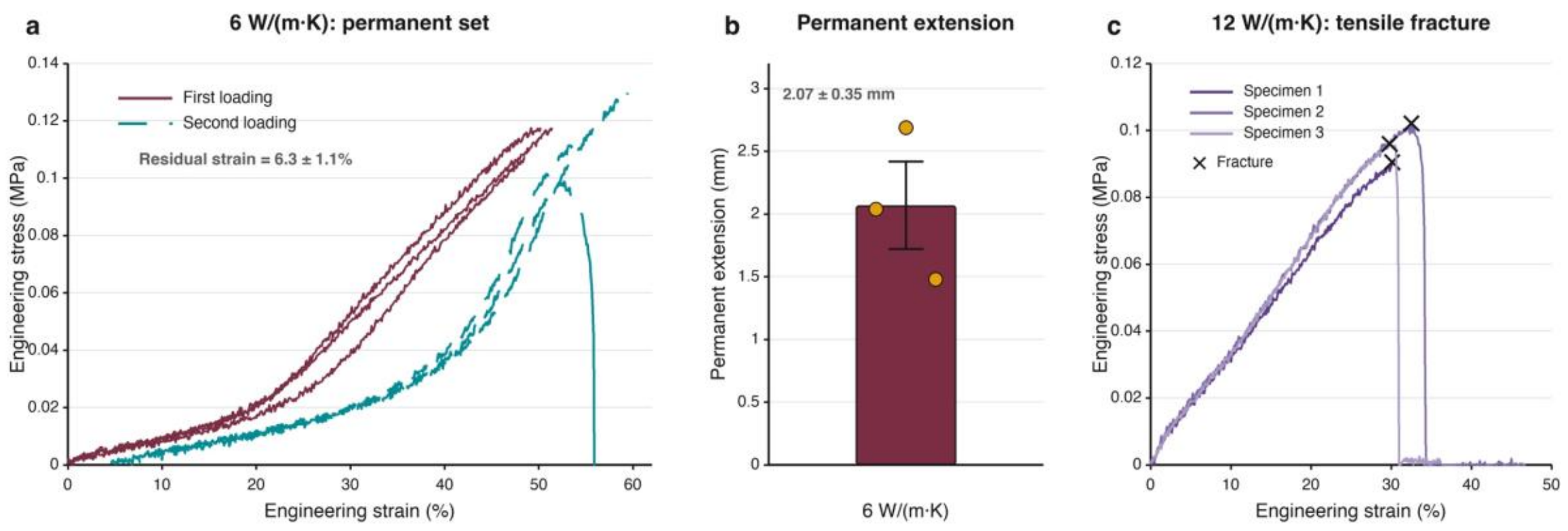

**Figure S32.** Tensile response and irreversible deformation of commercial thermal pads. (a) First- and second-loading engineering stress–strain curves of the 6 W/(m·K) pad. The second-loading strain was recalculated relative to each specimen's original gauge length; its positive offset indicates incomplete recovery. (b) Permanent extension of the 6 W/(m·K) pad after loading to approximately 50% engineering strain. The bar and error bar show the mean ± SE, and the points show individual specimens (n = 3). (c) First-loading engineering stress–strain curves of the 12 W/(m·K) pad. Crosses mark fracture; all three specimens fractured before a second loading could be performed.

**Table S2.** Benchmarking thermally conductive composite systems corresponding to Figure 4d: filler loading, alignment, and coupled anisotropic thermal–electrical-mechanical performance.

| Material system / filler | Processing | Filler loading | Alignment? | κ⊥ ($W\ m^{-1}\ K^{-1}$) | σ∥, in-plane ($S\ cm^{-1}$) | Modulus | Stretchable? | Scalability & reason | Ref. |
|---|---|---|---|---|---|---|---|---|---|
| **★ Ni–LM / Ecoflex (ABBA)** | Magnetic-field alignment (ABBA) | 12 vol% Ni + 2.6 vol% EGaIn | Yes—vertical ABBA | 7.5 | Insulating ($<10^{-14}$ S $cm^{-1}$) | ≈0.16 MPa | Yes | Good—permanent magnetic bars; no specialist field generator | **This work** |
| ‣ Boron nitride (BN)-based / hybrid | | | | | | | | | |
| BNNS / PVA/PDMS | Electrospinning + rolling + PDMS infiltration | 15.6 vol% BNNS | Yes—vertical BNNS | 2.0 | Insulating | NR | No | Limited — requires electrospinning equipment + multi-step infiltration | [S1] |
| BNNS / Polymer film | Electrospinning + vertical folding + pressing | 33 wt% BNNS | Yes—vertical fold | 16.3 (18 μm thick) | Insulating | NR | No (flexible film only) | Limited — requires electrospinning + precise folding; | [S2] |
| BN film / Silicone rubber | Stacking–cutting (roll-cut) | 37 vol% BN | Yes—vertical BN film | 19.1 | Insulating | 5.42 MPa (compressive; 37 vol% BN) | No (compressible, not stretchable) | Good — straightforward stacking-cutting; no specialist equipment needed | [S3] |
| BN / Silicone rubber | Shear orientation + layer stacking | 60 wt% BN | Yes—shear/layer | 7.62 | Insulating | NR | No (compressible, not stretchable) | Good — simple layer stacking; scalable to larger areas | [S4] |
| BN + Graphite / Silicone rubber | Stacking–cutting method | 40 vol% total (BN:Gf = 1:5) | Yes—stack/cut | 23.7 | Insulating | 4.85 MPa (compressive) | No (compressible, not stretchable) | Good — roll-cut demonstrated at cm scale; adaptable | [S5] |
| BNNS / PDMS | Bidirectional freeze casting/ ice-templating + PDMS infiltration | 18.7 vol% BNNS | Yes—freeze-cast | 3.8 | Insulating | NR | No (compressible PDMS) | Limited — freeze-casting apparatus required; slow batch process | [S6] |
| BNNS + CF / Silicone rubber | Bi-directional freeze casting | 55.6 wt% total (120 BN + 5 phr CF) | Yes—vertical network | 9.5 | Insulating ($5.3\times10^{-14}$ S $cm^{-1}$) | NR | No (flexible, not stretchable) | Limited — freeze casting limits throughput; mold-constrained batch size | [S7] |
| BNNS@$Fe_3O_4$ / Epoxy | Magnetic field alignment (vertical) | 30 wt% BNNS@$Fe_3O_4$ | Yes—magnetic | 14.55 | Insulating | NR | No (flexible film only) | Limited — multi-step nanoparticle synthesis | [S8] |
| FeCo/hBN / Polymer film | Electrostatic self-assembly + magnetic field | 80 wt% total (50 BN + 30 FeCo) | Yes—magnetic | 2.25 | Insulating | NR | No (flexible film only) | Limited — multi-step nanoparticle synthesis + self-assembly. | [S9] |
| x-BN / PVDF | Step-by-step orientation | 30 wt% x-BN | Yes—stepwise | 3.5 | Insulating | NR | No (semi-rigid PVDF) | Good — simple hot-press orientation; adaptable to film casting | [S10] |
| BNNS@$Al_2O_3$ / RTV-2 silicone rubber | Surface-modified hybrid filler; solution mixing + curing | 30 wt% total (BNNS:$Al_2O_3$ = 1:1) | No | 0.89 | Insulating ($1.74\times10^{-12}$, from $\rho_v$) | 0.84 MPa | Yes | Good — conventional mixing/curing; filler modification adds one step | [S11] |

| Material system / filler | Processing | Filler loading | Alignment? | κ⊥ (W $m^{-1}$ $K^{-1}$) | σ∥, in-plane (S $cm^{-1}$) | Modulus | Stretchable? | Scalability & reason | Ref. |
|---|---|---|---|---|---|---|---|---|---|
| BN–multilayer graphene / silicone rubber | Hybrid-filler preparation + rubber mixing/vulcanization | 10 wt% BN–MG hybrid (BN:MG = 1:10) | No | 0.69 | Insulating ($2.5\times10^{-12}$, from $\rho_v$) | NR | No (tensile stretchability not demonstrated) | Good — standard rubber compounding; hybrid preparation adds one step | [S12] |
| Vertically aligned BN / POE | Vertically oriented BN network + POE infiltration/hot pressing | 43.75 vol% BN | Yes—vertical network | 6.94 | Insulating ($\leq10^{-8}$) | 1.3 GPa | Yes | Limited — network orientation and infiltration are batch processes | [S13] |
| Chain-aligned BN / PDMS | BN chain alignment + PDMS casting/curing | 20 wt% BN | Yes—chain-like | 1.57 | Insulating ($\leq10^{-8}$) | 1.43 MPa | Yes | Good — casting/curing is scalable; alignment control is required | [S14] |
| 3D h-BN network / waterborne PU | In situ h-BN/WPU scaffold formation + hot pressing | 50 wt% h-BN | Yes—3D channels | 3.8 | Insulating ($\leq10^{-8}$; ) | 5.3 MPa | Yes | Limited — freeze-templating and network infiltration are multi-step | [S15] |
| h-BN@PDA / silicone rubber | PDA coating of h-BN + solution mixing/curing | 30 wt% h-BN@PDA | No | 0.95 | Insulating ($4.0\times10^{-12}$) | NR | No (tensile stretchability not demonstrated) | Good — conventional mixing/cure; PDA coating adds one step | [S16] |
| ▸ Carbon-based / hybrid | | | | | | | | | |
| CF / Fluorinated rubber | Electrostatic flocking (vertical) | 13.2 wt% CF | Yes—vertical CF | 23.3 | Conducting | 7.1 MPa | Yes | Limited — electrostatic flocking requires specialist setup; | [S17] |
| V-$Al_2O_3$@PCF / PVA | $O_2$ plasma + sol-gel $Al_2O_3$ + magnetic | 80 wt% $Al_2O_3$@PCF | Yes—magnetic | 22.5 | Semi-insulating (~$10^{-4}$ S $cm^{-1}$) | NR | No (rigid PVA film) | Limited — plasma treatment + sol-gel + magnetic alignment; 3-step process | [S18] |
| CF / Al2O3/Silicone rubber | Extrusion (shear flow alignment) | 88 wt% total (20 CF + 68 $Al_2O_3$) | Yes—shear/extrusion | 18.5 | Conducting | NR | No (flexible, compressible, not stretchable) | Good — extrusion is continuous and industrially compatible | [S19] |
| Ni-coated polyacrylonitrile CF/ Epoxy | Hot melt prepreg + Hand lay-up + press curing | 50 vol% Ni-coated CF | Yes—fiber/prepreg | 0.9 | Conducting (15 S $cm^{-1}$) | 70 GPa (rigid) | No (rigid epoxy) | Good — prepreg/layup/press-curing is industrially established. | [S20] |
| MWCNT/ Polyurethane acrylate | Ultrasonic-assisted emulsion polymerization | 1 wt% MWCNT | No | 0.25 | Conducting (0.9 S $cm^{-1}$) | — | No (flexible only) | Limited — CNT functionalization adds chemical complexity. | [S21] |
| CF + synthetic graphite + graphene/Polyamide 4.6 | Melt compounding | 20 wt% CF (κ⊥ formulation) | No directed alignment | 4.19 | Conducting (4 S $cm^{-1}$) | — | No (rigid polyamide) | Good — melt compounding; industrially scalable | [S22] |
| Exfoliated Graphite Nanoplatelets/Epoxy | Shear mixing + curing | 19.5 vol% GNP | No directed alignment | 11.2 | Conducting (4 S $cm^{-1}$) | — | No (rigid epoxy) | Good — uses commercial graphene powder and simple epoxy mixing/casting. | [S23] |

| Material system / filler | Processing | Filler loading | Alignment? | κ⊥ (W m⁻¹ K⁻¹) | σ∥, in-plane (S cm⁻¹) | Modulus | Stretchable? | Scalability & reason | Ref. |
|---|---|---|---|---|---|---|---|---|---|
| CNT/Epoxy | Mixing + curing | NR (simulation formulation) | No (simulation) | 1.269 | Conducting (43.7 S cm⁻¹; simulation input) | — | No (rigid epoxy) | Good — simple mixing; CNT dispersion can be challenging at scale | [S24] |
| NUL-graphene + CF / PVDF | Hot pressing | 13.8 wt% NUL-Gr/CF | Yes—vertical Gr/CF | 7.96 | Conducting (0.1 S cm⁻¹) | — | No (semi-rigid PVDF) | Good — hot pressing; scalable with industrial press | [S25] |
| GNF / PVDF | Hot pressing | 25 vol% GNF | Yes—flow reorientation | 10 | Conducting (1 S cm⁻¹) | 3 GPa | No rigid PVDF | Good — hot pressing; scalable | [S26] |
| Graphene Nanoplatelets/ PBAT/PLA | Melt compounding | 40 wt% GNP | No | 3.15 | Conducting (3.38 S cm⁻¹) | — | No (rigid biopolymer) | Good — melt compounding; industrially scalable | [S27] |
| PGO-coated wood / epoxy | Wood scaffold coating + epoxy infiltration/curing | NR (PGO-coated wood scaffold) | Yes—wood channels | 2.1 | Insulating (≈10⁻¹⁵) | 8.10 GPa | No (rigid scaffold) | Limited — bio-template coating and infiltration are multi-step | [S28] |
| PBO-grafted rGO aerogel + BN / PBO | Aerogel construction + PBO grafting + BN/PBO infiltration | 50 wt% hybrid filler | Yes—aerogel network | 4.81 | Conducting (21.1 S cm⁻¹) | NR | No (rigid PBO composite) | Limited — aerogel fabrication and grafting are multi-step | [S29] |
| Vertically oriented CF / PDMS | Vertically oriented CF preform + PDMS infiltration | 23.5 vol% CF | Yes—vertical CF | 12.26 | Conducting (≥10⁻⁵) | NR | No (tensile stretchability not demonstrated) | Good — preform/infiltration can scale; orientation must be controlled | [S30] |
| Flaky graphite + hollow glass / TPU | Close-packing design + compounding/hot pressing | 80 vol% total (30 graphite + 50 hollow glass) | Yes—packing-controlled | 6.55 | Conducting (≥10⁻⁵; qualitative) | NR | No (tensile stretchability not demonstrated) | Good — conventional compounding and hot pressing | [S31] |
| Vertically aligned graphene / silicone rubber | Coupled rotating magnetic + centrifugal fields | NR in accessible article metadata | Yes—vertical graphene | 12.76 | NR | NR | No (tensile stretchability NR) | Limited—coupled field and controlled centrifugation | [S32] |
| ‣ AlN / ceramic-based / hybrid | | | | | | | | | |
| AlN / PTFE | Mechanical rolling + IPA demulsification | 80 wt% AlN | Yes—rolling | 2.62 | Insulating | 0.73 MPa (compressive; <0.04 mm) | No (thin film, not stretchable) | Good — simple rolling process; compatible with roll-to-roll manufacturing | [S33] |
| AlN/rGO / PDMS foam | Freeze casting + PVA scaffold + PDMS infiltration | 50 wt% AlN/rGO | Yes—freeze-cast | 1.43 | Insulating | NR | No (PDMS foam, not stretchable) | Limited — freeze casting + scaffold removal + infiltration; batch size limited by furnace | [S34] |

| Material system / filler | Processing | Filler loading | Alignment? | κ⊥ (W m⁻¹ K⁻¹) | σ∥, in-plane (S cm⁻¹) | Modulus | Stretchable? | Scalability & reason | Ref. |
|---|---|---|---|---|---|---|---|---|---|
| BNNS + $Al_2O_3$ / PDMS | Sucrose-template + hot pressing | 35.7 wt% BNNS@$Al_2O_3$ | Yes—templated | 0.8 | Insulating | NR | Yes | Limited — hot pressing plus template removal may limit reproducibility | [S35] |
| AlN + BN / addition-cure silicone rubber | Hybrid-filler mixing + silicone molding/curing | 50 wt% total (AlN:BN = 4:1) | No | 0.554 | Insulating (≤$10^{-8}$; qualitative) | NR | No | Good — conventional mixing and addition curing | [S36] |
| SiC–BNNS / epoxy | SiC–BNNS heterostructure synthesis + epoxy mixing/curing | 20 wt% SiC–BNNS | No | 0.89 | Insulating ($4.31\times10^{-16}$) | NR | No (rigid epoxy) | Limited — heterostructure synthesis adds processing complexity | [S37] |
| BNNS@SiC nanowires / epoxy | BNNS-coated SiC nanowire synthesis + epoxy casting/curing | 20 wt% BNNS@SiC NW | No | 1.17 | Insulating ($1.93\times10^{-16}$, from $\rho_v$) | NR | No (rigid epoxy) | Limited — core–shell filler synthesis is multi-step | [S38] |
| $Al_2O_3$ + BNNS / epoxy | Hybrid-filler mixing + epoxy casting/curing | 10 wt% total (6 $Al_2O_3$ + 4 BNNS) | No | 0.47 | Insulating (≤$10^{-8}$; qualitative) | NR | No (rigid epoxy) | Good — standard mixing/casting; dispersion control is required | [S39] |
| Graded $Al_2O_3$ / silicone rubber | Controlled filler grading + silicone molding/curing | 96 wt% $Al_2O_3$ | No (graded, not aligned) | 8.4 | Insulating (<$10^{-14}$) | NR | No (compressible only) | Limited — graded architecture requires tight process control | [S40] |
| $Al_2O_3$ + VM-$SiO_2$ / silicone elastomer | Modified-silica preparation + vacuum mixing/molding/curing | 55 vol% $Al_2O_3$/VM-$SiO_2$ | No | 1.25 | Insulating (≤$10^{-8}$) | 3 MPa | No (compressible only) | Good — conventional elastomer processing; silica modification adds one step | [S41] |
| ▸ Metal-based / hybrid | | | | | | | | | |
| Ni / high-density polyethylene | Melt mixing | 30 vol% Ni | No | 1.99 | Conducting (83 S cm⁻¹) | 100 GPa (Young's; rigid) | No (rigid HDPE) | Good — standard melt mixing; industrially scalable | [S42] |
| Graphene nanoplatelets + aluminum/Epoxy | Shear mixing + casting | 50 vol% total (5.5 GNP + 44.5 Al) | No (alignment disrupted) | 2.62 | Conducting (0.03 S cm⁻¹) | — | No | Good — simple mixing + casting; scalable | [S43] |
| Cu nanowires / polyacrylate | Melt mixing | 0.9 vol% Cu NW | No | 2.46 | Conducting (0.04 S cm⁻¹) | 2.21 GPa | No (rigid polyacrylate) | Good — melt mixing; scalable but nanowire synthesis adds cost | [S44] |
| MXene + liquid metal / ANF–PVA | Mixed casting + alignment | 93.1 wt% total (LM + MXene; AP/MBLM13) | Yes—layered network | 1.5 | Conducting (42.65 S cm⁻¹) | 5.89 GPa | No (stiff aramid/PVA) | Limited — MXene synthesis + LM integration; multi-step, limited throughput | [S45] |
| Graphite + copper + silver-plated expanded graphite/ Epoxy | Mixing + curing | 10 phr total (≈9.1 wt%) | No | 1.7 | Conducting 0.05 (S cm-1) | — | No (rigid epoxy) | Good — standard mixing + curing; scalable | [S46] |

| Material system / filler | Processing | Filler loading | Alignment? | κ⊥ (W m⁻¹ K⁻¹) | σ∥, in-plane (S cm⁻¹) | Modulus | Stretchable? | Scalability & reason | Ref. |
|---|---|---|---|---|---|---|---|---|---|
| Ni-coated Cu NWs/Epoxy | Coating + magnetic alignment | 9 vol% Ni-coated Cu NW | Yes—magnetic | 2.9 | Insulating $5.59\times10^{-16}$ ( $S\ cm^{-1}$) | 3.5 GPa | No (rigid epoxy) | Limited — Ni coating adds synthesis complexity. | [S47] |
| Layered LMPa / epoxy | Layer-by-layer low-melting-point alloy/epoxy casting and cure | 20 vol% LMPA | Yes—layered | 1.23 | Semi-insulating ($5.0\times10^{-8}$ $S\ cm^{-1}$) | NR | No (rigid epoxy) | Limited — directional multilayer structuring adds processing steps | [S48] |
| Cu-coated carbon felt / silicone rubber | Cu plating on carbon felt + silicone infiltration/curing | NR (Cu-coated CFelt scaffold) | Yes—3D felt network | 7.323 | Conducting (≥$10^{-5}$) | NR | No | Limited — metal coating and felt infiltration are multi-step | [S49] |
| ▸ Liquid-metal-based / hybrid | | | | | | | | | |
| Ga LM foam + Ga microparticles/ Ecoflex | Sugar template + vacuum infiltration; | 45 vol% total (LM skeleton + Ga particles) | Yes—strain-induced | 4.8 (unstrained); 17.6 at 400% strain | Conducting ($1.1\times10^{4}$ under strain) | 0.3 MPa | Yes | Limited — scalable materials, but sacrificial templating and infiltration add complexity. | [S50] |
| Cu–EGaIn / TPE | Hot compression | 90 vol% Cu–EGaIn | No—isotropic network | 32.71 (compressed) | Conducting $1.18\times10^{4}$ ($S\ cm^{-1}$) | NR | Yes | Good — particle coating + hot compression; no template | [S51] |
| 3D EGaIn network / PDMS foam | Porous PDMS foam + vacuum LM infiltration; no alignment | NR (infiltrated 3D LM network) | No—3D isotropic | 4.05 at 60% compression | Conducting (14× increase at 60% compression) | NR | No | Good — simple foam fabrication and LM infiltration | [S52] |
| Ga-based LM / porous SEPS | Salt-derived SEPS scaffold; negative-pressure LM injection | NR (LM-filled porous SEPS) | No—isotropic foam | 8.76 (unstrained) | Conducting | ≤ 45 kPa | Yes | Limited porous-template fabrication + negative-pressure injection | [S53] |
| Ga-based LM / elastomer foam | Reformed sugar template; vacuum LM infusion | NR (LM-filled foam) | No—isotropic foam | 0.48 | Conducting 7.2 ($S\ cm^{-1}$ ) | NR | Yes | Limited — template and LM extraction/recovery | [S54] |
| 3D Gyroid Ga skeleton / Ecoflex | FDM printed ABS sacrificial mold; LM casting and infiltration | NR (cast 3D Ga gyroid) | Yes—architected gyroid | 21.2 (unstrained); 27.1 at 300% strain | Conducting $2.07\times10^{4}$ ( $S\ cm^{-1}$) | <100 kPa | Yes | Limited — 3D-printed sacrificial molds, LM casting, and mold dissolution | [S55] |
| Galinstan gauze network / Ecoflex | Gauze-templated LM network; vertical stacking; gel infiltration | 88 wt% LM | Yes—stacked gauze | 6.08 | Conducting | 0.56 MPa | No (flexible, compressible) | Good — gauze coating/stacking and silicone infiltration | [S56] |
| Thiomalic-acid-modified LM / cellulose aerogel | LM functionalization; vacuum backfill into through-thickness channels | NR (LM-filled cellulose channels) | Yes—cellular channels | 7.421 | Conducting 0.256 $S\ cm^{-1}$ | NR | No | Limited — chemical modification + LM functionalization/backfill | [S57] |

| Material system / filler | Processing | Filler loading | Alignment? | κ⊥ ($W\ m^{-1}\ K^{-1}$) | σ∥, in-plane ($S\ cm^{-1}$) | Modulus | Stretchable? | Scalability & reason | Ref. |
|---|---|---|---|---|---|---|---|---|---|
| Compressed graphene foam / LM/PDMS | Graphene foam compression + LM dispersion and PDMS infiltration; vertical alignment | ≈24 wt% total (4 c-GF + 20 LM) | Yes—vertical c-GF | 4.96 | Conducting | E NR | No (flexible only) | Limited — graphene-foam fabrication , compression, infiltration | [S58] |
| EGaIn + $Ti_3C_2T_x$ MXene / PEBA elastomer | Controlled stirring oxidation; MXene bridging; ssolution casting | NR (LM/MXene formulation not in metadata) | No—isotropic | 6.34 (isotropic) | Conducting | NR | No (flexible only) | Good — simple stirring and ssolution casting; no external alignment | [S59] |
| Galinstan droplets / PDMS | High-speed mechanical mixing; casting/spinning /laminating; thermal cure | 92.5 wt% (66.1 vol%) Galinstan | No—random droplets | 2.2 (92.5 wt% formulation) | Insulating (<$2\times10^{-9}$ S $cm^{-1}$) | 0.72 MPa | Yes | Good — simple mixing/casting | [S60] |
| EGaIn microdroplets / Ecoflex | Shear mixing + casting/curing; tensile strain elongates droplets in situ | 50 vol% EGaIn | Yes—strain-induced | 4.7 ± 0.2 (unstrained); 9.8 at 400% strain | Insulating | 100 kPa | Yes | Good — direct mixing and casting; no rigid alignment scaffold | [S61] |
| EGaIn droplets / Ecoflex | Vacuum centrifugal mixing; casting + oven cure; droplets left discrete | 80 vol% EGaIn | No—random droplets | 4.8 (isotropic/spherical droplets) | Insulating before activation | NR | Yes | Good — one-pot mixing/casting; stable after ambient/freshwater aging | [S62] |
| Ni-decorated Ga–In droplets / PVA | Ni coating + magnetic-field-induced one-step film molding/alignment | 40 wt% LM@Ni (LM:Ni = 5:2) | Yes—magnetic | 5.34 | Insulating $2.56\times10^{-10}$(S cm-1) | NR | No (flexible only) | Good — solution casting and magnetic alignment; recyclable PVA matrix | [S63] |
| LM-interconnected CNT/ANF film | Vacuum filtration + mechanical activation; layered CNT network | 90 wt% total (LM:CNT = 6:3) | Yes—layered CNT network | 1.1 | NR | NR | No (flexible only) | Good — vacuum filtration; scalable to sheet size | [S64] |

Notes: Reported filler loading corresponds to the formulation used for the tabulated κ⊥ where traceable; Source units (wt%, vol%, or phr) are retained; NR = not reported; 'NR (architecture)' denotes a network/scaffold paper that did not provide a defensible bulk wt% or vol% for direct comparison; Category header and row colors identify the filler family; Stretchability cells use blue for Yes and gray for No or not reported; Alignment includes deliberately induced field-, flow-, freeze-, channel-, strain-, or architecture-directed orientation; 'No' means no directed filler orientation; κ⊥ = through-plane thermal conductivity; σ∥ = in-plane electrical conductivity; Conductivity bands: insulating ≤$10^{-8}$, semi-insulating $10^{-8}$–$10^{-5}$, conducting >$10^{-5}$ S $cm^{-1}$.

**Video S1**. Side-view optical recording of rapid and repeatable magnetic-field-induced alignment of dry Ni particles in air.

**Video S2**. Top-view optical recording of Ecoflex infiltration into a preformed magnetically aligned Ni scaffold.

**Video S3**. Side-view optical recording Ecoflex infiltration into a preformed magnetically aligned Ni scaffold.

**Video S4.** Animated μ-CT of a composite fabricated using the AB method. Purple lines indicate the reconstruction bounding box (8.1 x 8.1 x 2.8 $mm^3$).

**Video S5.** Infrared thermal imaging of surface temperature evolution during tensile stretching of the composite.

**Video S6.** Thermal IR imaging comparing B, BA, AB, and ABBA composites under identical thermostatic hot-stage conditions.

**Video S7.** Heat dissipation of pristine Ecoflex, commercial thermal pad, and ABBA composite.

**Video S8.** Demonstration comparing the surface-temperature of pristine Ecoflex and ABBA substrates applied to gloved fingers